\def\arxiv{} 

\documentclass{article}

\usepackage{arxiv}

\usepackage[utf8]{inputenc} 
\usepackage[T1]{fontenc}    
\usepackage{hyperref}       
\usepackage{url}            
\usepackage{booktabs}       
\usepackage{amsfonts}       
\usepackage{nicefrac}       
\usepackage{microtype}      
\usepackage{lipsum}

\usepackage{xcolor}

\usepackage{caption}
\usepackage{subcaption}
\usepackage{textcomp}
\usepackage{siunitx}
\usepackage{longtable}
\usepackage[colorinlistoftodos]{todonotes}
\usepackage{pmboxdraw}
\usepackage{hyperref}
\usepackage{amsmath}
\usepackage[utf8]{inputenc}
\usepackage{spverbatim}
\usepackage{supertabular}
\usepackage{multicol}
\usepackage{booktabs}

\title{PadChest: A large chest x-ray image dataset with multi-label annotated reports}

\author{
  Aurelia Bustos \\
Department of Software and Computing Systems\\ 
University Institute for Computing Research\\
University of Alicante, Spain \\
  \texttt{aurelia@medbravo.org} \\
   \And
  Antonio Pertusa \\
Department of Software and Computing Systems\\ 
University Institute for Computing Research\\
University of Alicante, Spain \\
  \texttt{pertusa@ua.es} \\
    \And Jose-Maria Salinas \\ 
    Department of Health Informatics\\ 
    Hospital San Juan de Alicante, Spain \\
    \texttt{salinas\_josser@gva.es} \\
    \And 
    Maria de~la~Iglesia-Vayá \\
    Centre of Excellence in Biomedical Image \\
    Regional Ministry of Health, Valencia, Spain \\
    \texttt{delaiglesia\_mar@gva.es} \\
}

\begin{document}
\maketitle

\begin{abstract}
We present a labeled large-scale, high resolution chest x-ray dataset for the automated exploration of medical images along with their associated reports. This dataset includes more than 160,000 images obtained from 67,000 patients that were interpreted and reported by radiologists at Hospital San Juan Hospital (Spain) from 2009 to 2017, covering six different position views and additional information on image acquisition and patient demography. The reports were labeled with 174 different radiographic findings, 19 differential diagnoses and 104 anatomic locations organized as a hierarchical taxonomy and mapped onto standard Unified Medical Language System (UMLS) terminology. Of these reports, 27\% were manually annotated by trained physicians and the remaining set was labeled using a supervised method based on a recurrent neural network with attention mechanisms. The labels generated were then validated in an independent test set achieving a 0.93 Micro-F1 score. To the best of our knowledge, this is one of the largest public chest x-ray database suitable for training supervised models concerning radiographs, and the first to contain radiographic reports in Spanish. The PadChest dataset can be downloaded from \url{http://bimcv.cipf.es/bimcv-projects/padchest/}.
\end{abstract}


\graphicspath{{figs/}}

\section{Introduction}
\label{sec:intro}

Chest x-rays are essential for both the screening and the diagnosis of pulmonary, cardiovascular, bone and other thoracic disorders. The adequate interpretation of the radiographic findings requires medical training acquired over many years, with radiologists being the most qualified professionals in this fields. Due to increasing workload pressures, many radiologists today have to read more than 100 x-ray studies daily. Therefore, automated tools trained to predict the risk of specific abnormalities given a particular x-ray image have the potential to support the reading workflow of the radiologist. Those tools could be used to enhance the confidence of the radiologist or prioritize the reading list where critical cases would be read first. Decision support systems (DSS) designed as tools to assist in the clinical interpretation of chest x-rays would therefore fulfill an unmet need.

Deep learning techniques are currently obtaining promising results and perform extremely well in a variety of sophisticated tasks \citep{naturedl}, especially those related to computer vision, often equaling or exceeding human performance \citep{Goodfellow2016DeepLearning}. The application of deep neural networks to medical imaging and chest radiographs in particular, has become a growing area of research in recent years \citep{qin2018computer}. For instance, \cite{wang2017chestx}  trained a Convolutional Neural Network (CNN) to classify and localize 8 pathologies using the chest x-ray database (\textit{ChestX-Ray8}) which comprised 108,948 frontal-view x-ray images of 32,717 different patients. Using the same repository, \cite{rajpurkar2017chexnet} extended the annotations to 14 different pathologies (\textit{ChestX-Ray14}) and designed a model with a deeper CNN architecture to classify images as 14 pathological entities. This method was reported to obtain greater diagnostic efficiency in the detection of pneumonias when compared to that of radiologists. \citep{guan2018diagnose} proposed the attention guided CNN to help combine global and local information in order to improve recognition performance. Chest-XRay14 was also employed by \cite{wang2018tienet} to design a network architecture that combines text and image with attention mechanisms capable of generating text that describes the image, while \cite{jing2017automatic} introduced a hierarchical model of Recurrent Neural Networks (RNN) with which to generate long paragraphs from the images and obtain semantically linked phrases for the same purpose.  

Despite claims that they achieve and/or surpass physician-level performance, current deep learning models for the classification of pathologies using chest x-rays are proving not to be generalizable across institutions and not yet ready for adoption in real-world clinical settings \citep{zech2018variable}. Moreover, warnings of potential unintended consequences of their use are discussed by \cite{cabitza2017unintended}. 

It is unclear how to extend the significant success in computer vision using deep neural networks to the medical domain \citep{shin2017book}. Open questions concerning medical radiology datasets that still need to be addressed are: 
\begin{itemize}
    \item How to annotate the huge amount of medical images required by deep learning models \citep{shin2017book} and meet the required quality. Large-scale crowd-sourced hand-annotation which has proved successful in the general domain, e.g. \textit{ImageNet} \citep{deng2009imagenet}, is not feasible because of the medical expertise required to carry it. This, is compounded by the fact that the semantic interpretation and extraction of medical knowledge from the corpora of medical text in unstructured natural language remains a challenge \citep{Weng_2010} and is an area of active research \citep{bustos2018learning}.  
    \item The clinically relevant image labels that need to be defined and which criteria should be followed to annotate them \citep{shin2017book}. 
    \item How to deal with uncertainties in radiology texts. Medical data is characterized by uncertainty and incompleteness and machine learning decision support systems (ML-DSS) need to adapt to input data reflecting the nature of medical information, rather than at imposing an idea of data accuracy and completeness that does not fit patient records and medical registries, for which data quality is far from optimal. In this respect, \cite{cabitza2017unintended} advises caution as regards the unintended consequences of adopting ML-DSS that demise context and ignore the fact that observer variability obeys not only interpretative deficiencies but also intrinsic variability in the observed phenomena. 
    \item How to effectively control for potential confounding factors, such as the presence of tubes, catheters, the quality of the image as assessed by radiologists, patient position, etc. and unbalanced entity prevalence, which models learn to exploit as predictive features to the detriment of clinical radiological patterns. 
\end{itemize}

There are a number of publicly available chest x-ray datasets that can be used for image classification and retrieval tasks. The National Institute of Health of America (NIH) repository \citep{wang2017chestx} contains 112,120 frontal-view chest x-rays, corresponding to 30,805 different patients, and multi-labeled with 14 different thoracic diseases \citep{rajpurkar2017chexnet}. The Korean Institute of Tuberculosis dataset \citep{ryoo2014activities} consists of 10,848 DICOMs, of which 3,828 show tuberculosis abnormalities. The Indiana University dataset \citep{demner2015preparing} comprises 7,470 frontal and lateral chest x-ray images, corresponding to 3,955 radiology reports with disease annotations, such as cardiac hypertrophy, pulmonary edema, opacity, or pleural effusion. The JSRT dataset \citep{shiraishi2000development} consists of 247 x-rays of which 154 have been labeled for lung nodules (100 malignant). \cite{van2006segmentation} also provides masks of the lung area for the evaluation of segmentation performance. The Shenzhen dataset \citep{jaeger2014two} has a total of 662 images belonging to two categories (normal and tuberculosis). 

With regard to the annotation methods applied, \cite{shin2016interleaved,shin2017book} used a large dataset that included 780,000 documents from 216,000 images comprising CT, MR, PET and other image modalities, and automatically mined categorical semantic labels using a non-parametric topic modeling method. The resulting annotations were judged to be non-specific. In order to increase disease specificity, the authors have matched frequent pathology types using a disease ontology and semantics. This method, however, assigned specific disease labels to around only 10\% of the dataset. 
In \textit{ChestX-Ray-8}, \cite{wang2017chestx} used MetaMap \citep{aronson2010overview}, DNorm \citep{leaman2015challenges} and custom negation rules applied to a syntactic parser in order to label the presence or otherwise of 8 entities, which was further expanded to 14 entities in 125,000 images (\textit{Chest-XRay14}), also by applying analogous methods. They validated the image labeling procedure against 3,800 annotated reports of x-ray images from OpenI - Indiana DB- \citep{demner2015preparing}. 

Outstanding questions, such as those mentioned above, still remain unaddressed by the medical image datasets currently available. This problem is compounded by the fact that the, medical annotations used as a ground-truth are reduced to a small number of entities and, on many occasions, owing to the inherent limitations of the natural language processing (NLP) techniques applied to automate their extraction, these annotations contain omissions, inconsistencies and are not validated by physicians. 

In this work, we propose a dataset called PadChest (PAthology Detection in Chest radiographs) which is, to the best of our knowledge, one of the largest and most exhaustively labeled public chest x-ray dataset and the only one to contain the source report excerpts, which are written in Spanish. The labels of the images are mapped onto standard Unified Medical Language System (UMLS) terminology and can therefore be used regardless of the language. They cover the full spectrum of thoracic entities, contrasting with the much more reduced number of entities annotated in previous datasets. Moreover, entities are localized with anatomical labels, images are stored in high resolution, and extensive information is provided, including the patient's demography, type of projection and acquisition parameters for the imaging study, among others. 

In contrast to previous approaches that relied solely on automatic annotation tools, in this work the labeling required for the baseline ground-truth was carried out manually by trained physicians. While tools to annotate medical text in Spanish are less extensively developed and tested than in English, the most compelling reason for opting for an annotation carried out manually by physicians was to maximize the reliability of the baseline ground-truth. 

Using a laboriously developed ground-truth as a baseline, we propose a supervised annotation method based on deep neural networks and labeling resolution rules in order to extract labels from the remaining reports (73\% of the samples). This pipeline is designed for large-scale exhaustive annotation of Spanish chest x-ray reports and aims to overcome some common NLP challenges in the medical domain. For instance, the proposed method deals with anaphora resolution in which findings are mentioned in different sentences or even in different studies for the same patient and whose interpretation depends upon another expression in that context. This method also deals with co-reference resolution, in which words for locations are related to the entities that they are describing, and also with hedging statements, in which uncertainty, probability and indirect expressions are to be learned by deep neural models.

Moreover, in an effort to incorporate uncertainty into the ground-truth, we differentiate radiographic findings from differential diagnoses, acknowledging the existence of two distinct image classification problems. Radiographic findings are completely observable in the images and we, therefore hypothesize that state-of-the-art deep learning models could achieve reliable results for this task. However, differential diagnoses are characterized by intrinsic uncertainty and a highly multidimensional context which is not included in the image. The expectation that automated methods could obtain similar results for differential diagnoses to those of doctors using only x-rays as input may, therefore, be unrealistic. This is because diagnoses such as ``pneumonia" are not based solely on the x-ray image and depend to a great extent on external factors, such as laboratory tests, physical examinations, symptoms and signs, temporal clues and clinical judgment, which may additionally vary among practitioners.

The main objectives of the proposed dataset are: 

\begin{itemize}
\item
To broaden the scope of radiographic diagnoses and findings that trainable models can learn from chest x-ray images, including respiratory, cardiac, infectious, neoplastic, bone and soft tissue diagnoses, the positions of nasogastric tubes, endotracheal tubes, central venous catheters, electrical
devices, etc.
\item To increase the disease detection capabilities of trained models by providing the localizations of the entities as regions of interest.
\item To make available all relevant metadata, along with all the entities described by radiologists, so as to help control potential confounders in predictive models. For example projection types that dictate the adequate interpretation of radiographic findings and patient positioning have been identified as confounding factors in deep learning models. Indeed, the first task for radiologists is to identify the type of projection before reading the x-ray in order to correctly interpret the findings. 
\item To help advance automatic clinical text annotations in Spanish. To the best of our knowledge, this is the first publicly available chest x-ray dataset containing excerpts from radiology reports in this language. Although the excerpts from the report are provided in Spanish, the labels are mapped onto biomedical vocabulary unique identifier (CUIs) codes, thus making the dataset usable regardless of the language. 
\item Training text models with the proposed deep learning methods would make it possible to help automatically label other large-scale x-ray repositories in Spanish speaking health institutions, using the source code provided. 
\end{itemize}

The remainder of the paper is as follows. Section \ref{sec:methods} describes the methodology employed to build the dataset, including the manually labeled subset and the automatic labeling of the remaining reports using deep neural networks. Section \ref{sec:evaluation} shows the evaluation results of the automatic-labeling methods described in the previous section, while Section \ref{sec:dataset} details the statistics of the dataset. Finally, Section \ref{sec:conclusions} addresses the discussion, conclusions and future work.

\section{Material and methods}
\label{sec:methods}

\begin{figure*}
\centering
  \includegraphics[width=1\linewidth]{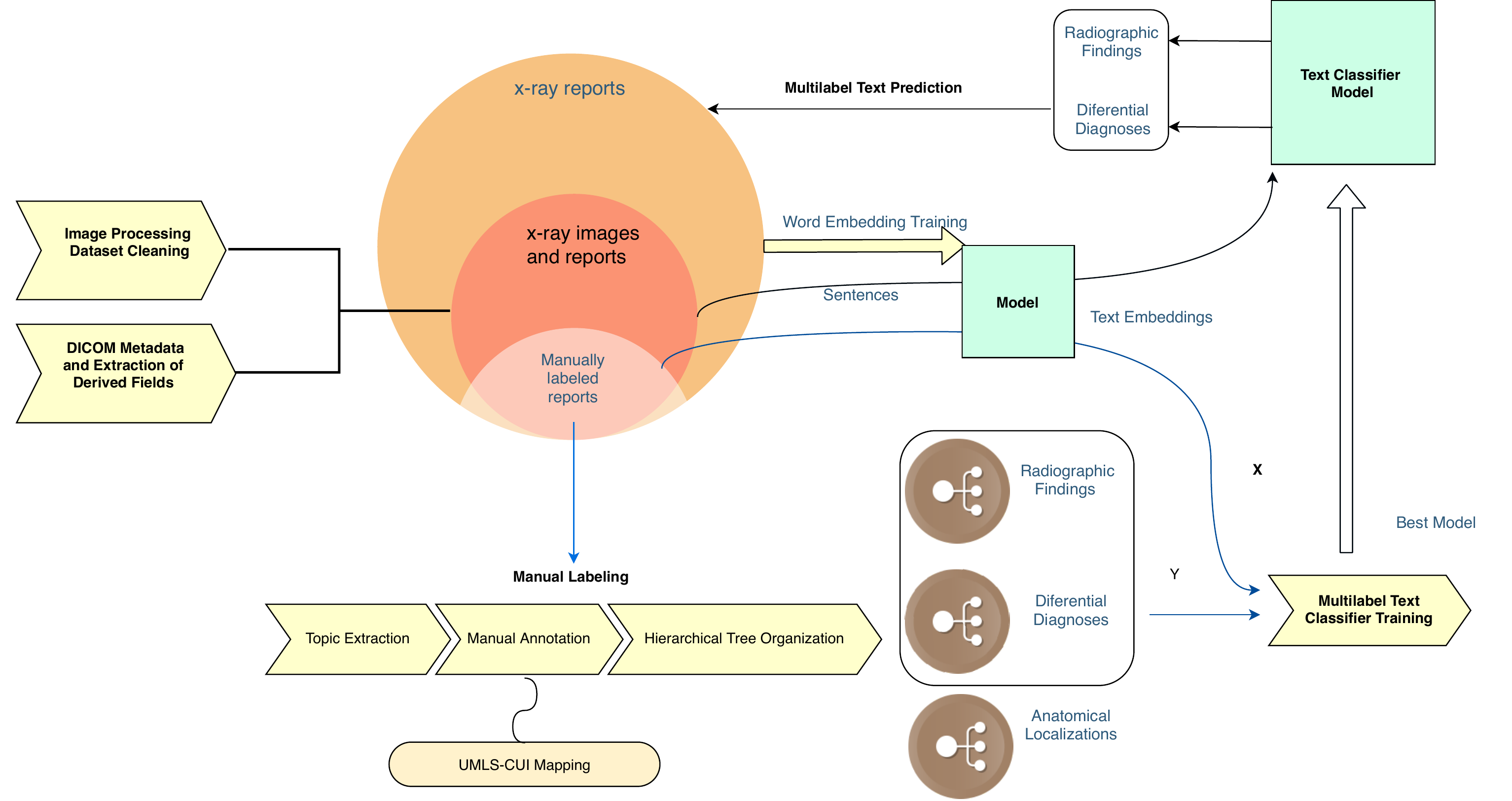}
  \caption{Dataset Building Pipeline: The PadChest dataset consists of 206,222 x-ray reports (large circle), 109,931 (middle circle) of which had their 160,868 corresponding images in DICOM format and were acquired from the years 2009 to 2017. A subsample of 27,593 reports (pale oval region) from the years 2014 to 2017 were manually labeled and further used to train a multi-label text classifier based on neural networks in order to annotate the remaining dataset with radiographic findings and differential diagnoses. Anatomical localizations were extracted using regular expressions. Different diagnoses, radiographic findings and anatomical localizations were mapped onto the NLM Unified Medical Language System (UMLS) controlled biomedical vocabulary unique identifiers (CUIs) and organized into semantic concept trees. 
  }
  \label{fig:Pipeline}
\end{figure*}

The PadChest dataset consists of all the available chest-x rays that had been interpreted and reported by 18 radiologists at the \textit{Hospital Universitario de San Juan, Alicante} (Spain) from Jan 2009 to Dec 2017, amounting to 109,931 studies and 168,861 different images, as shown in Tab. \ref{tab:datasetSize}. 

This project was approved by the institutional research committee, and both the images and the associated reports were made anonymous and de-identified by the Medical Image Bank of the Valencian Community at the Department of Universal Health and Public Health Services (BIMCV-CSUSP) and the Health Informatics Department at San Juan Hospital.

The PadChest dataset can be downloaded from the repository of the medical imaging bank (BIMCV - PADCHEST\footnote{ \url{http://bimcv.cipf.es/bimcv-projects/padchest/}}), enabled by the Medical Image Bank of the Valencian Community (BIMCV). The BIMCV has launched various projects regarding population medical images, whose objective is to develop and implement an infrastructure with a massive storage capacity following the R$\&$D Cloud CEIB architecture \citep{Salinas12}. One of the missions of this bank is to promote the publication of scientific knowledge as open data by its affiliated health institutions. 

PadChest contains image files adding up to 1 TB, a csv file with 33 fields for each study and an instruction file containing field descriptions, examples and search information for efficient image retrieval. An example of a dataset study with two projections can be found in \ref{app:DatasetExample}, along with its associated labels and additional information fields.

The methodology employed to build PadChest comprises the following main steps: 
\begin{itemize}
\item Pre-processing of the images and DICOM metadata extraction.
\item Pre-processing of the reports.
\item Manual medical annotations using a hierarchical taxonomy of radiographic findings, differential diagnoses and their anatomic locations.
\item Automatic labeling of the remaining studies.
\end{itemize}

\subsection{Pre-processing of the images and DICOM metadata extraction}

The images were processed by rescaling the dynamic range using the DICOM window width and center, when available. They were not resized to avoid the loss of resolution. 

Some images were initially excluded when: 1) DICOM pixel data were not readable; 2) the photometric interpretation was missing or \texttt{MONOCHROME1}; 
3) the modality was missing or it was Computed Tomography (CT) rather than x-rays;
4) the projections were lateral horizontal, oblique or trans-thoracic; 5) the anatomic image protocol was other than that of the chest ( e.g humerus, abdomen, ..), and 6) the study report was missing or the radiography interpretation was not identifiable. 

In addition to the standard DICOM metadata, the information on image projection and radiographic positioning was extracted. These data were found in different DICOM fields in non-structured free text (Position View, Patient Orientation, Series Description and Code Meaning). The projections principally identified after excluding obliques were antero-posterior (AP), postero-anterior (PA) and lateral (L). The different body positions were erect, either standing or sitting, decubitus or lying down, supine or lying on back, lateral decubitus right lateral and left lateral. In addition, different protocols were identified based on different clinical scenarios: standard standing PA and L, mobile x-ray (for patients unable to stand) in either AP erect in bed or AP supine, pediatric protocol for patients up to 3 years old, lordotic views, ribs and sternum modality views.

\begin{figure*}
\centering
\begin{subfigure}{.19\textwidth}
  \centering
  \includegraphics[width=.9\textwidth]{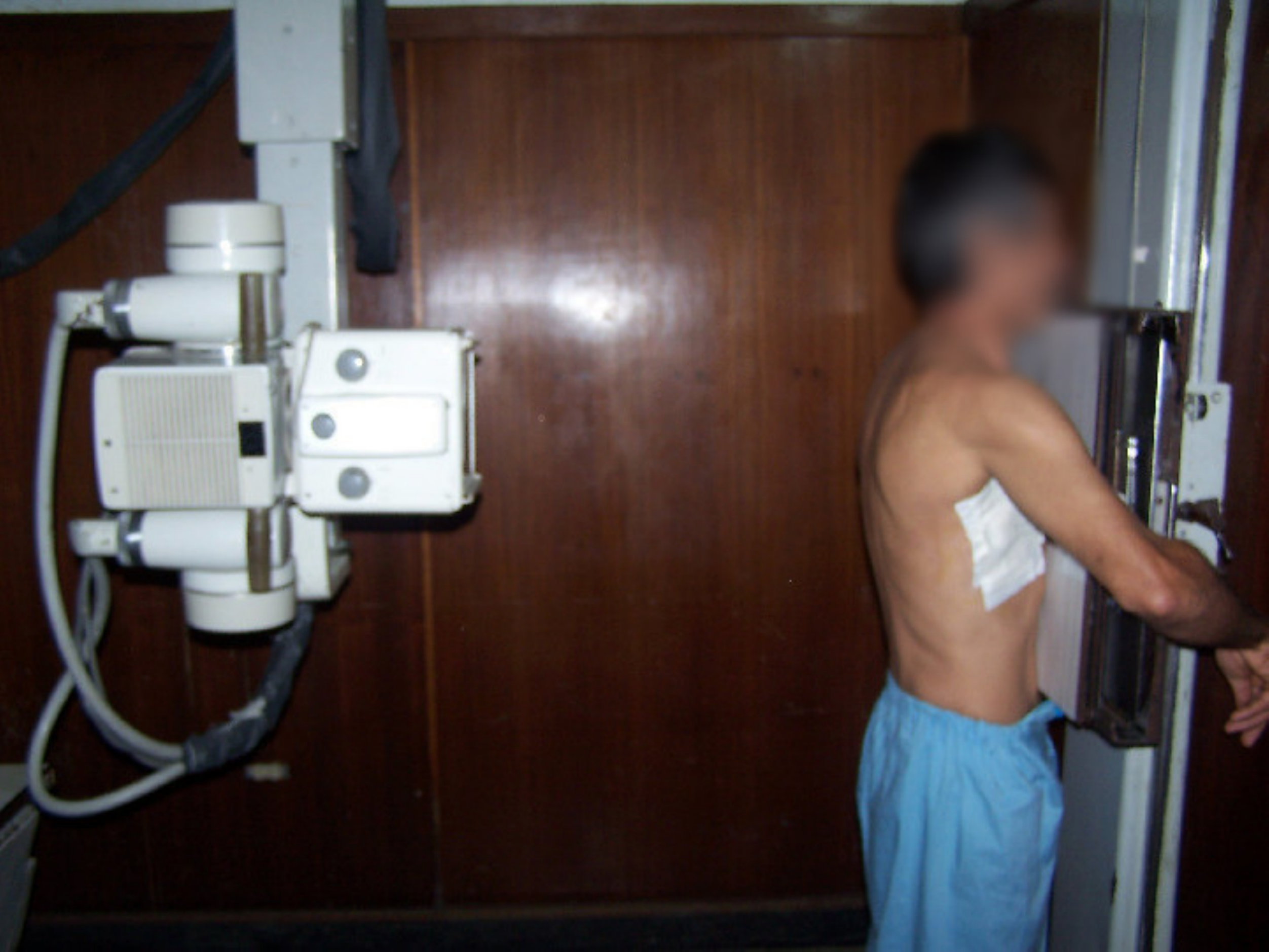}
  \caption{P-A}
  \label{fig:PAMan}
\end{subfigure}%
\begin{subfigure}{.19\textwidth}
  \centering
  \includegraphics[width=.9\textwidth]{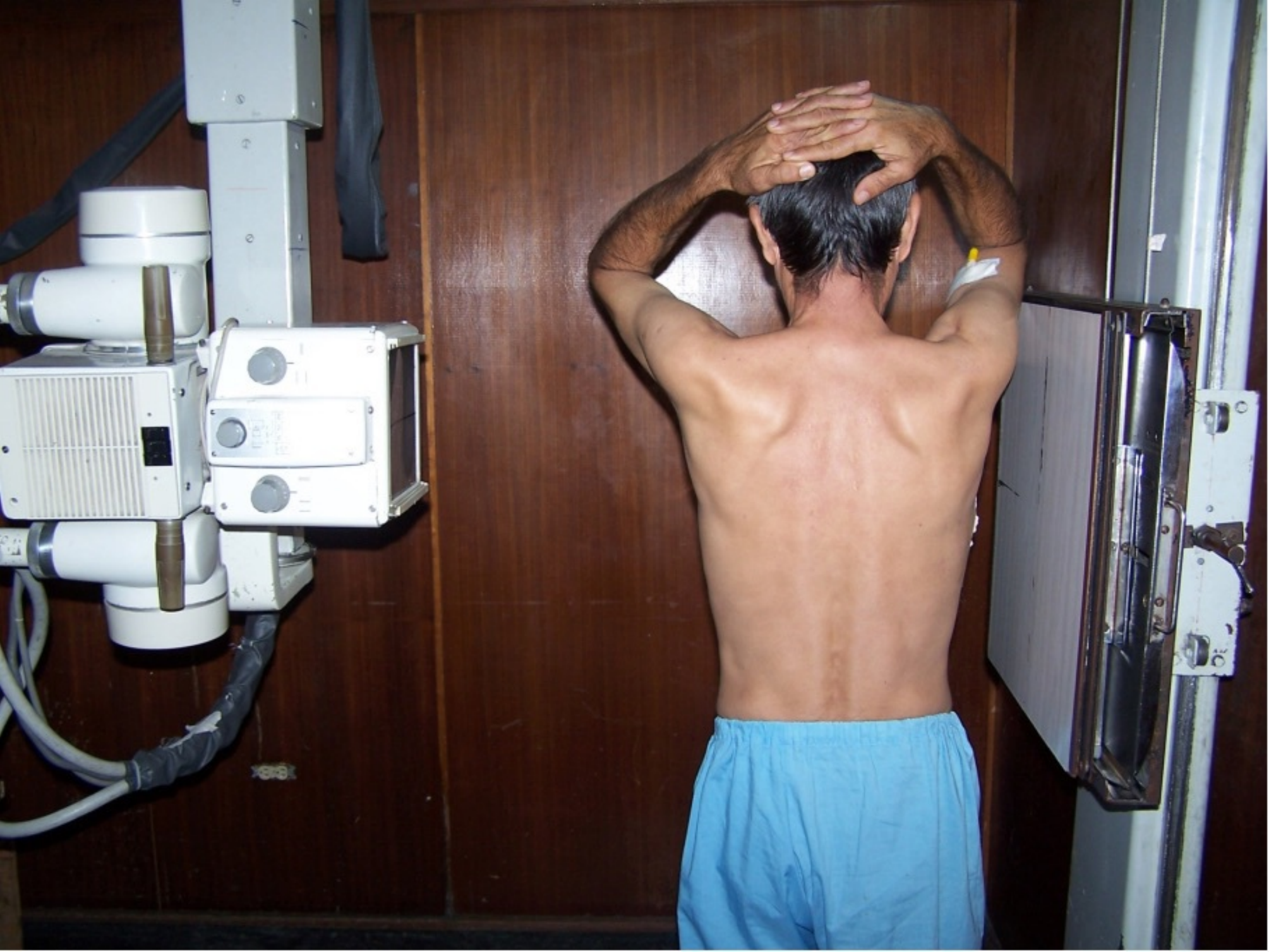}
  \caption{Lateral}
  \label{fig:LateralMan}
\end{subfigure}
\begin{subfigure}{.19\textwidth}
  \centering
  \includegraphics[width=.9\textwidth]{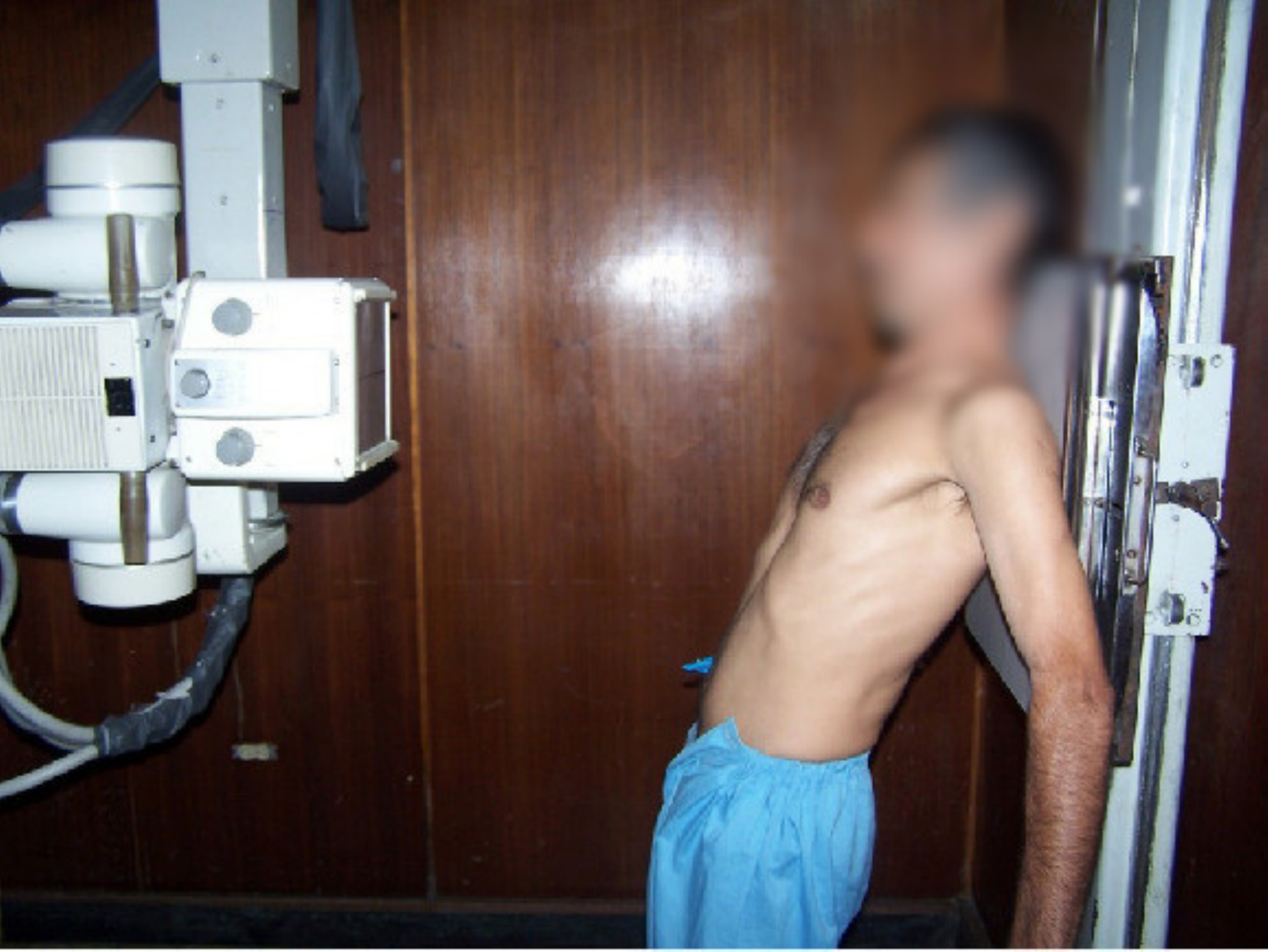}
  \caption{Lordotic}
  \label{fig:LordoticMan}
\end{subfigure}
\begin{subfigure}{.19\textwidth}
  \centering
  \includegraphics[width=.9\textwidth]{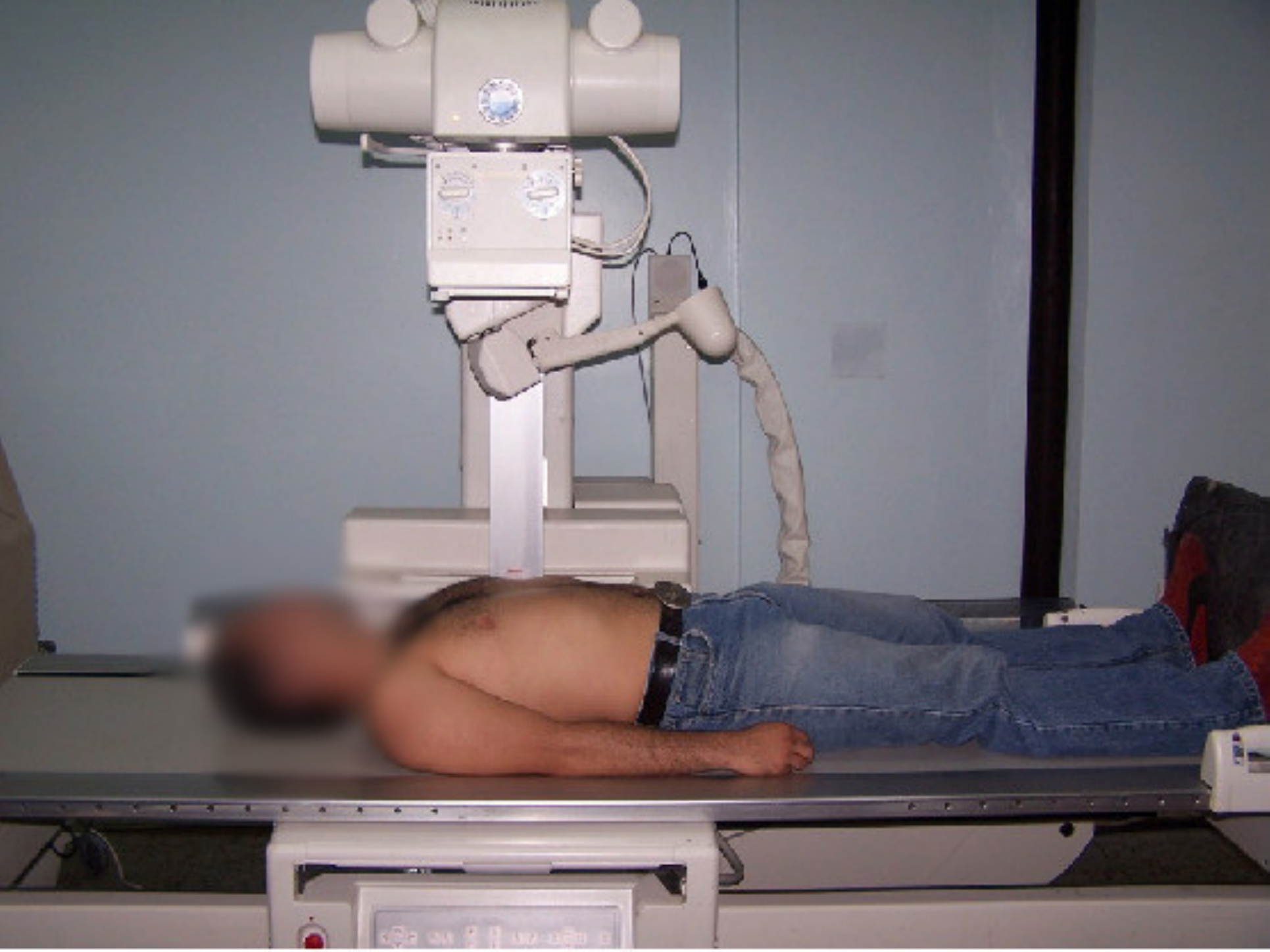}
  \caption{A-P supine}
  \label{fig:DorsalDecMan}
\end{subfigure}
\begin{subfigure}{.19\textwidth}
  \centering
  \includegraphics[width=.9\textwidth]{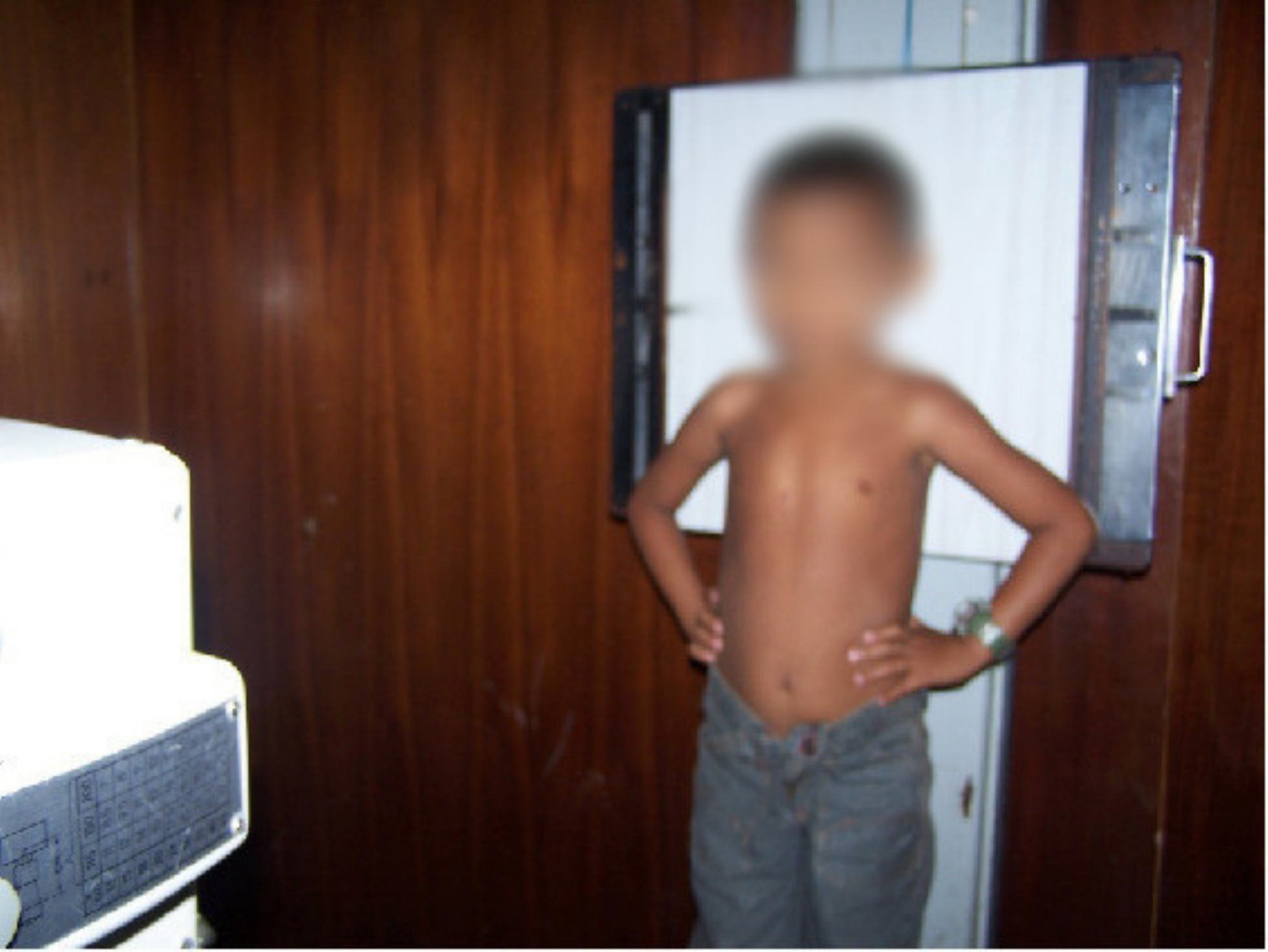}
  \caption{A-P}
  \label{fig:APMan}
\end{subfigure}
\\
\centering
\begin{subfigure}{.19\textwidth}
  \centering
  \includegraphics[width=.9\linewidth]{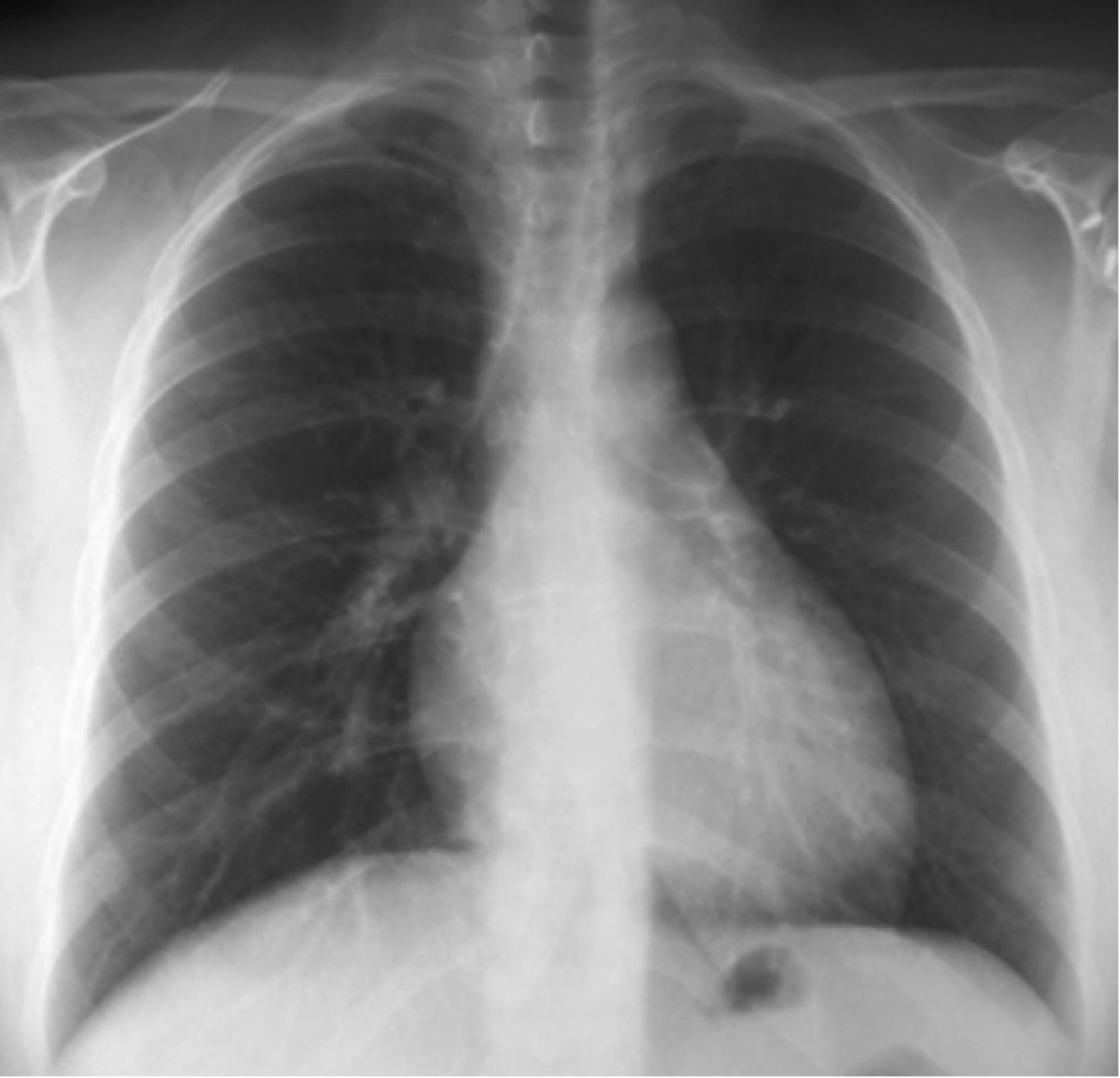}
  \caption{P-A}
  \label{fig:PA}
\end{subfigure}%
\begin{subfigure}{.19\textwidth}
  \centering
  \includegraphics[width=.9\linewidth]{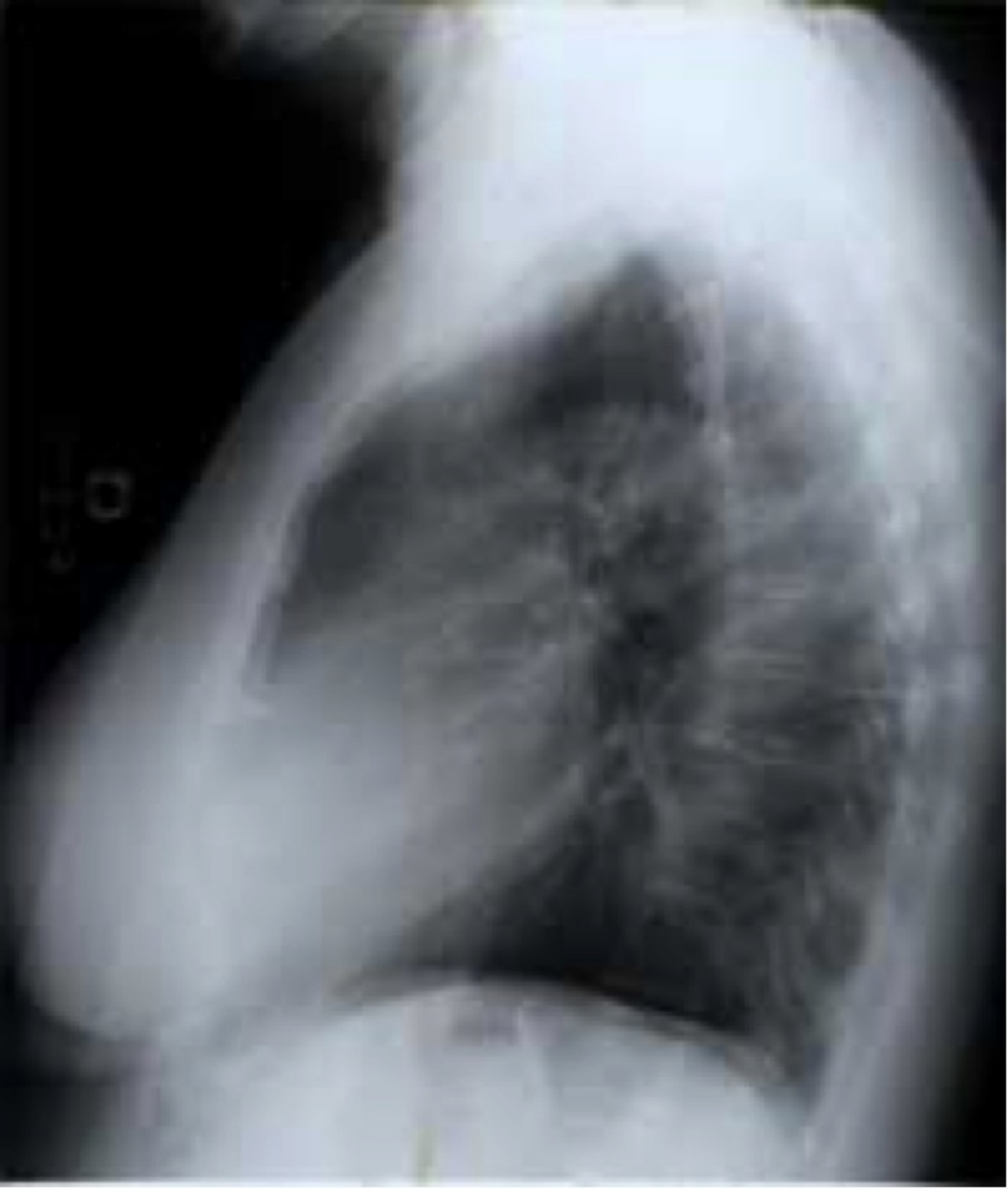}
  \caption{Lateral}
  \label{fig:Lateral}
\end{subfigure}
\begin{subfigure}{.19\textwidth}
  \centering
  \includegraphics[width=.9\linewidth]{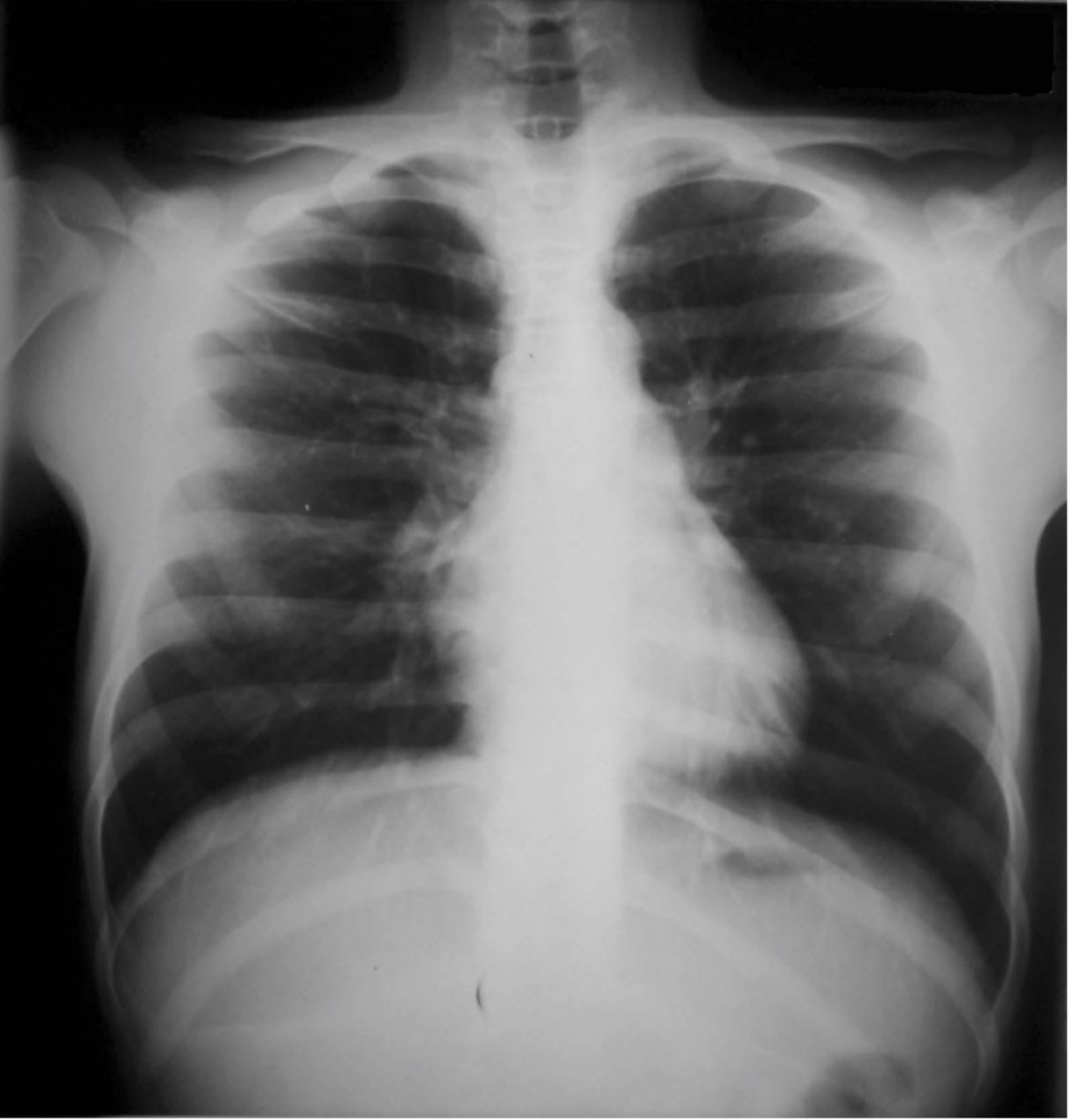}
  \caption{Lordotic}
  \label{fig:Lordotic}
\end{subfigure}
\begin{subfigure}{.19\textwidth}
  \centering
  \includegraphics[width=.9\linewidth]{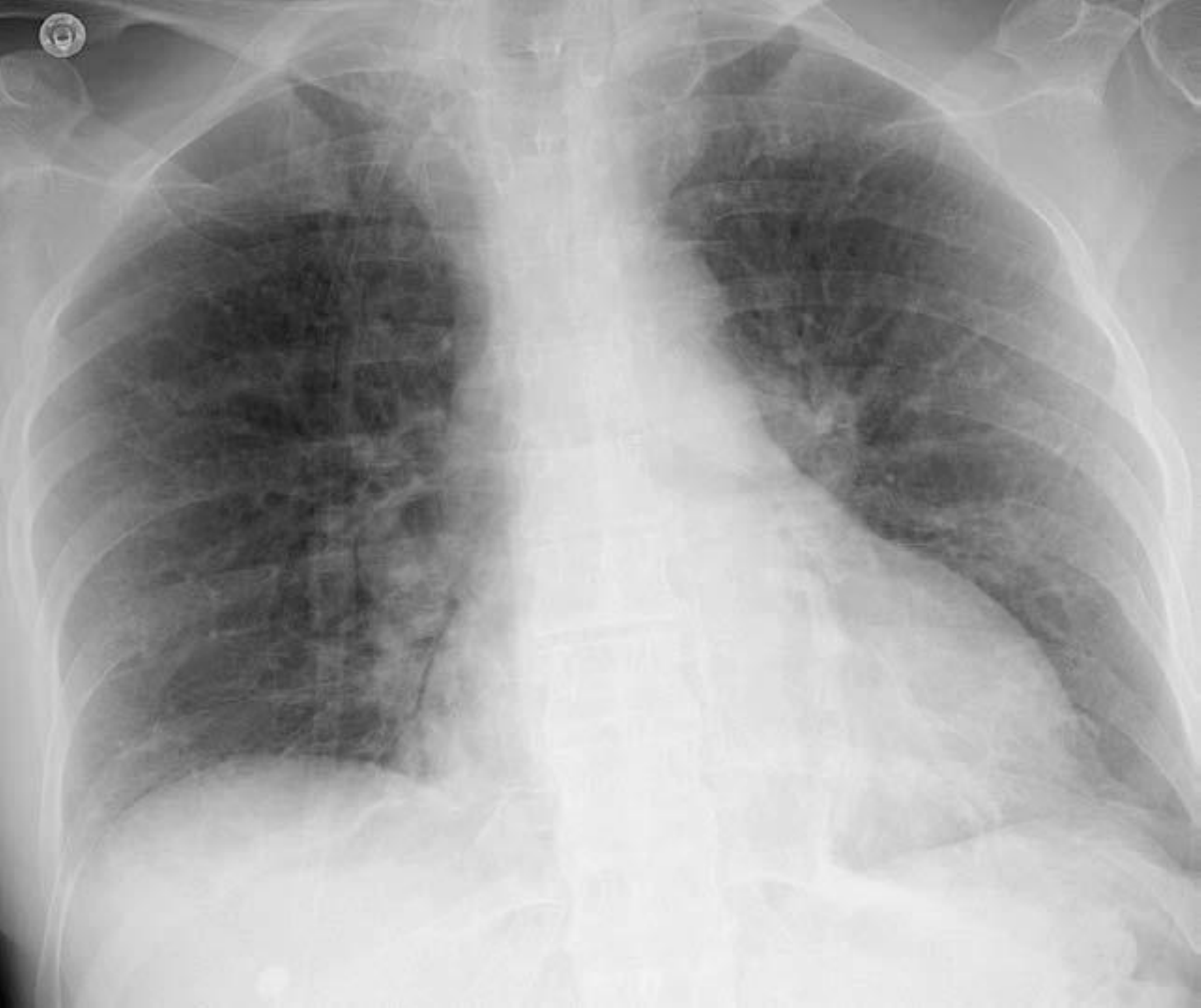}
  \caption{A-P supine}
  \label{fig:DorsalDec}
\end{subfigure}
\begin{subfigure}{.19\textwidth}
  \centering
  \includegraphics[width=.9\linewidth]{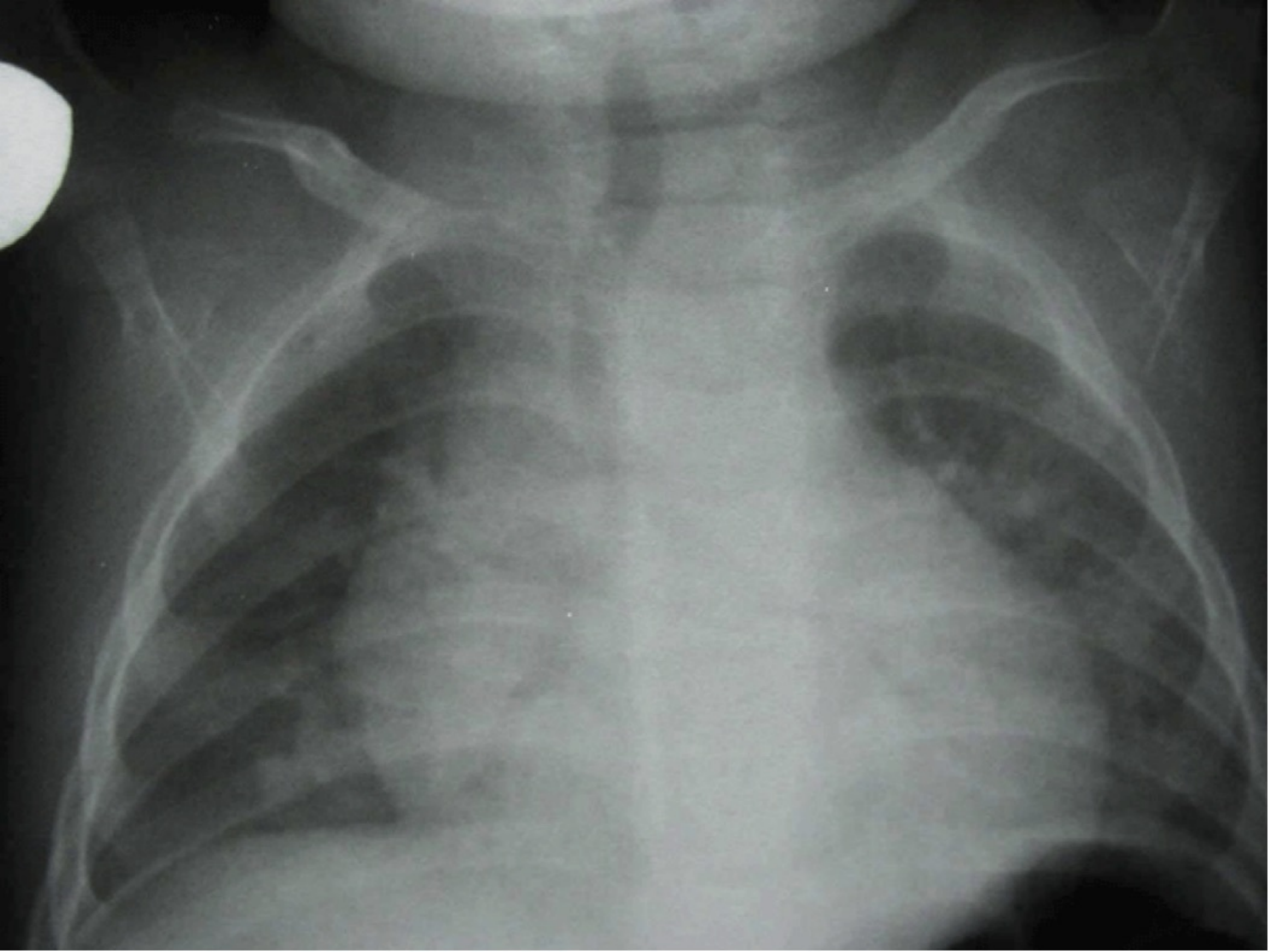}
  \caption{A-P}
  \label{fig:AP}
\end{subfigure}
\caption{Common chest x-ray projections.}
\label{fig:commonProjections}
\end{figure*}

\begin{figure}
\centering\includegraphics[width=.7\linewidth]{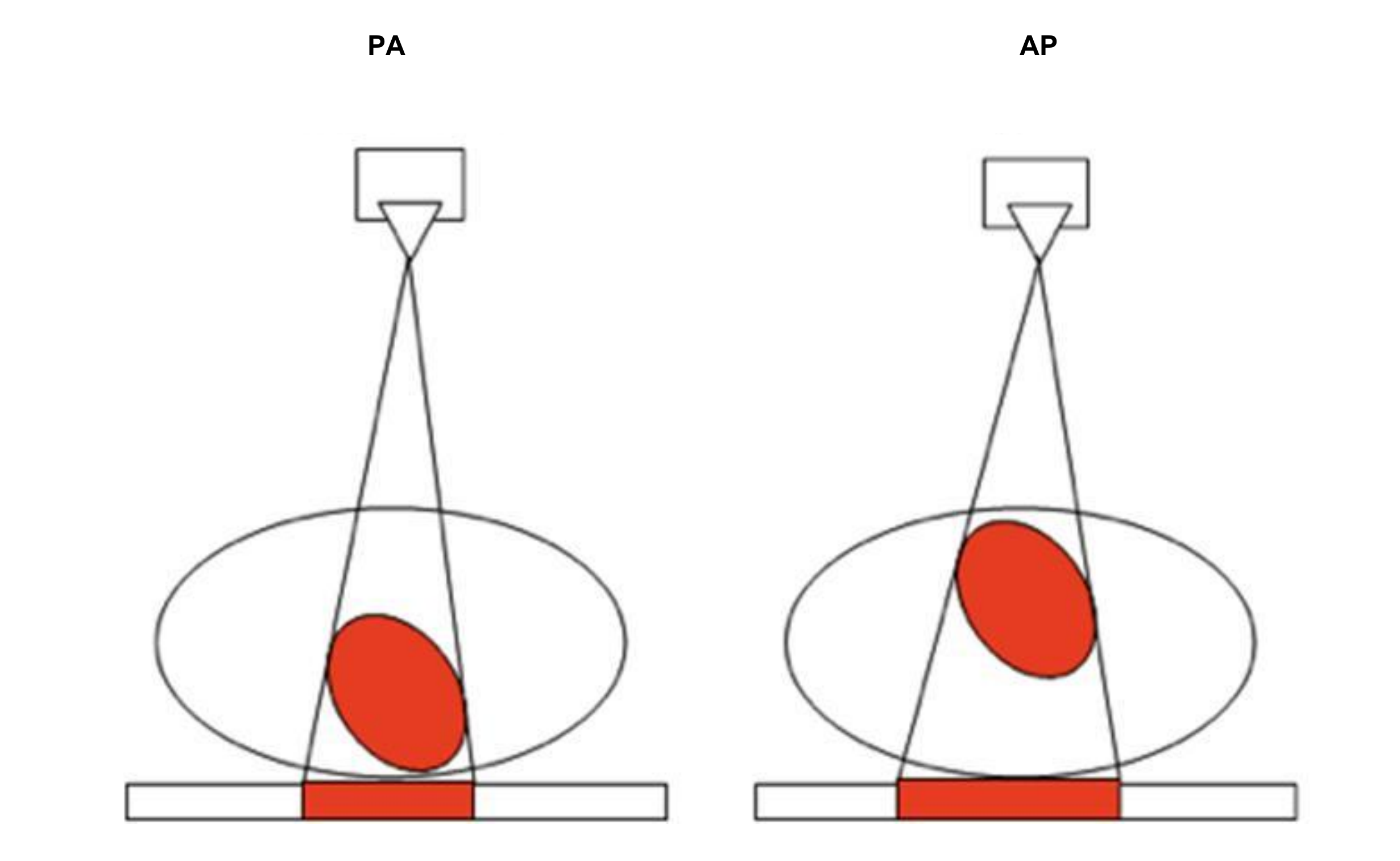}
\caption{Heart size projection in Postero-Anterior (PA) vs Antero-Posterior (AP). The projection of the heart (red silhouette) illustrates that anatomical ratios depend on the plane distance to the x-ray source.} 
\label{fig:APvsPA}
\end{figure}

\begin{figure*}
\centering
\begin{subfigure}{.5\textwidth}
  \centering
  \includegraphics[width=.9\linewidth]{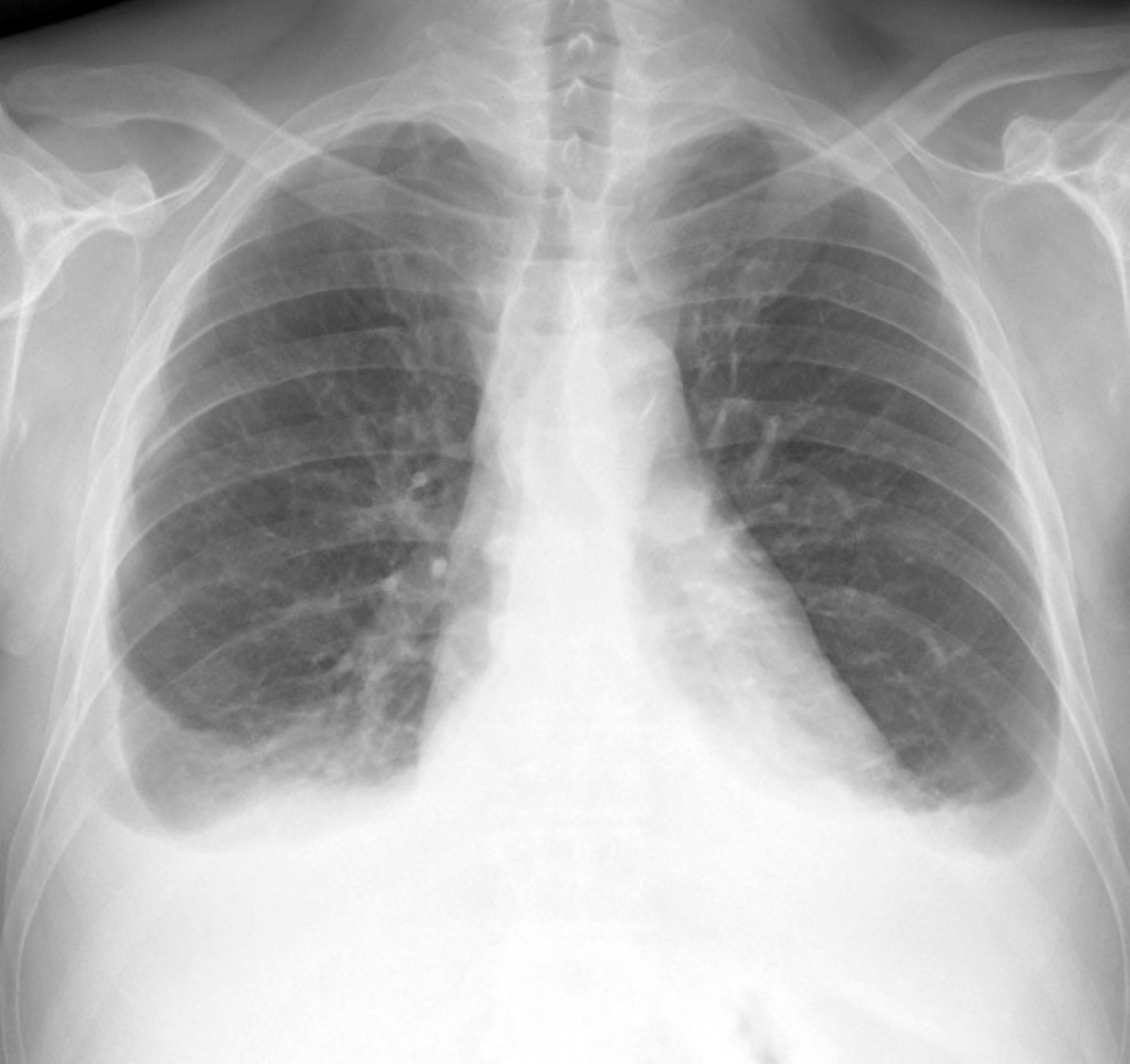}
  \caption{P-A}
  \label{fig:PleuralEffusionPA}
\end{subfigure}%
\begin{subfigure}{.5\textwidth}
  \centering
  \includegraphics[width=.9\linewidth]{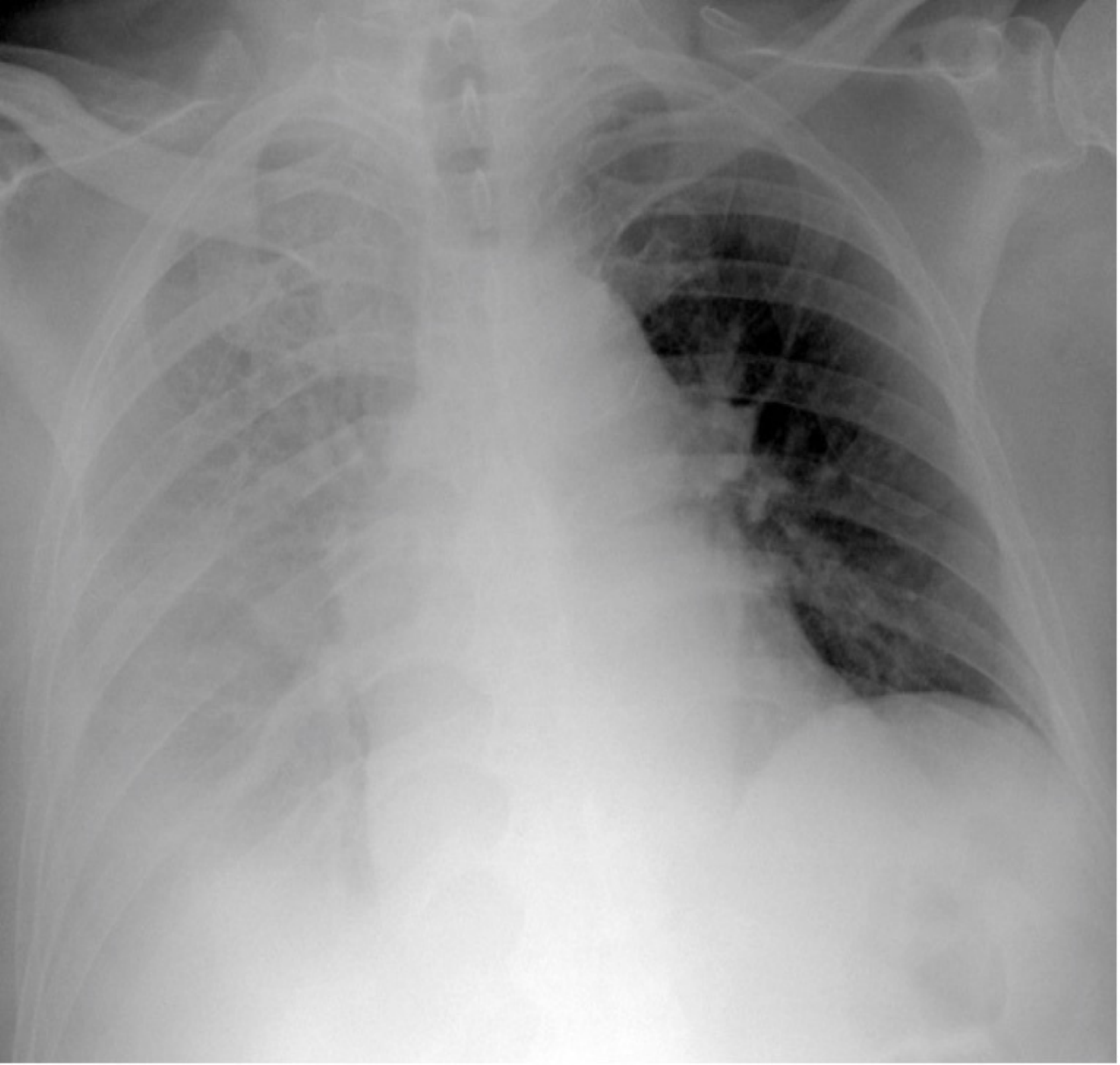}
  \caption{Decubit}
  \label{fig:PleuralEffusionDecubitAP}
\end{subfigure}
\label{fig:PleuralEffusion}
\caption{Pleural effusion in different projections: A bipedestation projection (a) shows the meniscus sign in which the fluid accumulates in the subpulmonary region, ascends through the thoracic wall and through the paramedian zone. In decubit projection (b) there is no meniscus sign. As the liquid goes to the most declining area there is a diffuse increase in hemithorax density and a loss of the net limit of the diaphragm with occupation of the pulmonary vertex by apical cap, costo-phrenic angle blunting and a thickening of the smaller fissure.}
\end{figure*}

The projection information is highly relevant for diagnosis. For example, AP views, which are commonly used in pediatric patients, show an enlarged heart silhouette 
(Fig. \ref{fig:AP}) that should not be interpreted as cardiomegaly, but merely the expected large-depth ratio of reversed organ observation (Fig. \ref{fig:APvsPA}). Another illustrative example is the distinct pattern that pleural effusions have in the standing position (Fig. \ref{fig:PleuralEffusionPA}), in which a typical meniscus sign is commonly found as opposed to decubit projections (Fig. \ref{fig:PleuralEffusionDecubitAP}). 
Given that the number of different projections is unbalanced (for instance, PA followed by lateral projections typically comprise the majority of chest x-rays), there is the risk that none of the other projections will have sufficient instances with which to train models capable of discriminating pathological from non-pathological patterns in the context of the projection. 

There are particular radiological landmarks that differentiate projections, which radiologists are trained to identify. For instance, in the case of PA projections, these landmarks are the presence of air in the gastric chamber and the scapulae projected outside the lung fields. Although these features can be learned, models trained in unbalanced datasets with a poor representation for different projections may not have sufficient instances to properly learn those patterns. An illustrative example is when the heart enlargement in AP projections is attributable only to the effect of the projection, while the trained model erroneously predicts cardiomegaly. 

Given that there were numerous combinations of projections, positions and protocols and all this information was not uniformly reported in DICOM meta-data, we decided to group them into 6 main classes: standard PA, standard L, AP vertical or erect, AP horizontal or supine, pediatric, and rib views.

For those images without DICOM information on the type of projection (20,367 samples), we used a pre-trained model available at \url{https://github.com/xinario/chestViewSplit} and implemented a custom method to preprocess and load PadChest images in the expected format. This method uses a pre-trained ResNet-50 \citep{resnet} CNN model initialized with the ImageNet \citep{deng2009imagenet} weights and trained with fine-tuning on chest x-rays. Note that these automatically labeled projection samples should not subsequently be used to train another classifier with radiographs as inputs and projections as outputs. These subsets of projections were the only ones obtained from the images using an automatic classifier, unlike all the other data provided in PadChest, which can be used to train any classifier on input images, as they were obtained from the reports.

\subsection{Preprocessing of the reports}
\label{S:Preprocessing}

The text dataset consisted of 206,222 study reports. There was only one report for each study, and if a study had two projections, such as PA and L, then the radiography description, therefore, included both results in the same report. Not all study reports had their x-ray images available, and conversely not all available x-rays had corresponding reports. 

The ground truth was obtained from raw text reports but was not usable ``as is" for the following reasons: 
\begin{itemize}
\item The reports were in unstructured free-text and did not use standardized dictionaries or medical codes.
\item Although the radiography description was consistently present in the report, other sections, such as patient's history, study reason, and diagnostic impression, were not systematically included in the report and, when present, those sections where not clearly delimited or did not follow a consistent pattern. Particularly, content format largely varied over the years.
\item In several cases, the radiography description contained other projection types, such as sinus and abdominal types, whose interpretation was included together with the chest x-rays in the same report.
\end{itemize}

The following pipeline was consequently implemented\footnote{Source code available at \url{https://github.com/auriml/Rx-thorax-automatic-captioning/blob/master/report_preprocessor.ipynb}} in order to obtain a final curated labeled dataset:

\begin{enumerate}
\item Lowercased text and removal of accents. 
\item The identification and extraction of the text containing the radiography description of chest x-rays in each report was carried out using an iterative incremental approach based on regular expressions and a repeated random sampling review to improve and/or add more regular expressions. A total of 167 (0.6\%) reports were excluded because the radiography description was not retrievable with the chosen expressions or because the section was empty.  
\item Removal of non alpha-numerical characters excluding dots and spaces.
\item Removal of stopwords, with the exception of \texttt{['sin', 'no', 'ni', 'con']}, using Spanish stopword removal from NLTK \citep{nltk}.
\item Spanish stemming to remove morphological affixes from words, leaving only word stems using NLTK.
\end{enumerate}

The complete report dataset had 501,840 sentences, amounting to 3.8 million words. The vocabulary size before preprocessing was 14,234 unique words, and was downsized to 9,691 different tokens after stopword removal (310 words) and stemming. 


The mean number of tokens per sentence was 7.1 
and the median was 5 tokens per sentence. 

\subsection{Medical codes annotation}
\label{S:2}
A total of 27,593 reports were manually annotated 
following a hierarchical taxonomy described below. 
The manually labeled dataset was subsequently used to train and validate a multi-label text classifier, thus enabling the automatic annotation of the remaining 82,338 reports. 

\subsubsection{Topic extraction}

\begin{figure}[t]
\centering\includegraphics[width=.6\linewidth]{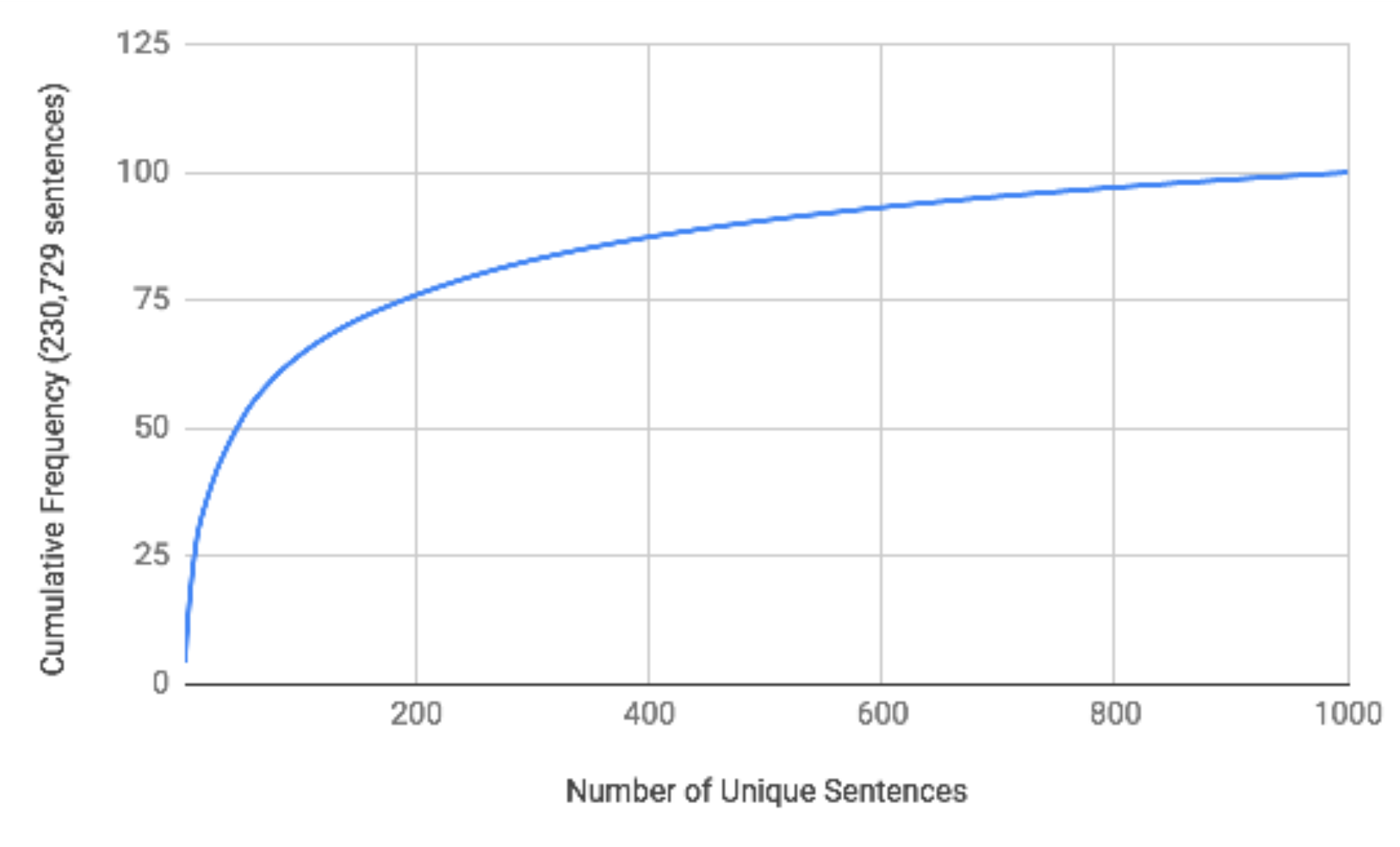}
\caption{Pareto curve of sentence redundancy. The cumulative frequency of the 1,000 most repeated sentences in the entire set of x-ray reports amounted to 230,729 sentences (the total number of sentences in the entire dataset was 500,000).}
\label{fig:Pareto}
\end{figure}

In order to help the practitioner label the dataset with radiographic findings and diagnoses, an automatic process was performed on the pre-processed reports so as to extract a series of topics (concepts) that were subsequently used to speed-up the manual labeling process.

For this goal, first, the preprocessed radiography reports were split into sentences and unique sentences were ordered by frequency. 

Of 145,134 unique sentences from a total of 500,000 sentences, the 1,000 first most repeated ones comprised 230,729 (46\%). Fig. \ref{fig:Pareto} shows a Pareto curve of the 1,000 most frequent sentences. Note that 20\% of them covered up to 76\% of 230,729 sentences. In order to exploit this redundancy, the strategy employed was to manually label at sentence level.

Topic extraction discovers keyphrases and concepts in each sentence based on the frequency and linguistic patterns in the text.
In order to facilitate the manual annotation task, the topics of sentences were automatically extracted and then presented to physicians for evaluation. For example, the list of phrases from the topic associated with the Chronic obstructive pulmonary disease (COPD) is: ``epoc sign radiolog atrap aere sugest marc dorsal import inter cardiomegalia aereo lev significativos prominent vascular apical torax escoliosis derecho". 

Three different unsupervised clustering methods with which to extract the topics from the reports were evaluated:  
\begin{enumerate}
\item Non-negative Matrix Factorization (NMF), \cite{NMFSklearn}. 
First, term frequency–inverse document frequency (tf-idf) statistics were extracted to reflect how important a word is in the corpus.An NMF model was then fitted with the tf-idf features using two different objective functions: the Frobenius norm, and the generalized Kullback-Leibler divergence. 

\item Latent Dirichlet Allocation (LDA), \cite{LDA}. In this case, tf (raw term frequency) features were extracted to fit an LDA model.

\item k-Means. First, Doc2Vec \citep{le2014distributed} representations were extracted, which is a generalization of Word2Vec \citep{word2vec} for learning vectors from documents. The following parameters were used to learn the representations: Learning rate 0.025, size of word vectors 300, size of the context window 10, number of epochs 55, minimum . number of word occurrences 5, number of negative sampled 5, loss function negative sampling, sampling threshold $10^{-3}$. Once the representations had been extracted, the k-Means algorithm was used to cluster these vectors into 20 topics as shown in Tab.  \ref{tab:topics}.

\end{enumerate}

The output from these methods is a list of topics, each represented as a list of terms. The results were evaluated qualitatively for semantic performance, and homogeneity, and the 20 topics obtained with Doc2Vec and k-Means were chosen by physician consensus. Grouping sentences by topic increased efficiency, thus allowing physicians to find batches of sentences, in which the same labels could be propagated seamlessly, more easily. 

\begin{table}
    \centering
    \begin{footnotesize}
    \begin{tabular}{lp{6.2cm}}
        \hline
        \textbf{Topic}&\textbf{Semantic group}\\
        \hline
        1 & hypoexpansion basal, laminar atelectasis, bronchiectasis, pleural effusion ,pneumonia, infiltrates \\
        2 & pleural effusion both sides, bronchovascular markings, gynecomastia, interstitial pattern bilateral, pulmonary edema, respiratory distress, heart insufficiency \\
        3 & tracheostomy tube, endotracheal tube, NSG tube, chest drain tube \\
        4,12 & exclude sentences without radiolographical findings mainly mentioning additional imaging test recommended in the follow-up\\
        5 & scoliosis, kyphosis, vertebral degenerative changes, osteopenia, osteoporosis \\
        6 & COPD signs, air trapping, chronic changes, hyperinflated lung both sides, flattened diaphragm, emphysema \\
        7 & vertebral compression \\
        8 & NSG tube, hiatal hernia \\
        9,15,16, 18 & normal study \\
        10 & calcified granuloma, nipple shadow \\
        11 & callus rib fracture, humeral prosthesis, osteosynthesis material, cervical rib \\
        13 & unchanged \\
        14 & central venous catheter, reservoir central, pacemaker \\
        17 & cardiomegaly, vascular hilar enlargement, pulmonary hypertension, hilar congestion, vascular redistribution \\
        19 &  surgery, sternotomy, diaphragmatic eventration, hemidiaphragm elevation, costophrenic angle blunting \\
    \end{tabular}
    \end{footnotesize}
    \caption{Sentence topics: Sentences were automatically assigned to 20 topics with the sole purpose of organizing unique sentences by semantic logic. This would, therefore serve as a tool with wich to help the physician in the manual extraction of labels. Each of these topics captured different medical semantic groups that facilitated the manual task of label extraction by ordering the semantically close sentences.}
    \label{tab:topics}
\end{table}

\subsubsection{Manual labeling}
A total of 22,120 unique sentences corresponding to 27,593 study reports were labeled and manually reviewed by trained physicians. Medical entities were extracted as per physician criteria and mapped onto Unified Medical Language System (UMLS \citep{umls}) controlled biomedical vocabulary unique identifiers. Medical entities whose exact meaning could not be found in any term of the UML metathesaurus, but that were relevant for annotation purposes were also extracted using the physician's labels but without assigning a code to them. Four additional labels with special coding rules were explicitly defined in order to maintain consistency in the following cases: 

\begin{itemize}
\item Sentences describing normality, i.e. those that either reported complete resolution, did not describe radiographic findings or negated their presence were labeled as ``Normal". 
\item Sentences that qualified images as suboptimal or with deficient technique as reported by radiologists were labeled as ``Suboptimal".  
\item Sentences that were not interpretable or that did not mention any radiographic findings or diagnoses were labeled as ``Exclude". 
\item Sentences mentioning that there were no changes in a medical entity described previously but not included in the report were labeled as "Unchanged" unknown entity. This procedure allows to distinguish pathologic from normal studies even when the medical entity is not mentioned and only a temporal reference is given. This case is frequent on temporal series of studies done on hospitalized patients to control the evolution of a pathological finding. It is important to remark that before training a model to predict pathologies from images using PadChest, labels of type ``Unchanged" are intended to be replaced by labels reported for the same patient, if available in a prior study. Otherwise, the label should be removed as it could not be learned from the image. 
\end{itemize}

All the sentences were assigned to multiple labels or at least one. When a list of potential diagnoses was reported, all differential diagnoses were labeled. Similarly, when different radiographic findings were described all of them were labeled.

As a result, both the sentences and hence the reports (as a sequence of sentences) were multi-label. Finally, the following consecutive rules were applied to ensure consistency when obtaining the final unique set of multi-labels for each report:

\begin{enumerate}
\item The set of unique labels was obtained as the union operation of all labels assigned to each report. 
\item The report was assigned a single label ``Normal" only if in addition to that label, there were no other labels regarded as a  radiographic finding or diagnosis.
\item The report was assigned a single label ``Exclude" if this was the only label. 
\end{enumerate}

\subsubsection{Organization of labels in hierarchical taxonomies}
\label{S:3}

The list of unique manually annotated labels was organized into three hierarchical trees (see \ref{app:hierarchies}) in order to facilitate the exploitation and retrieval of images by semiological groups, differential diagnoses and anatomic locations. The hierarchical organization criteria and content of the three resulting taxonomy trees were reviewed by a radiologist with 20 years of clinical experience. 

The main purpose of the hierarchic organization is to allow comprehensive retrieval of the studies grouped by higher hierarchical levels, thus enabling the construction of partitions using different criteria when compiling training sets for machine learning techniques. This is particularly relevant in order to control balance and granularity when deciding the classes to be inferred by classification methods. 

The hierarchies are multi-axial, that is, the same term can be found in different branches, and child-parent relationships should satisfy that a child entity ``is a" parent entity. For instance, ``chronic tuberculosis" is  ``tuberculosis". Following these criteria, each medical concept was classified and assigned a node of the radiographic finding versus the differential diagnosis tree, and spatial concepts were assigned a node in the anatomical locations tree.

The criteria applied to each of the three hierarchical trees were:

\begin{itemize}
\item Radiographic Findings Tree (\ref{app:rxFinding}): Findings were defined as any medical entity or diagnosis that radiologists can assert based solely on the interpretation of a chest x-ray image without any additional knowledge of the patient's clinical information. Medical entities that can be directly diagnosed by interpreting the image are, for example, those affecting the bones (e.g, fractures, degenerative diseases, etc) or those involving alterations in air distribution (e.g. pneumothorax, subcutaneous emphysema, pneumoperitoneum, etc).
\item Differential Diagnosis Tree (\ref{app:DD}): Diagnoses, as opposed to radiographic findings, are those entities that radiologists can propose only as a possible list of differential diagnoses because the patient's clinical information and/or additional studies, such as other radiology tests or laboratory analyses, are required to interpret the radiographic pattern and suggest a diagnosis with certainty. The more a radiologist knows about the patient (clinical, exploration and laboratory data), the more accurate the diagnosis is. For example, an alveolar pattern is a radiographic finding that would prompt a very long list of differential diagnoses (including both infectious and non-infectious diseases such as lung edema, respiratory distress, etc.), but whenever it is present in a patient with fever, cough, leukocytosis, high CRP levels and crackles in the localization of the alteration, then it will almost certainly be pneumonia. This additional information is not always provided to the radiologist, who consequently lists the most probable differential diagnoses in the report. Up to 30\% of false positive and 30\% of false negative diagnoses of pneumonia based on the presentation and chest X-ray findings were reported by \cite{claessens2015early}. In this work, we did not access the patients' health record to confirm diagnoses, and those entities were, therefore, regarded as differential diagnoses.  

\item Topological Tree (\ref{app:localizations}): 
This tree groups anatomical or spatial qualifiers in which radiographic findings and diagnoses are located. These spatial concepts include body parts, organs or organ components, chest anatomical regions, structures, and type of tissues. These concepts were first identified by means of regular expressions (see \ref{app:regexLoc}) and then assigned to labels that were hierarchically organized in the topological tree. 
\end{itemize}

\subsection{Automatic labeling of the remaining reports}
The reports that were not labeled in the previous stage (75\%) were automatically tagged using a deep neural network classifier trained with the manually labeled data. We, therefore, used the manually annotated set of x-ray reports as training set. The goal was to produce a large dataset of annotated reports to be used as output of an image classifier trained with the x-rays.

We define the task of annotating the medical entities as a multi-label text classification problem in which, for each sentence $i$, the goal is to predict $y_{i,\ell} \in {0,1}$ for all $\ell \in \boldsymbol{\mathcal{L}}$, where $\boldsymbol{\mathcal{L}}$ is the label space that includes the different radiographic findings (\ref{app:rxFinding}) and differential diagnoses (\ref{app:DD}). 

Our method does not employ any structured data or external information beyond the manually labeled data. The descriptive statistics of the training set and ground truth are summarized in Tab. \ref{tab:DescStatisticsManLab}.

\begin{table}[t]
\centering
\begin{footnotesize}
\begin{tabular}{llll}
\hline
\textbf{Parameter} & \textbf{Value} \\
\hline
Number of training sentences& 20,439\\
Number of validation sentences& 2,271\\
Vocabulary size& 9,691 \\
Average (min-max) number of tokens per sentence& 7.1 (1-56) \\
Average (min-max) number of labels per sentence& 1.33 (1-9 ) \\ 
Label space $|\boldsymbol{\mathcal{L}}|$ & 193\\
\hline
\end{tabular}
\end{footnotesize}
\caption{Descriptive statistics of the report sentences used to train the models. All sentences were manually annotated by trained physicians. Note that the reports consist of a sequence of sentences.}
\label{tab:DescStatisticsManLab}
\end{table}

In this method, each input instance corresponds to a sentence (denoted as $\boldsymbol{X} = [\boldsymbol{x}_1, \boldsymbol{x}_2, .., \boldsymbol{x}_N]$) composed of a sequence of pretrained word-embeddings $\boldsymbol{x}_i$ of dimension $d_e$ for each of its $N$ tokens, and its extracted labels encoded in a one-hot vector $\boldsymbol{y}$ of dimension $|\boldsymbol{\mathcal{L}}|$. Word-embeddings of size $d_e = 100$ were trained on the full corpus of Spanish reports with FastText \citep{grave2017bag} in order to obtain the embeddings with the following parameters: Learning rate 0.025, size of the context window 5, number of epochs 55, minimum number of word occurrences 5, number of negative sampled 5, negative sampling loss function, sampling threshold $10^{-4}$, length of char n-gram (min-max) $3-6$. We favored FastText over Word2Vec method so as to be able to encode unseen words that may potentially appear at inference time.

The four evaluated models are:
\begin{itemize}
\item A 1-Dimensional Convolutional Neural Network (CNN). 
\item A Recurrent Neural Network (RNN).
\item A 1-Dimensional Convolutional Neural Network with attention mechanism (CNN-ATT).
\item A Recurrent Neural Network with attention mechanism (RNN-ATT).
\end{itemize}

The CNN and RNN models were regarded as baseline models in order to compare them with the same architectures including attention mechanisms. The topologies of these systems are shown in Figs \ref{fig:CNN_ATT} and \ref{fig:RNN_ATT} respectively.
All neural models were implemented using PyTorch \citep{paszke2017automatic} and the Ignite \footnote{\url{https://pytorch.org/ignite/}} library was used to help compact the programming code for common tasks such as training loop with metrics, early-stopping and model checkpointing.

\subsubsection{CNN}
\label{CNN}

In the baseline CNN, after inputing the text as a matrix of concatenated word-embeddings, the model computed a base representation of each sentence denoted as matrix $\boldsymbol{H} \in \mathbb{R}^{31x128}$ containing the horizontal concatenation of the n-grams spatial localizations with 128 features each. The sigmoid activations of a final linear layer outputted the $|\boldsymbol{\mathcal{L}}|$ binary classification labels. 

\begin{figure*}[t]
\centering
\includegraphics[width=1\linewidth]{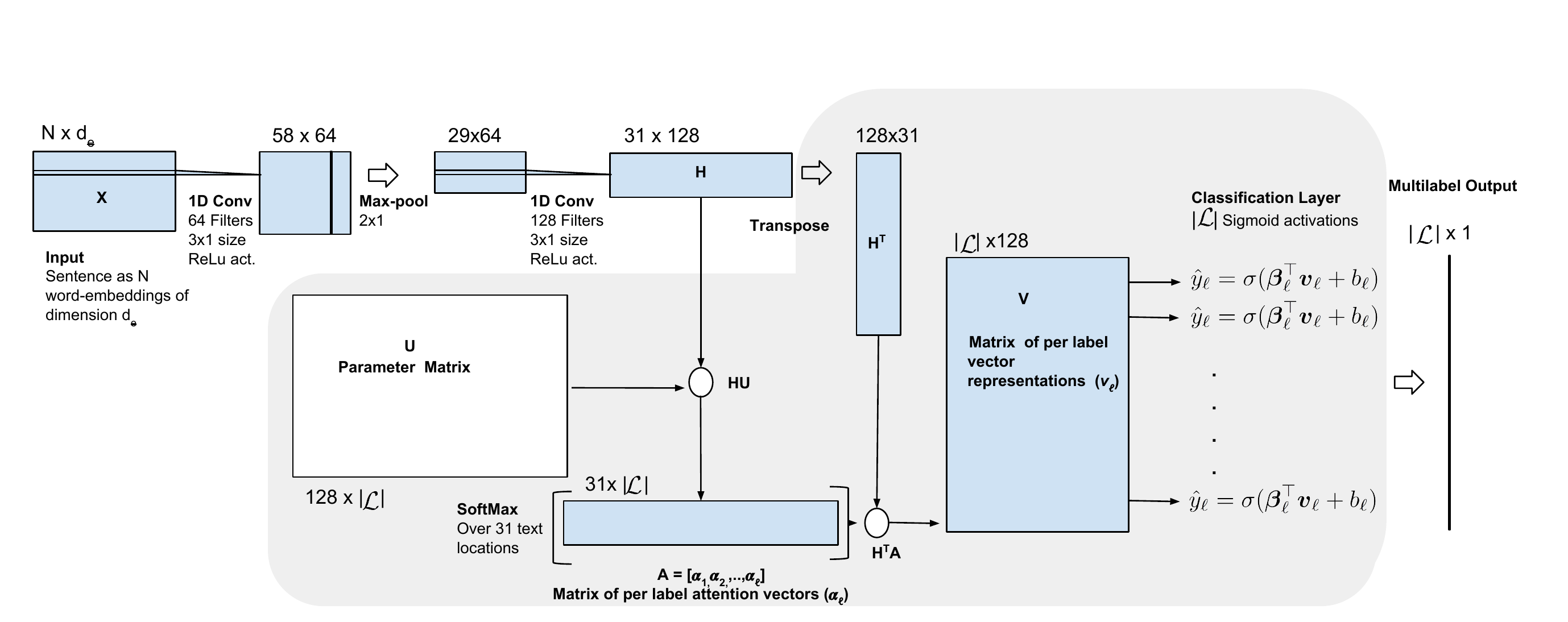}
\caption{Diagram of the CNN topology with attention mechanism highlighted in gray. Input is a representation of the text as a matrix of concatenated word-embeddings, and outputs are the most likely labels $\boldsymbol{\mathcal{L}}$.}  
\label{fig:CNN_ATT}
\end{figure*}

As shown in Fig. \ref{fig:CNN_ATT}, each sentence in the CNN is represented as a matrix of $N=56$ word-embeddings of $d_e=100$ dimensions. The first convolutional layer contains 64 filters of dimension 3x1 with stride 1 and ReLu activation function, followed by a 2x1 max pooling layer with stride 1. There is a second convolutional layer containing 128 filters with the same dimensions, stride and activation function as the first layer. In this baseline CNN model, the attention module (highlighted in gray in Fig. \ref{fig:CNN_ATT}) is replaced with a $|\boldsymbol{\mathcal{L}}|$ fully connected layer with multi-label sigmoid activations.

\subsubsection{RNN}

In the baseline RNN, after inputing the text as a variable length sequence of word-embeddings, the model computed a base representation of each sentence, denoted as matrix $\boldsymbol{H} \in \mathbb{R}^{Nx256}$ containing the sequence of hidden states, in which each hidden state represents one token at one time step encoded with 256 features each. As occurred with the CNN, the sigmoid activations of a final lineal layer outputted the $|\boldsymbol{\mathcal{L}}|$ binary classification labels.

\begin{figure*}[t]
\centering
\includegraphics[width=1\linewidth]{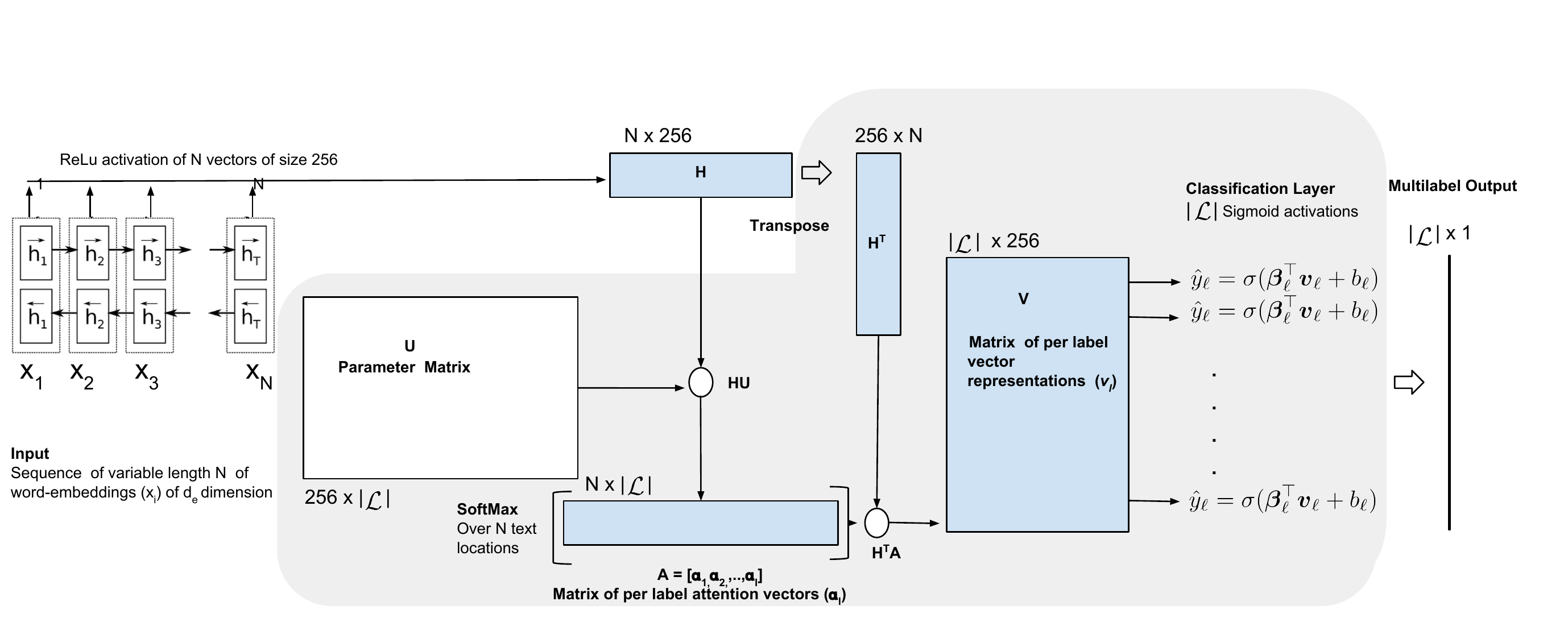}
\caption{Diagram of the RNN topology with attention mechanism highlighted in gray. Input is a representation of the text as a sequence of variable-length word-embeddings, and outputs are the most likely labels $\boldsymbol{\mathcal{L}}$.} 
\label{fig:RNN_ATT}
\end{figure*}

As shown in Fig. \ref{fig:RNN_ATT},each sentence in the RNN is inputed at each time step as a sequence of variable length of $N$ word-embeddings of $d_e=100$ dimensions. There are two LSTM bi-directional layers containing 128 hidden units each. In this baseline RNN model, the attention module (highlighted in gray in Fig. \ref{fig:RNN_ATT}) is removed and the hidden state of the last time step of the RNN is used as the input for a final $|\boldsymbol{\mathcal{L}}|$ fully connected layer with multi-label sigmoid activations.

\subsubsection{Attention model}
\label{attention}

The attention mechanism, which is based on the work by \cite{mullenbach2018explainable}, learns different text representations for each label. The underlying idea is that each snippet containing important information correlated with a label could be anywhere in the text and would differ among different labels.

In both the CNN-ATT and the RNN-ATT model, the final linear layer was replaced with an attention feed-forward module that first calculates attention vectors $\boldsymbol{\alpha}_\ell$ as a distribution of putative locations for each label $\ell$ in the sentence. As illustrated in Figs. \ref{fig:CNN_ATT} and \ref{fig:RNN_ATT}, attention vectors are computed as:  

\begin{equation}
\boldsymbol{\alpha}_\ell = \mathrm{SoftMax}(\boldsymbol{H} \boldsymbol{u}_\ell)
\end{equation}

\noindent where $\boldsymbol{u}_\ell$ is a vector 
parameter for label $\ell$ and $\mathrm{SoftMax}(\boldsymbol{x}) = \frac{\mathrm{exp}(\boldsymbol{x})}{\sum_{\forall{i}} \mathrm{exp}(\boldsymbol{x}_i)}$, where $\mathrm{exp}(\boldsymbol{x})$ is the element-wise exponentiation of the vector $\boldsymbol{x}$ as described in \cite{mullenbach2018explainable}.

In a second step, we apply the attention vectors $\boldsymbol{\alpha}_\ell \in \boldsymbol{A}$ to the base representation of a sentence in order to compute a matrix of 128-dimensional weighted feature vectors $\boldsymbol{v}$ for each label, defined as 
$\boldsymbol{v}_\ell = \boldsymbol{H}^\top\boldsymbol{\alpha}_\ell$ where $\boldsymbol{v}_\ell \in \boldsymbol{V}$. 

Finally, following the approach of \cite{mullenbach2018explainable}, after applying a linear layer consisting of $|\boldsymbol{\mathcal{L}}|$ vectors of weights $\boldsymbol{\beta}_l$ and a scalar bias $b_\ell$, a sigmoid transformation is used to compute the probability $\hat{y}_\ell$ for each label.  

\begin{equation}
\hat{y}_\ell = \sigma(\boldsymbol{\beta}^\top_\ell \boldsymbol{v}_\ell + b_\ell)
\end{equation}

The training procedure minimized the binary cross entropy loss $L$ in the case of all four topologies:  

\begin{equation}
L_{BCE}(\boldsymbol{X},\boldsymbol{y}) = - \sum_{\forall{l}} ({y_\ell \log(\hat{y}_\ell) + (1-y_\ell) \log (1-\hat{y}_\ell)})
\end{equation}

and the L2-norm of the model weights, using Adam optimizer \citep{kingma2014adam} for the CNN and CNN-ATT models and RMSprop \citep{tieleman2014rmsprop} for the RNN and RNN-ATT models.

\section{Evaluation of automatic labeling}
\label{sec:evaluation}

The following metrics were used to assess the performance of the automatic labeling method: 
\begin{itemize}
\item Accuracy: The accuracy score provides the fraction 
of labels that were correctly detected. 
In this work, as we are dealing with multi-label classification, we obtain the average accuracy per sample: If all the predicted labels for an input sentence strictly match the true set of labels, then the accuracy for that sample is 1, and is otherwise 0.0. The average accuracy of all the samples in the test set is reported as the overall accuracy.  
\item MacroF1:
Calculates metrics for each label, and finds their unweighted mean. This does not take label imbalance into account, as it places more emphasis on rare label predictions. 
\begin{equation}
\mathrm{MacroR} =  \frac{1}{|\boldsymbol{\mathcal{L}}|} \sum_{\ell=1}^{|\boldsymbol{\mathcal{L}}|}\frac{\mathrm{TP}_\ell}{\mathrm{TP}_\ell + \mathrm{FN}_\ell}
\end{equation}
\begin{equation}
\mathrm{MacroP} =  \frac{1}{|\boldsymbol{\mathcal{L}}|} \sum_{\ell=1}^{|\boldsymbol{\mathcal{L}}|}\frac{\mathrm{TP}_\ell}{\mathrm{TP}_\ell + \mathrm{FP}_\ell}
\end{equation}
\begin{equation}
\mathrm{MacroF1} = 2 \cdot\frac{\mathrm{MacroR} \cdot \mathrm{MacroP}}{\mathrm{MacroR} + \mathrm{MacroP}} 
\end{equation}
\item MicroF1: Calculates metrics globally by counting the number of true positives, false negatives and false positives. 
\begin{equation}
\mathrm{MicroR} = \frac{\sum_{\ell=1}^{|\boldsymbol{\mathcal{L}}|} \mathrm{TP}_\ell}{\sum_{\ell=1}^{|\boldsymbol{\mathcal{L}}|}{(\mathrm{TP}_\ell + \mathrm{FN}_\ell)}}
\end{equation}
\begin{equation}
\mathrm{MicroP} = \frac{\sum_{\ell=1}^{|\boldsymbol{\mathcal{L}}|} \mathrm{TP}_\ell}{\sum_{\ell=1}^{|\boldsymbol{\mathcal{L}}|}{(\mathrm{TP}_\ell + \mathrm{FP}_\ell)}}
\end{equation}
\begin{equation}
\mathrm{MicroF1} = 2 \cdot\frac{\mathrm{MicroR} \cdot \mathrm{MicroP}}{\mathrm{MicroR} + \mathrm{MicroP}} 
\end{equation}
\item WeightedF1: Calculates metrics for each label, and finds their average, weighted by support (the number of true instances for each label). This alters Macro to account for label imbalance and it may result in a score that is not between precision and recall. 
\end{itemize}

The four models were trained and validated in the same random partition and early stopping was employed in order to maintain the best model leaving the max number of epochs up to 500. The hyper-parameters used to train the models are summarized in Tab \ref{tab:parameterTuning}. The accuracy metrics obtained in the multi-label annotation of radiographic findings and differential diagnoses on text are summarized in Tab. \ref{tab:AccMetrics}.

\begin{table*}[t]
\centering
\begin{footnotesize}
\begin{tabular}{llll}
\hline
\textbf{Parameter} & \textbf{Values} & \textbf{CNN} & \textbf{RNN} \\
\hline
Batch size & $256, 512, 1024, 2048$ & 1024 & 1024\\ 
L2 Penalty & 0, $10^{-3}$, $10^{-2}$ & 0& 0 \\
Learning rate& $10^{-4}$, $5\cdot 10^{-4}$, $10^{-3}$, $10^{-2}$ & $10^{-4}$ & $10^{-2}$ \\ 
Filter size & $[2,10]$ &3&  \\
Number of filters second conv layer & 64, 128, 256 & 128&\\ 
Dropout probability & $[0.1,0.8]$ & 0.4& 0.4\\
Number of hidden LSTM units & 64, 128, 256, 512 && 128\\ 
Number of hidden LSTM layers & $[1,3]$& &2\\ 
\hline
\end{tabular}
\end{footnotesize}
\caption{Hyperparameter ranges evaluated and best values chosen for the CNN and RNN based models selected by grid-search and manual fine tuning using MicroF1 as performance measure in the validation set.}
\label{tab:parameterTuning}
\end{table*}

\begin{table*}[ht]
\centering
\begin{footnotesize}
\begin{tabular}{lccccc}
\hline
\textbf{Model} & \textbf{Epochs} & \textbf{Accuracy}& \textbf{MacroF1}& \textbf{MicroF1} & \textbf{WeightedF1}\\
\hline
\textbf{Validation Set}\\
CNN & 94 & 0.706 & 0.384 & 0.837 &0.814 \\
CNN-ATT & 230& 0.806& 0.458& 0.902 &0.886  \\
RNN & 78 &0.853& 0.483& 0.913 & 0.903 \\
RNN-ATT & 41 & \textbf{0.864} & \textbf{0.491} & \textbf{0.924} & \textbf{0.918}\\
\hline
\textbf{Test Set}\\ 
RNN-ATT & 41 & 0.857 & 0.491 & 0.939 & 0.926 \\ 
\hline
\end{tabular}
\end{footnotesize}
\caption{MicroF1 of all the models in a validation set of 2,271 sentences with a training set of 20,439 sentences. Best results are shown in bold type. The best model from among the four different architectures  was further tested in an independent random sample of 500 sentences drawn from the PadChest Dataset and manually labeled.}  
\label{tab:AccMetrics}
\end{table*}

\begin{figure*}[t]
\centering
\begin{subfigure}{.5\textwidth}
 \centering
 \includegraphics[width=.95\linewidth]{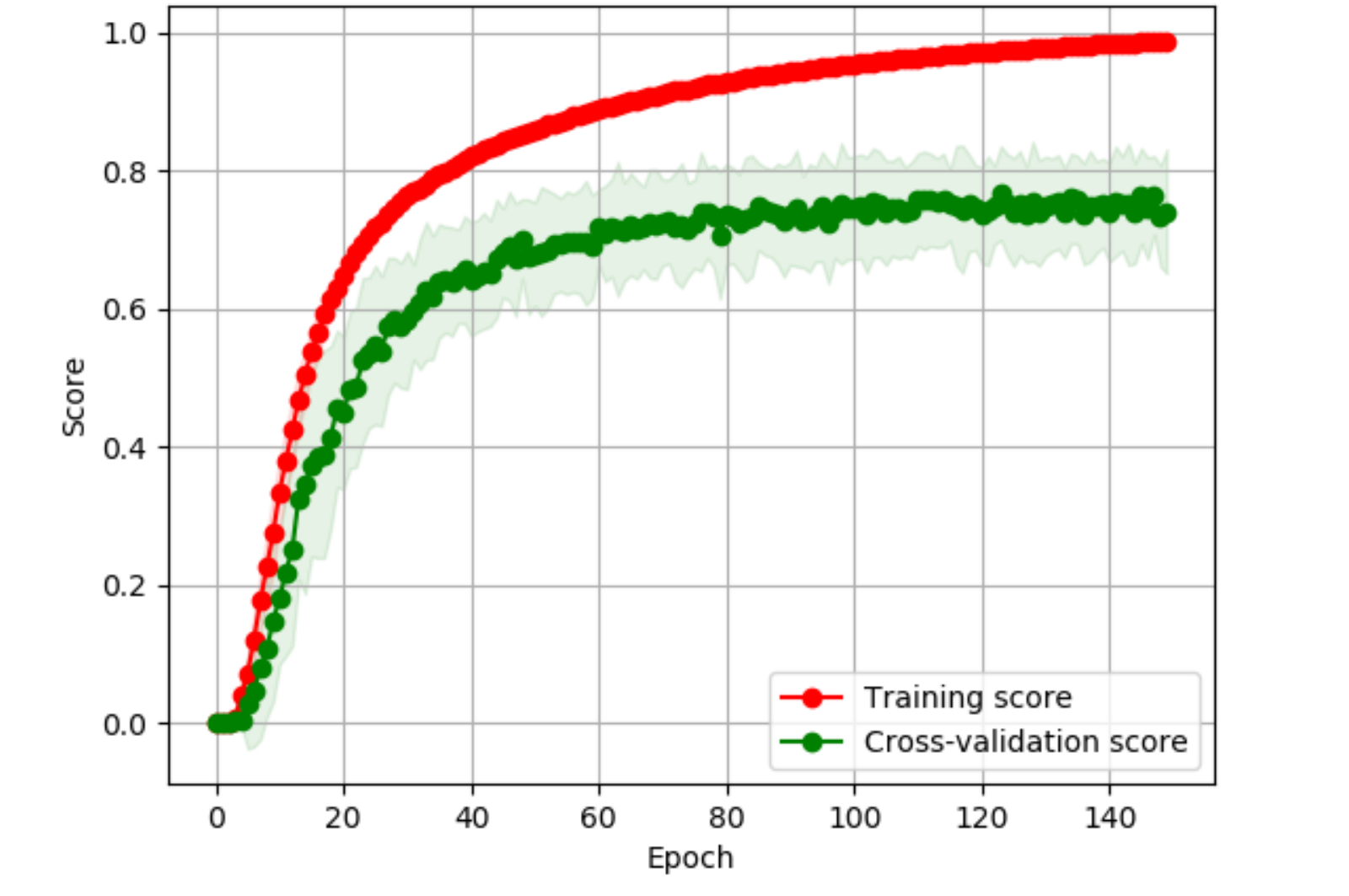}
 \caption{CNN}
 \label{fig:CNN_Curve}
\end{subfigure}%
\begin{subfigure}{.5\textwidth}
 \centering
 \includegraphics[width=.95\linewidth]{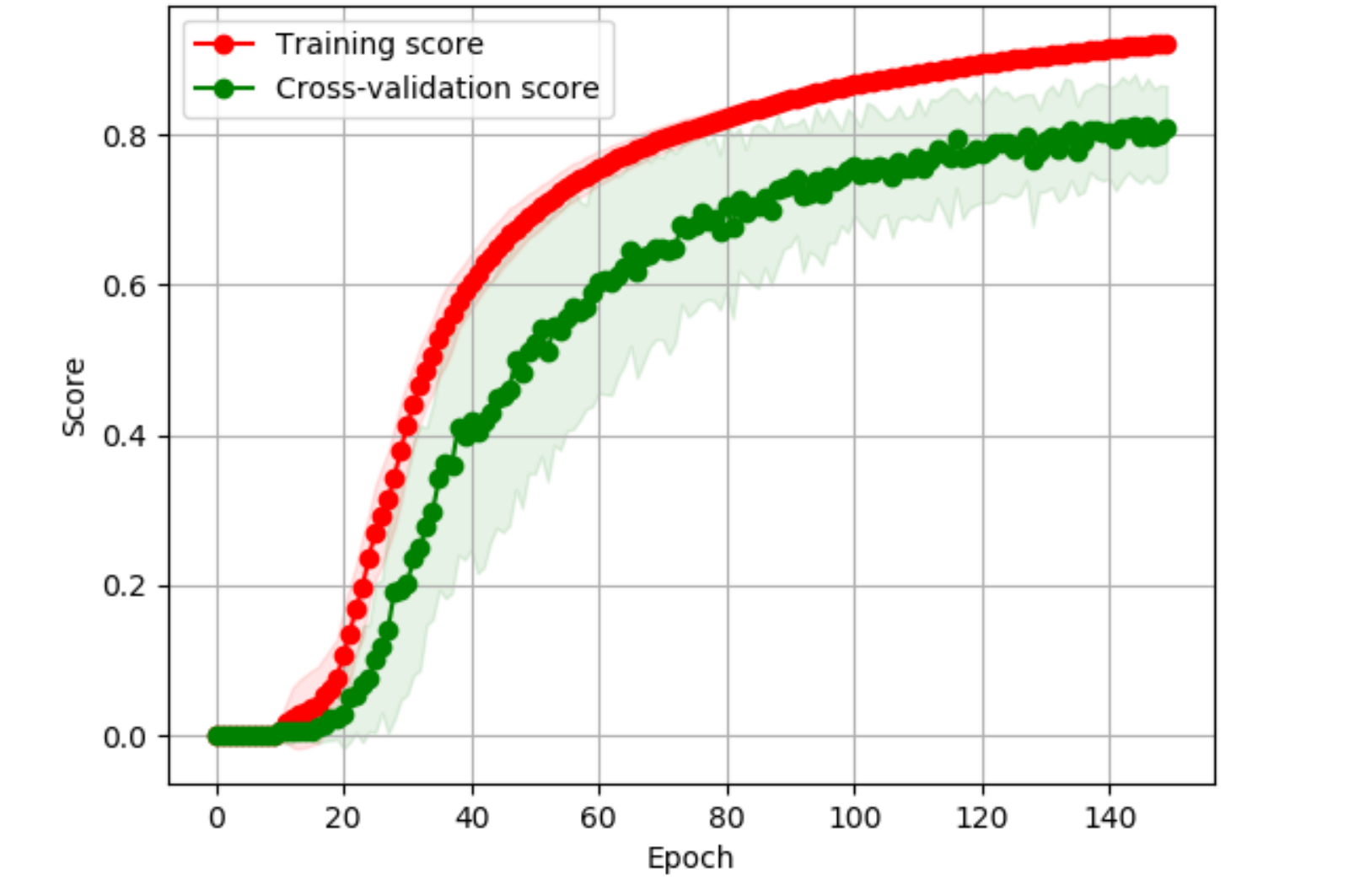}
 \caption{CNN-ATT}
 \label{fig:CNN_ATT_Curve}
\end{subfigure}%
\\
\begin{subfigure}{.5\textwidth}
 \centering
 \includegraphics[width=.95\linewidth]{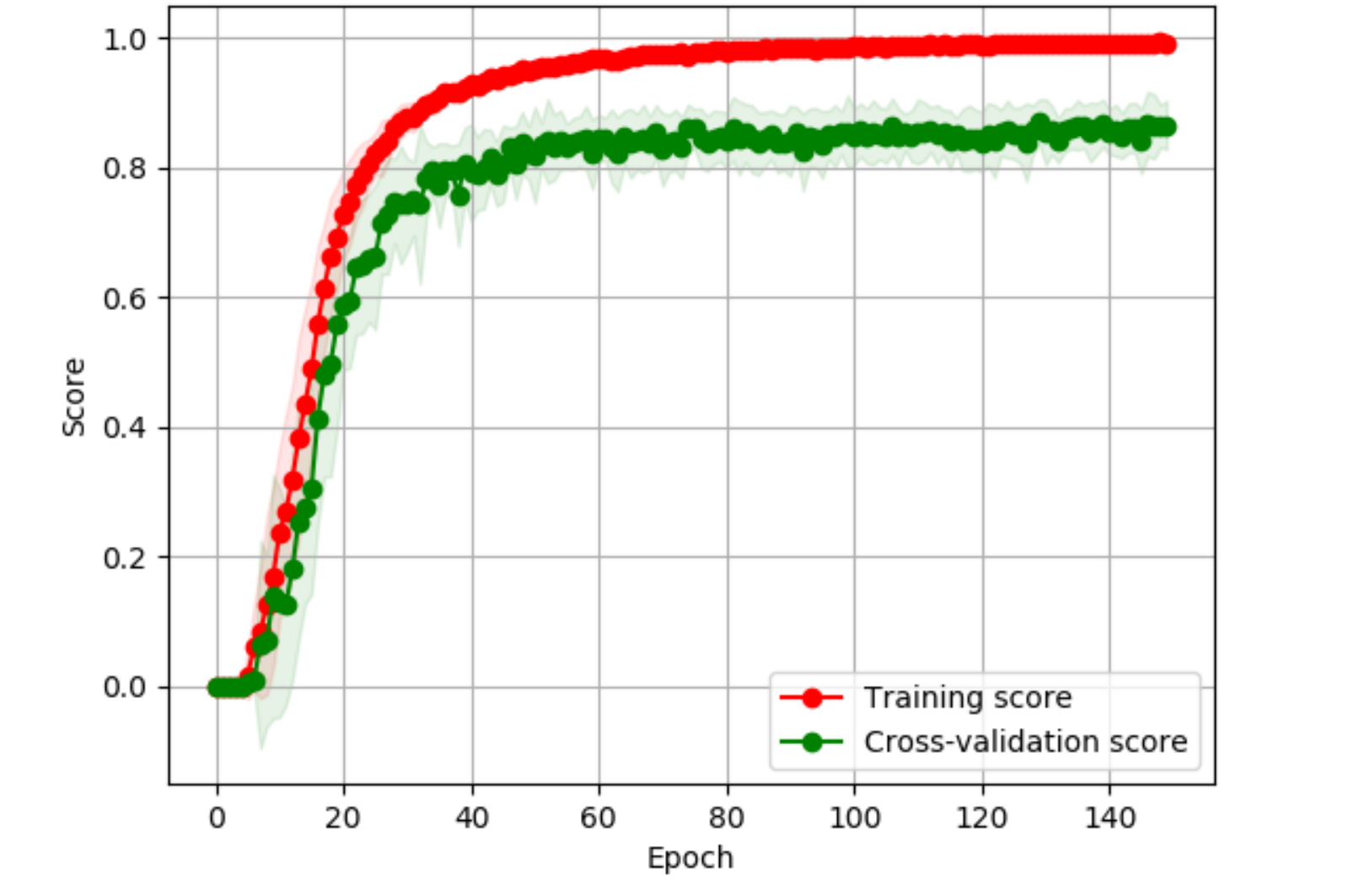}
 \caption{RNN}
 \label{fig:RNN_Curve}
\end{subfigure}%
\begin{subfigure}{.5\textwidth}
 \centering
 \includegraphics[width=.95\linewidth]{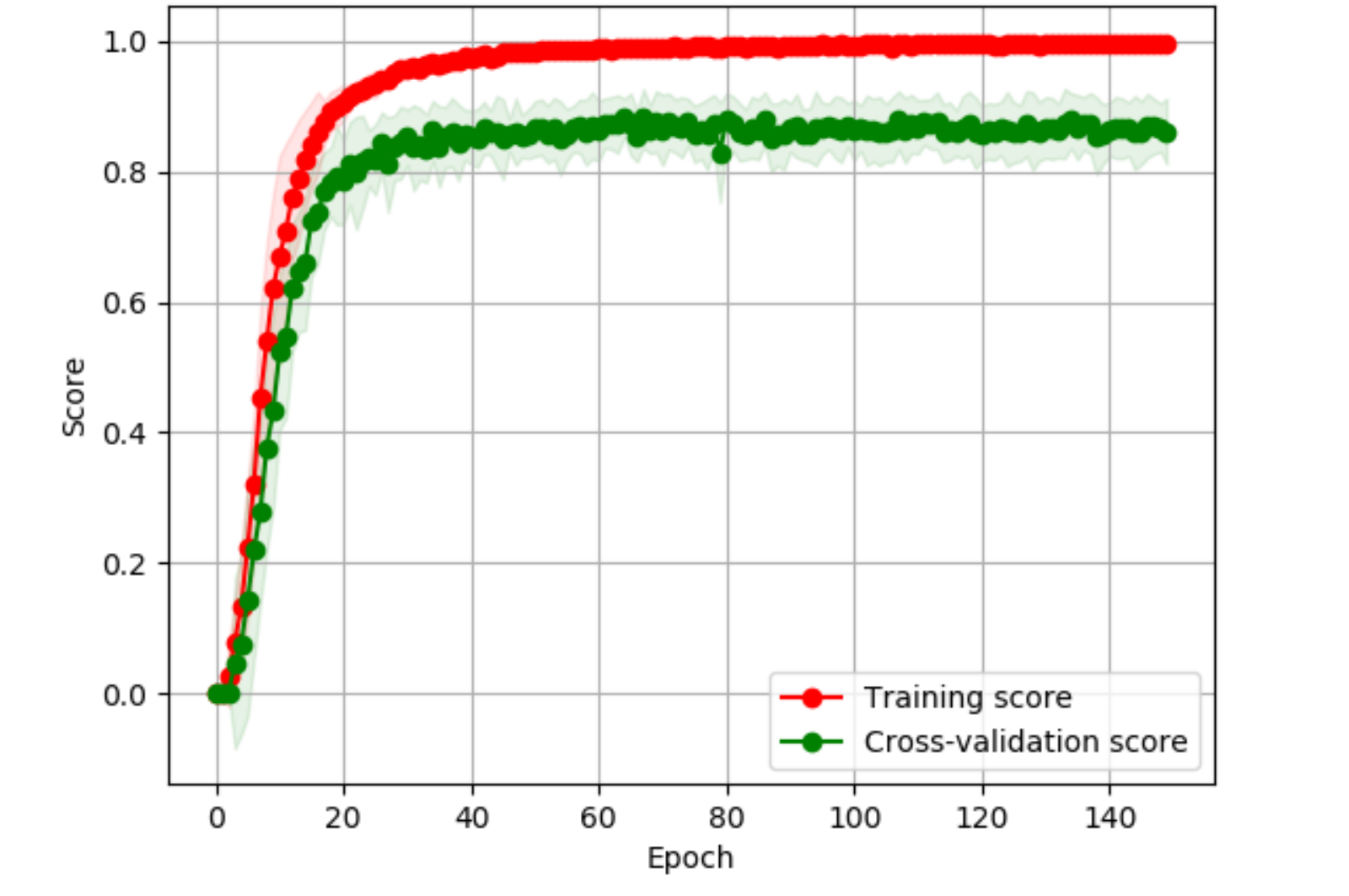}
 \caption{RNN-ATT}
 \label{fig:RNN_ATT_Curve}
\end{subfigure}%
\caption{Learning curves: Each curve shows the average and standard deviation (shaded regions) 
of the MicroF1 precision score in 150 epochs for each model after cross-validation. The training and validation set contained 20,439 and 2,271 samples respectively.} 
\label{fig:acccurves}
\end{figure*}

Cross-validation experiments using $k$-folds with $k=11$ were performed in order to attain accuracy curves and compare the learning pattern up to 150 epochs for the four models (CNN, CNN-ATT, RNN and RNN-ATT). The results are shown in Fig. \ref{fig:acccurves}.
These experiments allow us to understand the dispersion in accuracy attributable to the differences in the data distribution from the training and validation sets. 

We draw the following conclusions:
\begin{itemize}
\item The RNN outperformed the CNN.
\item Attention mechanisms increased the performance of both the CNN and the RNN models with the following particularities: The gain in the overall MicroF1 was more pronounced in the CNN-ATT model, but it had a slower but steadier increase when compared to the plain CNN. Conversely, the accuracy gain of the RNN-ATT versus the plain RNN was less appreciable but the model learned faster 
in early epochs. 
\item Accuracy suffers from a higher variance in the CNN-ATT as can be seen from the  wider standard deviation interval compared to that of the RNN-ATT.
\item For all four methods, in an attempt to reduce the variability in validation metrics, we experimented reducing the size of the training set in favor of a larger validation set but observed that the model's predictive performance deteriorated as regards both the training and validation metrics, particularly in the CNN architecture. We consequently, hypothesize that this architecture seems to benefit more from adding a larger number of instances, but this could not be proven given the available size of the labeled dataset. 
\end{itemize}

A random sample of 500 sentences drawn from the partition of the PadChest dataset and not used to train or validate the model was further labeled manually and used as an independent test set. In order to test the suitability of the trained model as regards its generalization to the whole PadChest dataset, this sample test was drawn from x-reports belonging to a different period (2009 to 2013) with respect to those of the training and validation sets (2014-2017).

The best model from among the four different architectures was further tested in this test set, obtaining a MicroF1 score of 0.93, as shown in Tab. \ref{tab:AccMetrics}.  This result was similar to those achieved in the validation set, illustrating its suitability for use in the automatic annotation of the PadChest dataset samples that were not labeled manually. The RNN-ATT model was consequently used to extract the annotations for the remaining PadChest dataset, as shown in Tab. \ref{tab:datasetSize}.

\section{Dataset overview}
\label{sec:dataset}

\begin{table*}[t]
\begin{footnotesize}
    \centering
    \begin{tabular}{lp{13cm}l}
        \textbf{ Name } & \textbf{Description } \\
        \hline
        ImageID & Image identifier \\
        ImageDir & Zip folder containing the image  \\
        StudyID & Study identifier \\
        PatientID & Patient's code  \\
        PatientBirth & Year in format \texttt{YYYY}  \\
        Projection & Classification of the 5 main x-ray projections: \texttt{PA} (Postero-Anterior standard), \texttt{L} (Lateral), \texttt{AP} (Antero-Posterior erect or vertical), \texttt{AP-horizontal} (Antero-Posterior horizontal), \texttt{COSTAL} (rib views) \\
        Pediatric & \texttt{PED} if the image acquisition followed a pediatric protocol\\
        MethodProjection & The method applied in order to assign a projection type: based on a manual review of DICOM fields or based on the classification output of a pre-trained ResNet50 model for images without DICOM information about the projection. \\
        ReportID & Integer identifier of the report\\
        Report & A text snippet extracted from the original report containing the radiographical interpretation. The text is preprocessed, while the words are stemmed and tokenized. Each sentence is separated by `.' \\
        MethodLabel & The method applied for manual labeling, manually by physicians (\texttt{Physician}) or supervised (\texttt{RNN\_model}) \\
        Labels & A sequence of unique labels extracted from each report\\
        Localizations & A sequence of unique anatomical locations extracted from each report. Each anatomical location is always preceded by the token \texttt{loc} \\
        LabelsLocalizationsBySentence & Sequences of labels followed by its anatomical locations.         Each single sequence corresponds to the labels and anatomical locations extracted from a sentence and repeats the pattern formed of one label followed by no or many locations for this label \texttt{[ label, (0..n) loc name ]}. The sequences are ordered by sentence order in the report \\
        LabelCUIS & A sequence of UMLS Metathesaurus CUIs corresponding to the extracted labels in the Labels field \\
        LocalizationsCUIS & A sequence of UMLS Metathesaurus CUIs corresponding to the extracted anatomical locations in the Localizations field\\
        \hline
    \end{tabular}

    \end{footnotesize}

    \caption{Dataset Fields: All additional processed fields that are different from original DICOM fields. Additional information on UMLS Metathesaurus CUIs can be found at  \url{https://uts.nlm.nih.gov/home.html}.}
    \label{tab:dataset_fields}
\end{table*}

\begin{table*}[t]
\begin{footnotesize}
    \begin{tabular}{l l p{11.5cm}l}
        \textbf{Name} & \textbf{DICOM tag} & \textbf{Description } \\
        \hline
        StudyDate\_DICOM  & 0008,0020 & \texttt{YYYY-MM-DD} Starting date of the study \\
        PatientSex\_DICOM & 0010,0040 & Sex of the patient. \texttt{M} male, \texttt{F} female, \texttt{O} other \\
        ViewPosition\_DICOM & 0018,5101 & Radiographic view of the image relative to the subject's orientation \\
        Modality\_DICOM & 0008,0060 &  Type of equipment used to acquire the image data \\
        Manufacturer\_DICOM & 0008,0070 & Manufacturer of the equipment that produced the images \\
        PhotometricInterpretation\_DICOM &  0028,0004 & Intended interpretation of the pixel data represented as a single monochrome image plane. \texttt{MONOCHROME1}: the minimum sample value should be displayed in white. \texttt{MONOCHROME2}: the minimum sample value is intended to be displayed in black \\
        PixelRepresentation\_DICOM & 0028,0103 & Data representation of the pixel samples, unsigned integer or 2's complement \\
        PixelAspectRatio\_DICOM & 0028,0034 & Ratio of the vertical size and horizontal size of the pixels in the image specified by a pair of integer values, in which the first value is the vertical pixel size and the second value is the horizontal pixel size\\
        SpatialResolution\_DICOM & 0018,1050 & The inherent limiting resolution in mm of the acquisition equipment \\
        BitsStored\_DICOM & 0028,0101 & Number of bits stored for each pixel sample  \\
        WindowCenter\_DICOM & 0028,1051 & Window width for display  \\
        WindowWidth\_DICOM & 0028,1050 & Window center for display. \\
        Rows\_DICOM & 0028,0010 & Number of rows in the image  \\
        Columns\_DICOM & 0028,0011 & Number of columns in the image \\
        XRayTubeCurrent\_DICOM & 0018,1151 & X-ray Tube Current in mA  \\
        ExposureTime\_DICOM & 0018,1150 & Duration of x-ray exposure in msec \\
        Exposure\_DICOM & 0018,1152 & Exposure expressed in mAs, calculated from Exposure Time and x-ray Tube Current \\
        ExposureInuAs\_DICOM & 0018,1153 & Exposure in $\mu$As 	 \\
        RelativeXRayExposure\_DICOM & 0018,1405 & Relative x-ray exposure on the plate \\
    \end{tabular}
\end{footnotesize}

    \caption{DICOM dataset fields: DICOM® (Digital Imaging and Communications in Medicine) is the international standard employed to transmit, store, retrieve, print, process, and display medical imaging information. Detailed  descriptions of DICOM standard fields can be found in \cite{DICOM}.}
    \label{tab:dataset_fields_DICOM}
\end{table*}

The dataset consists of 160,868 labeled chest x-ray images from 69,882 patients, acquired in a single institution between 2009 and 2017 (see Tab.  \ref{tab:datasetSize}). The patients had a mean of 1.62 chest x-ray studies performed at different time points (from 1 to 119). Each study contains one or more images corresponding to different position views, mainly P-A and lateral, and is associated with a single radiography report describing the results of all position views in a common text section. The mean of the chest-x ray images  was 2.37 per patient. The dataset generated provides two types of fields for each chest-x ray image: those fields with the suffix DICOM  \ref{tab:dataset_fields_DICOM} contain the values of the original field in the DICOM standard and the remaining fields \ref{tab:dataset_fields} enrich the PadChest dataset with additional processed information.   

\begin{table}
\centering
\begin{small}
\begin{tabular}{l r r r}
 & \textbf{Patients} & \textbf{Reports}&\textbf{Images} \\
\hline
\textbf{Initial Set}&  69,882
 & 115,678  & 168,171
 \\
\hline
\textbf{Excluded Set}&  4,744 & 5,864  & 7,303 \\
Photometric interpretation && &5,652\\
Modality && & 871\\
Study Report && &705\\
Position View&& &42\\
Protocol && &23\\
Pixel Data && &10\\

\hline
\textbf{Ground Truth } &   &   &   \\
Manual Report Labeling &  24,491 & 27,593  & 39,039  \\
Automatic Report Labeling & 43,134 & 82,338 &  121,829 \\
\hline
\textbf{Final Labeled Set} & 67,625 &109,931 & 160,868
 \\
\hline
\end{tabular}
\end{small}
\caption{PadChest global statistics. 
}
\label{tab:datasetSize}
\end{table}

\begin{table}[ht]
\centering
\begin{small}
\begin{tabular}{l r}
& \textbf{Images} \\
\hline
\textbf{Projection and Positioning} 
\\
PA & 96,010\\ 
L  & 51,124\\ 
AP-Horizontal & 14,355 \\
AP-Vertical & 5,158\\
Costal & 631\\
Pediatric & 274\\
\hline
\textbf{Gender} & \\
Male & 80,923\\
Female & 79,923\\
\hline
\end{tabular}
\end{small}
\caption{Descriptive statistics: Categorical variables.}
\end{table}

\begin{table}
\centering
\begin{small}
\begin{tabular}{l r r r}
 & \textbf{Mean \textpm\ \SI{} Std} & \textbf{Min} & \textbf{Max}\\
\hline
Year of Birth &  1953 \textpm\ \SI{20} & 1904 & 2017 \\
Age &  58.5 \textpm\ \SI{20} & 0 & 105 \\
Studies per patient &  1.62 \textpm\ \SI{1.66} & 1 & 116 \\
Images per patient &  2.37 \textpm\ \SI{2.34} & 1 & 119 \\
Images per study &  1.46 \textpm\ \SI{0.53} & 1 & 11 \\
Rows per image &  2254 \textpm\ \SI{739} & 118 & 4280 \\
Columns per image &  2299 \textpm\ \SI{748} & 20 & 4280 \\
Spatial Resolution &  0.16 \textpm\ \SI{0.02} & 0.1 & 0.2\\ 
\hline
\end{tabular}
\end{small}
\caption{Descriptive statistics: Continuous variables.}
\label{tab:datasetVar}
\end{table}

\begin{figure*}
\centering
\begin{subfigure}{.33\linewidth}
  \centering
  \includegraphics[width=1\linewidth]{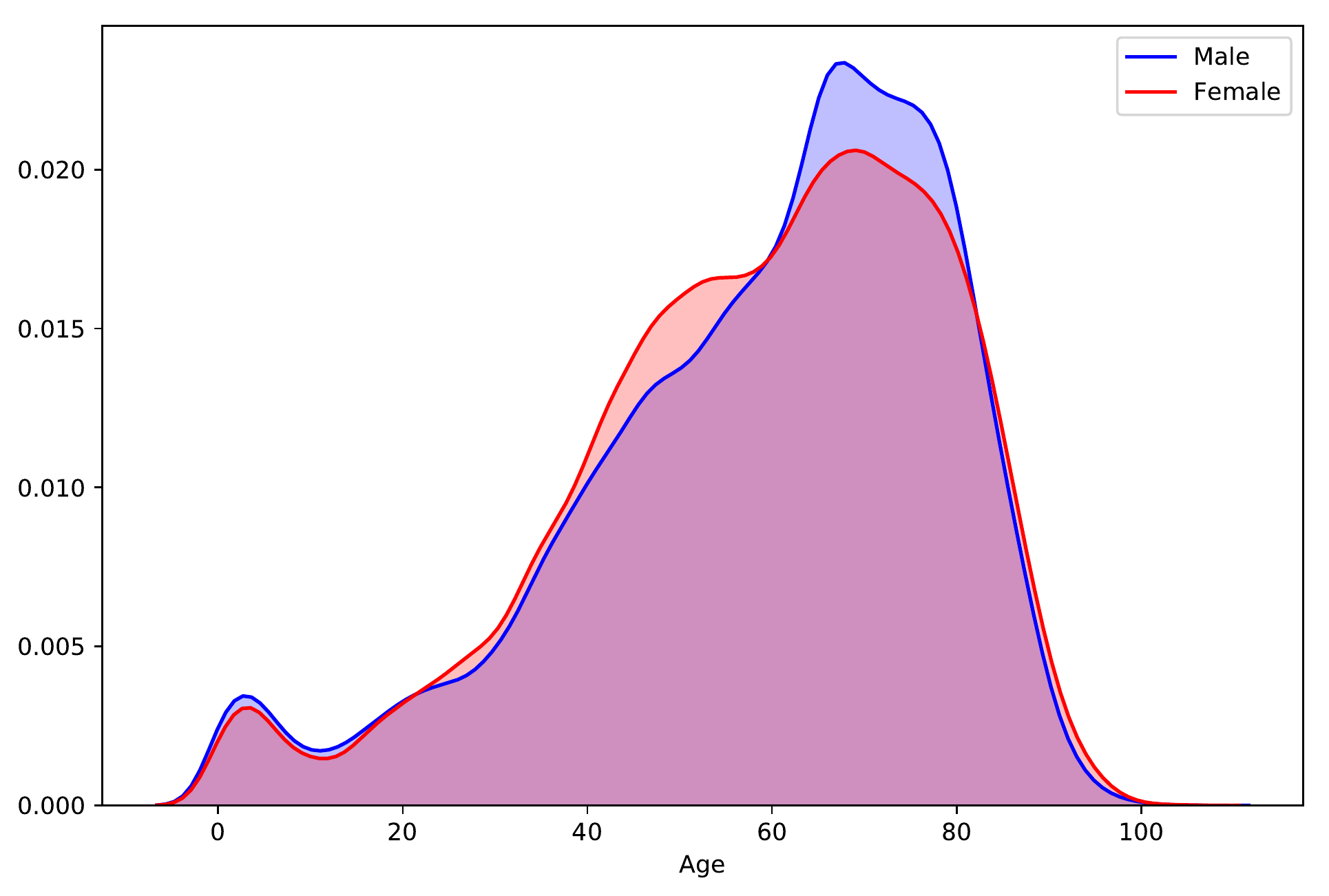}
  \label{fig:AgeDistributionbySex}
    \vspace{-0.4cm}
  \caption{Ages by gender}
\end{subfigure}
\begin{subfigure}{.33\linewidth}
  \centering
  \includegraphics[width=1\linewidth]{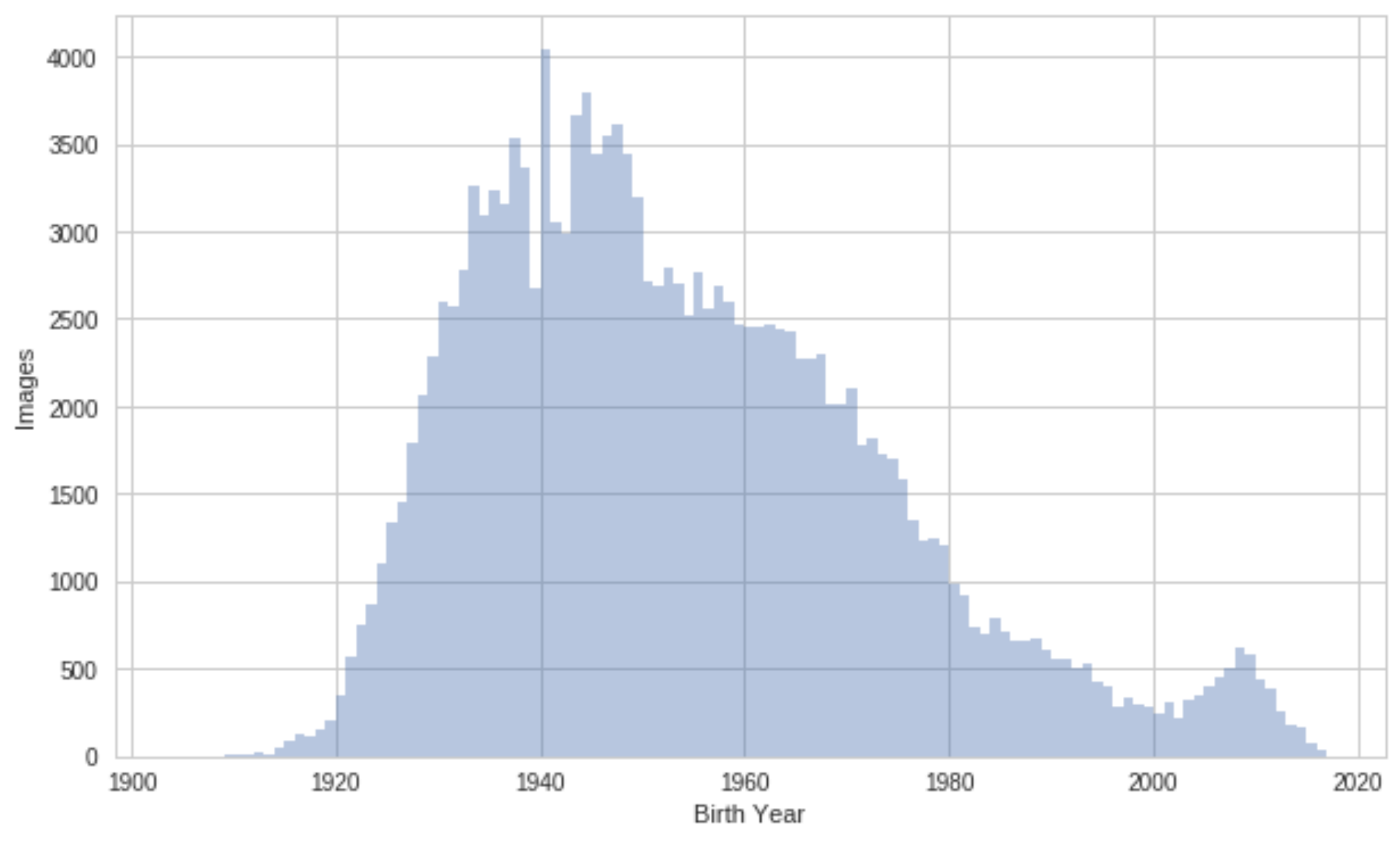}
  \label{fig:BirthYears}
  \vspace{-0.4cm}
  \caption{Birth year} 
\end{subfigure}
\begin{subfigure}{.33\textwidth}
  \centering
  \includegraphics[width=1\linewidth]{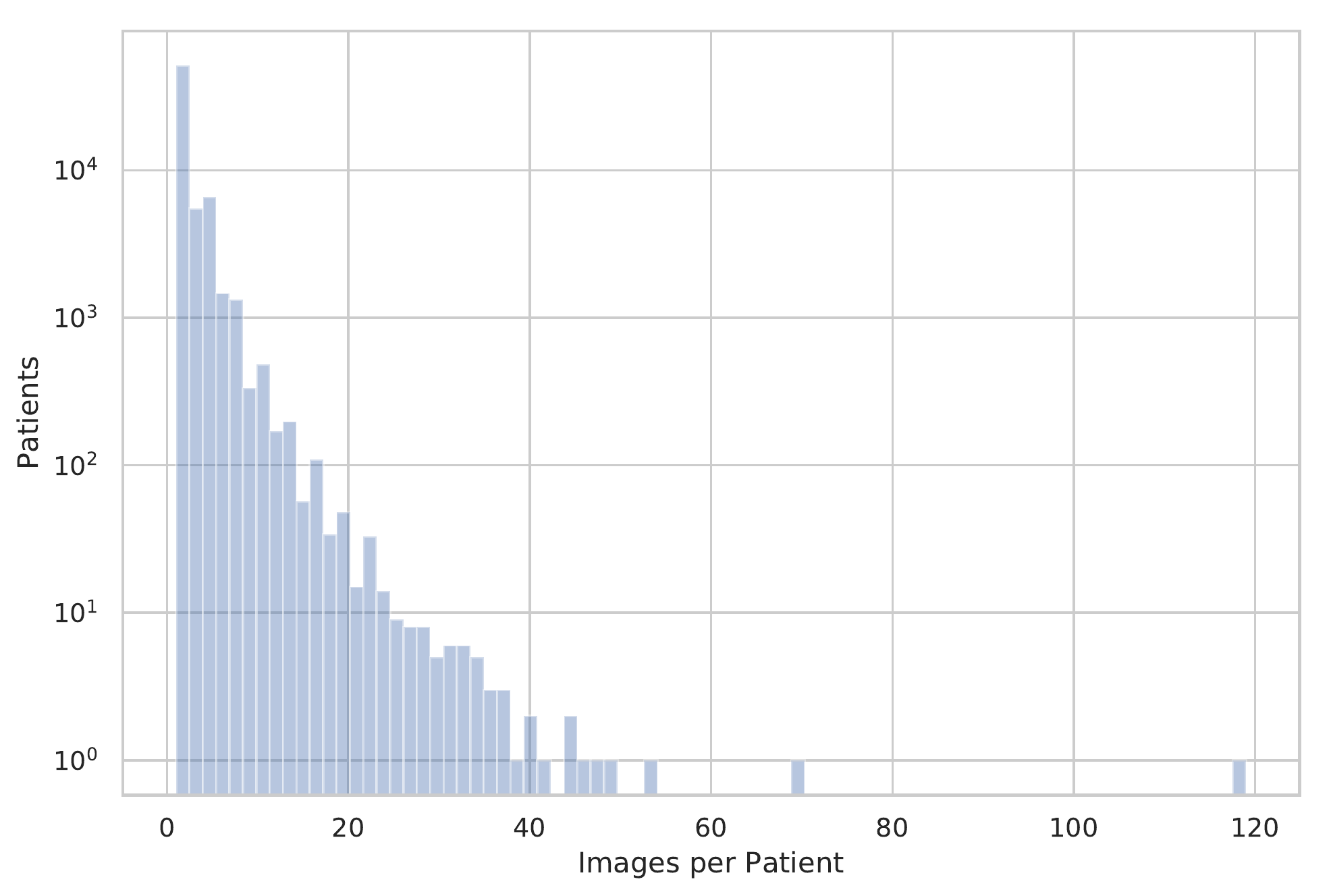}
  \label{fig:StudyPatient}
    \vspace{-0.4cm}
  \caption{Images per patient} 
\end{subfigure}
\begin{subfigure}{.33\textwidth}
  \centering
  \includegraphics[width=1\linewidth]{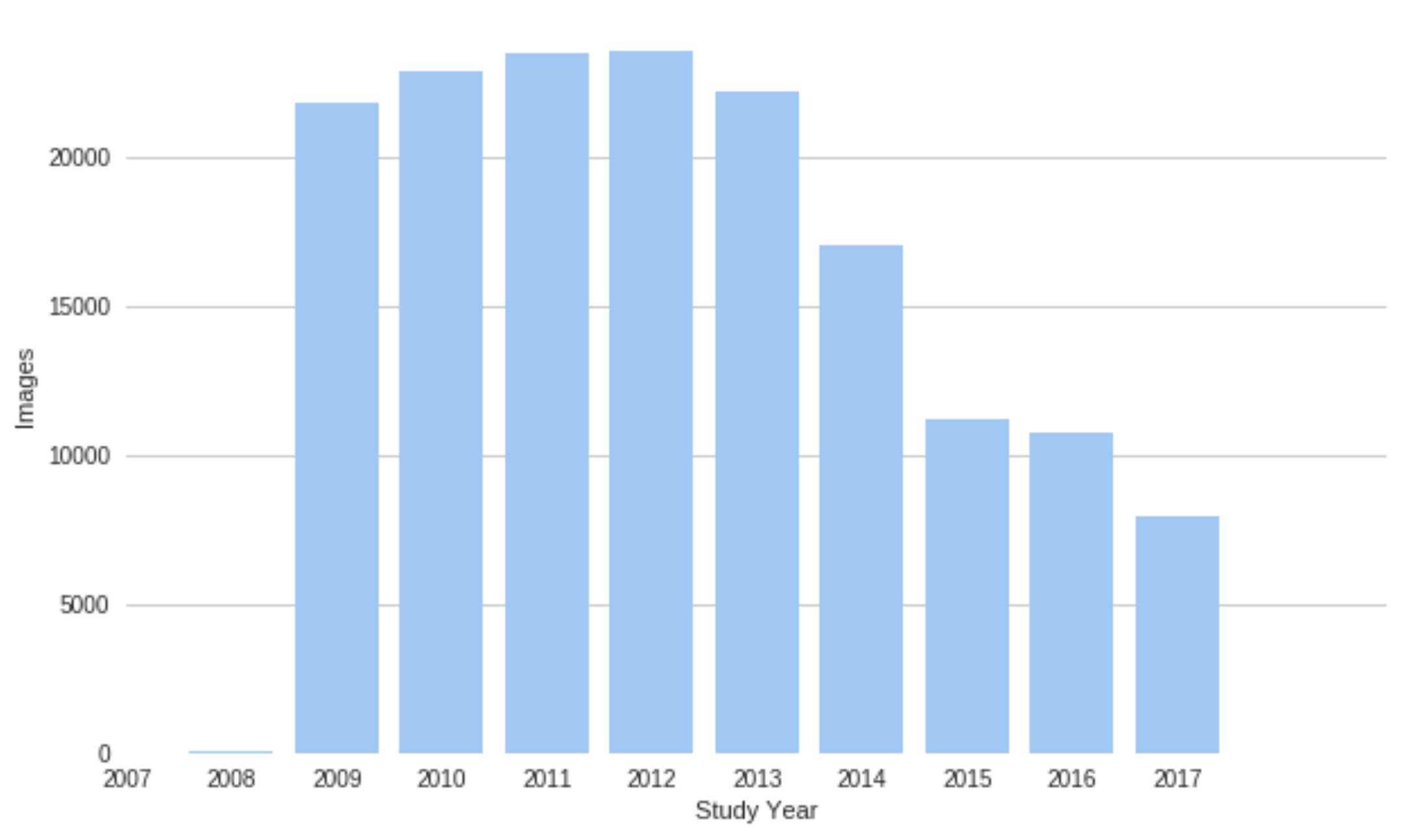}
  \label{fig:StudyYears}
    \vspace{-0.4cm}
  \caption{Year of study} 
\end{subfigure}
\begin{subfigure}{.33\textwidth}
  \centering
  \includegraphics[width=1\linewidth]{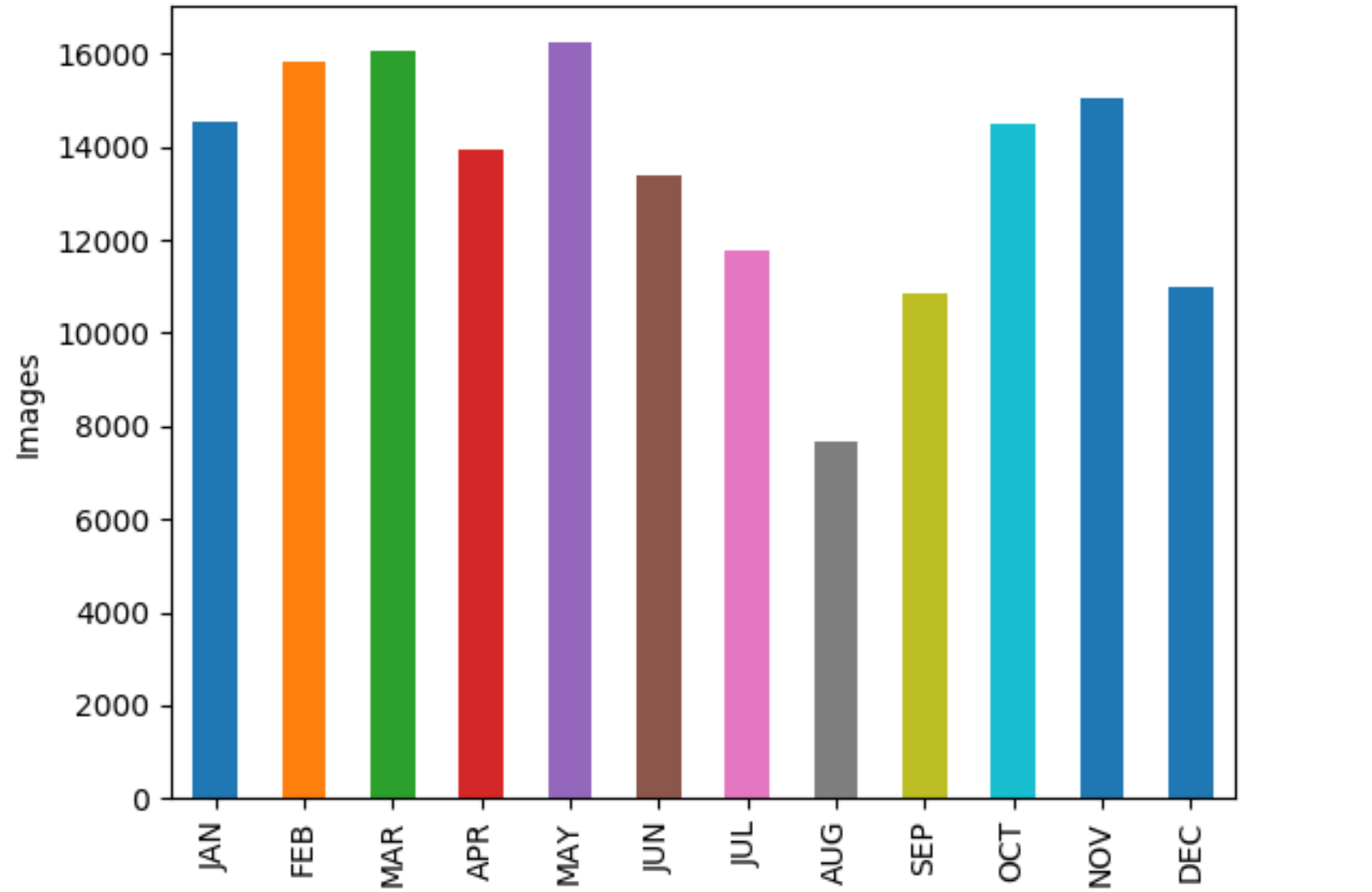}
  \label{figs:StudyMonthHistogram}
    \vspace{-0.4cm}
  \caption{Month of study}
\end{subfigure}
\begin{subfigure}{.33\textwidth}
  \centering
  \includegraphics[width=1\linewidth]{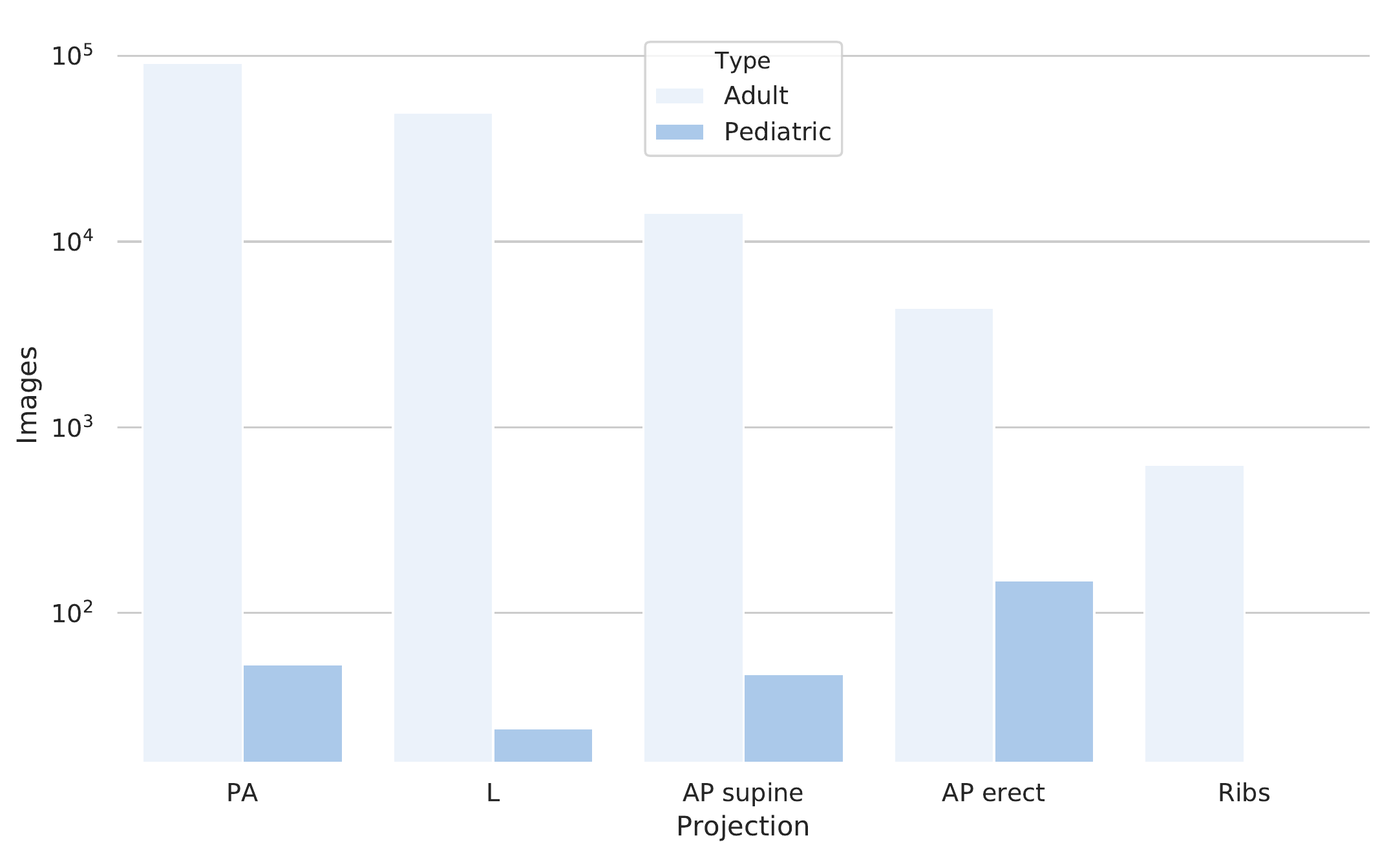}
  \label{fig:PositionView}
    \vspace{-0.4cm}
  \caption{Position view}
\end{subfigure}
\caption{Dataset descriptive statistics. a) Kernel density estimation of the probability distribution of images by patient's age and gender. b)  Statistics of the year of birth. c) Number of different images per patient. d), e) and f) Image year, month and type of projection:  Postero-anterior (PA), Lateral (L), Antero-posterior (AP) supine and erect histograms, respectively. Note that in the case of the subset of images with projection not retrievable  (UNK) from DICOM, a CNN model was used to classify them into L and PA.}
\label{fig:DescriptiveStatistics1}
\end{figure*}

\begin{figure*}
\begin{subfigure}{.33\textwidth}
  \centering
  \includegraphics[width=1\linewidth]{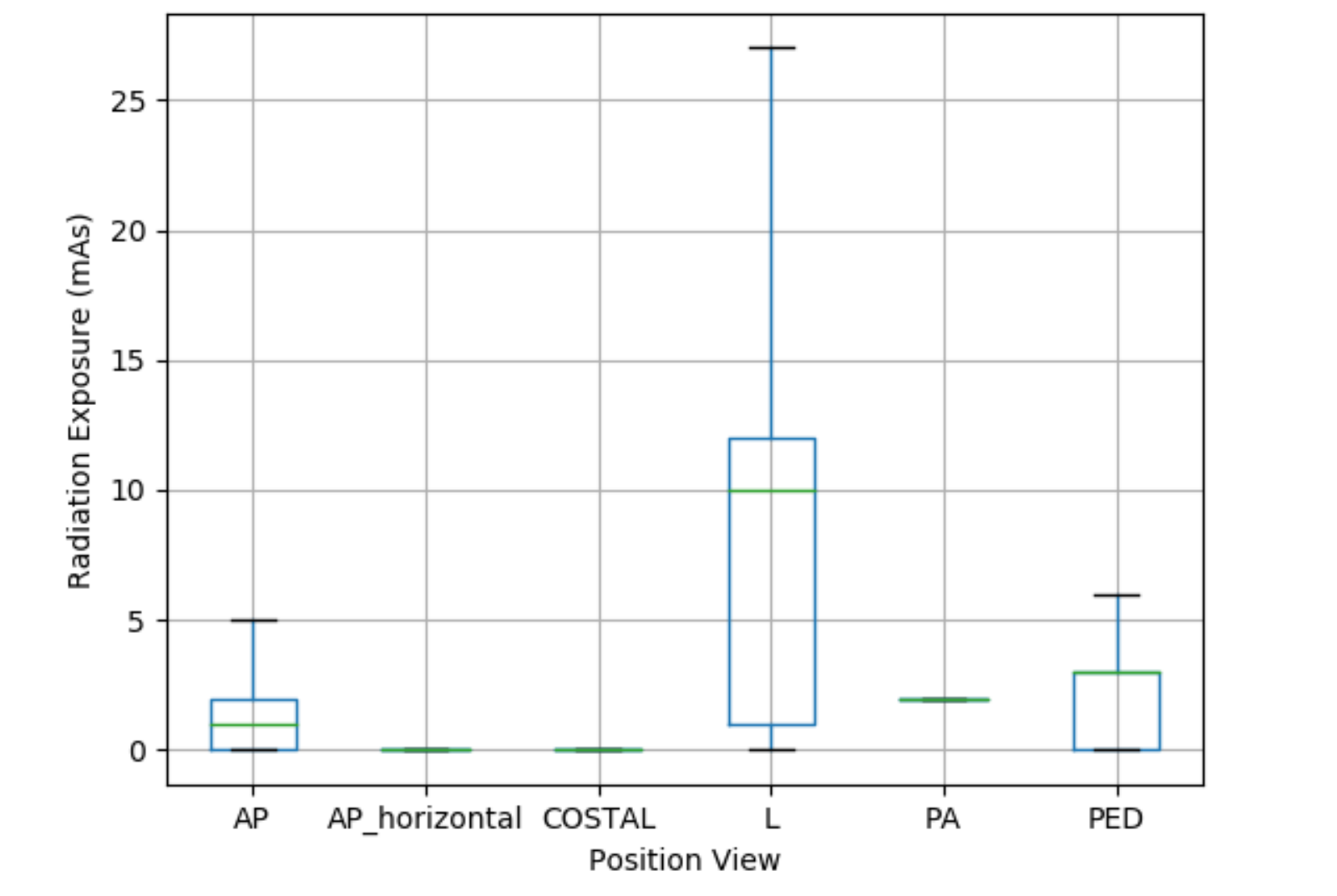}
  \label{fig:RadiationExposureByPositionView}
      \vspace{-0.4cm}
  \caption{Radiation exposure by projection}
\end{subfigure}
\begin{subfigure}{.33\textwidth}
  \centering
  \includegraphics[width=1\linewidth]{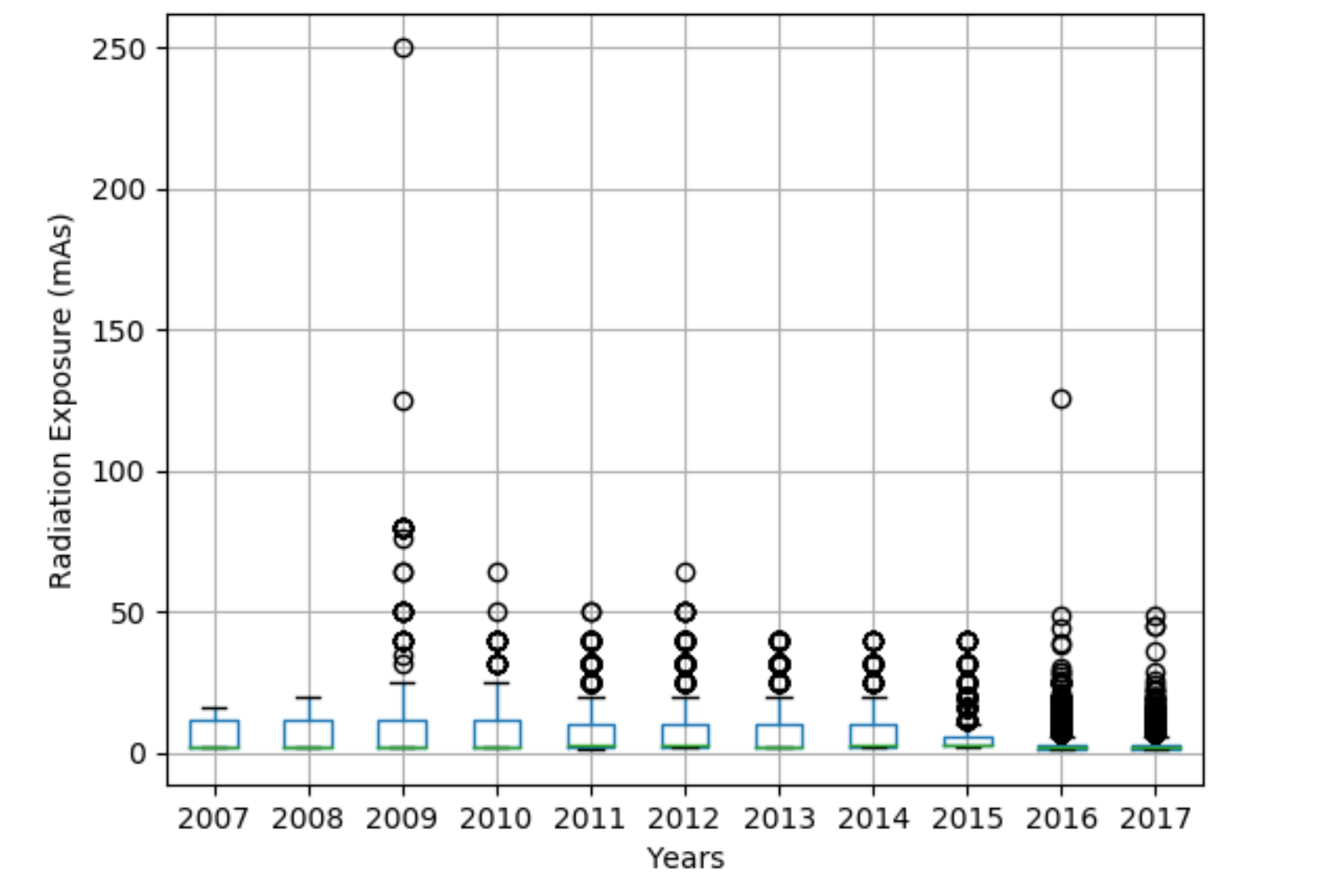}
  \label{fig:RadiationExposureByYear}
      \vspace{-0.4cm}
  \caption{Radiation exposure by year}
\end{subfigure}
\begin{subfigure}{.33\textwidth} 
  \centering
  \includegraphics[width=1\linewidth]{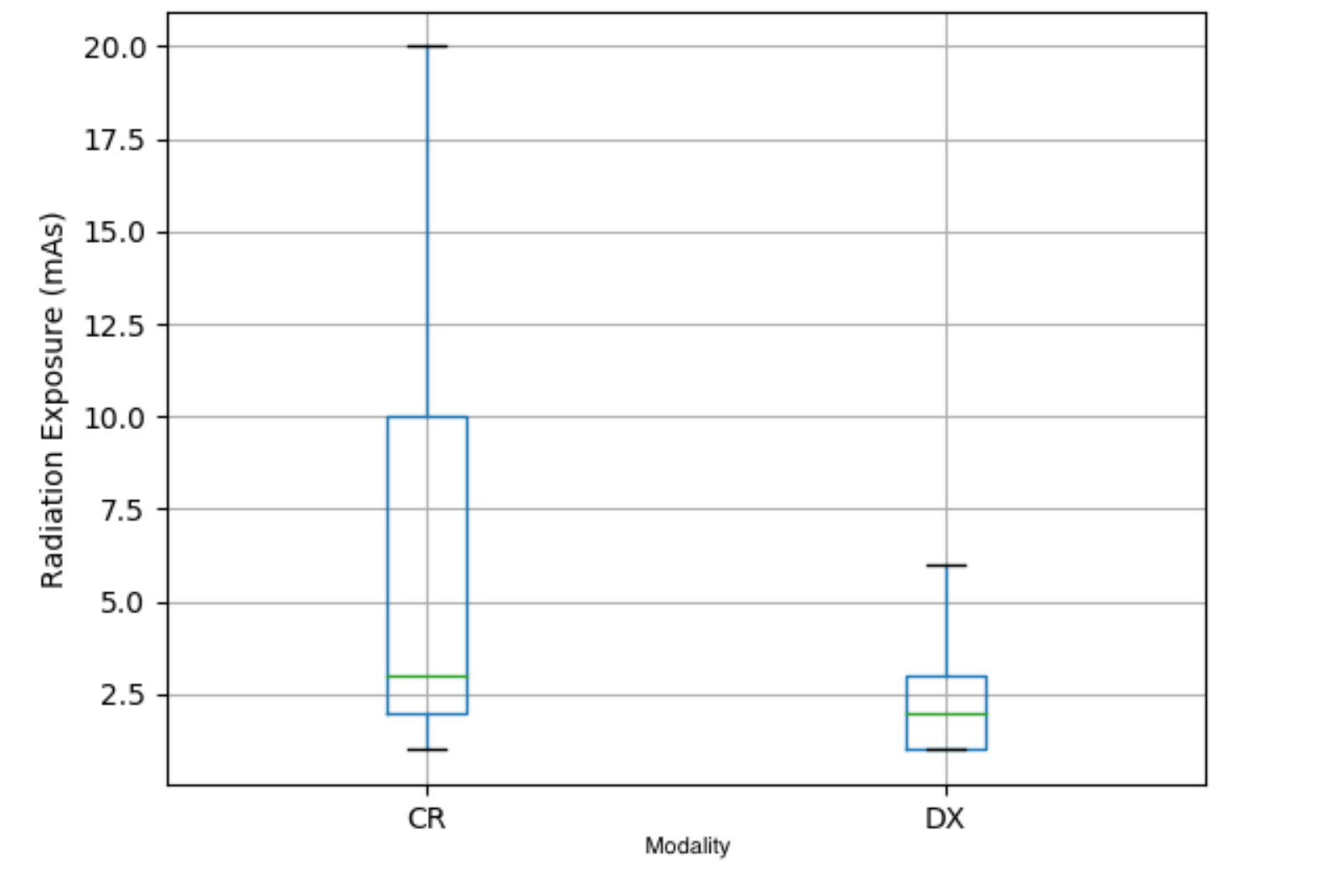}
  \label{fig:RadiationExposureByModality}
    \vspace{-0.4cm}
  \caption{Radiation exposure by modality}
\end{subfigure}
\caption{Dataset radiation statistics. Radiation exposure was plotted against a) projections, b) years of study and c) modality (computed radiography CR versus digital radiography DX).}
\label{fig:DescriptiveStatistics2}
\end{figure*}

\subsection{Patient Characteristics}

The patients’ ages range from 0 to 105 years, with a mean of 58.5 and a median of 62. The distribution of number of images by age (see Fig. \ref{fig:DescriptiveStatistics1}) is skewed toward older ages with a long tail for ages under 40, and there are also differences according to gender in the age interval from 43 to 60 years, in which more x-rays correspond to women, contrasting with the age range from 71 to 75 in which more correspond to men. In the final labeled dataset, 80,923 images correspond to women and 79,923 to men. The median birth year of the population was 1953, with a standard deviation of 20 within a range from 1904 to 2017, as shown in Tab. \ref{tab:datasetVar}.  

As shown in Fig. \ref{fig:DescriptiveStatistics2}, the radiation exposure (mAs), which is a derived field calculated from the x-ray tube current (mean 250) and the exposure time (mean 31, median 7, max 1,000), largely varied as regards modality (higher in computer CX than in digital radiographies DX), position view (higher in lateral than in frontal views) and year, and had a decreasing trend over time. 

\subsection{Chest X-ray Diagnoses, Radiographic Findings and Anatomical Locations}
\begin{figure}
\centering
  \includegraphics[width=.5\columnwidth]{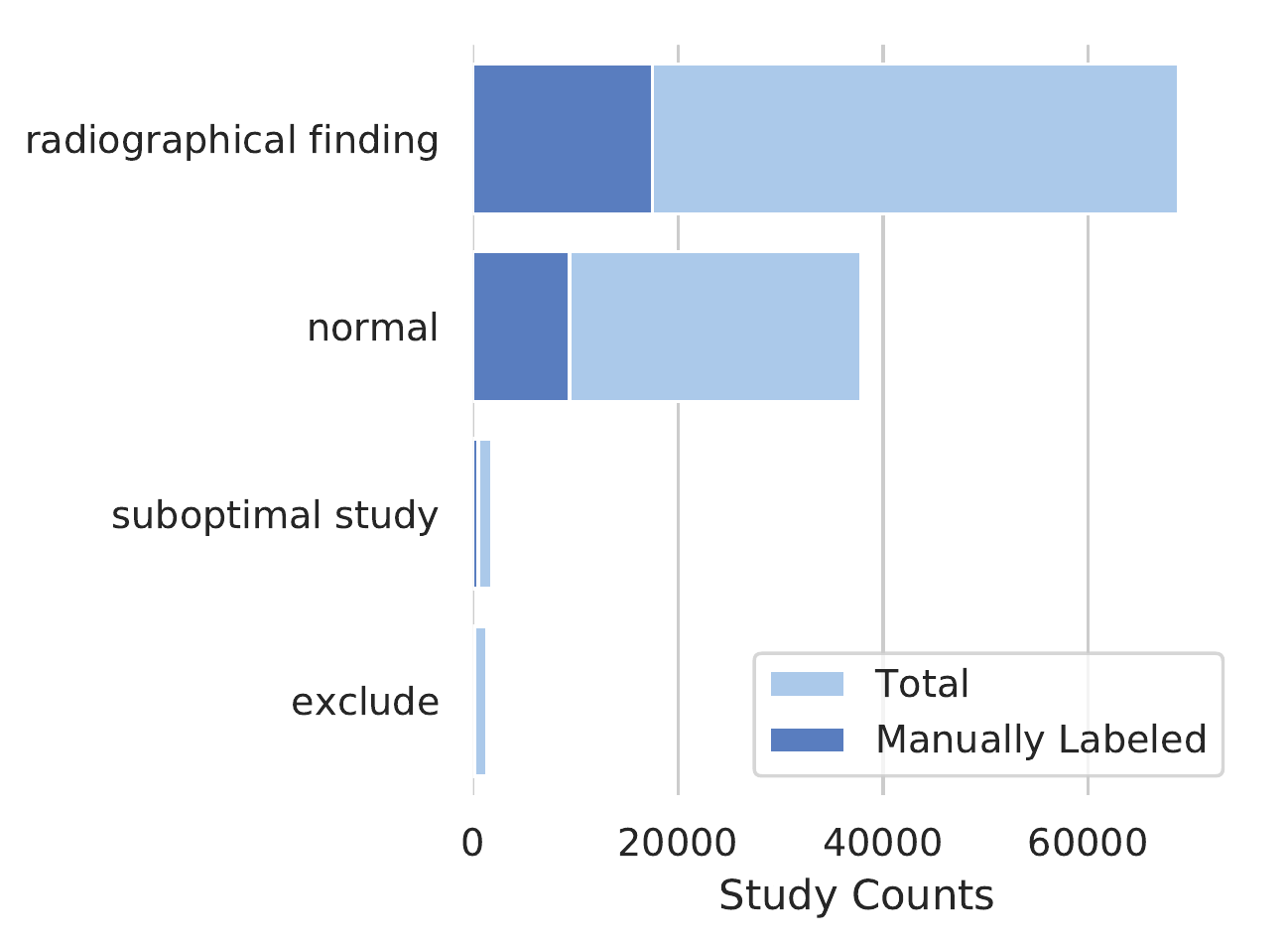}
  \caption{Chest x-ray report types. Data are shown for both the manual (dark color) and automatically labeled dataset (light color). } 
  \label{fig:ReportTypes}
\end{figure}

There are 68,855 studies (17,513 manually annotated by physicians) 
with radiographic findings and/or differential diagnoses, and 37,871 normal studies (of which 9,404 are manually labeled). Of these studies, 1.8\% were reported to have sub-optimal technique and 1.2\% (0.6\% manual) were excluded because no information could be extracted from the reports, as shown in Fig. \ref{fig:ReportTypes}.

\begin{figure}
  \centering
  \includegraphics[width=.8\columnwidth]{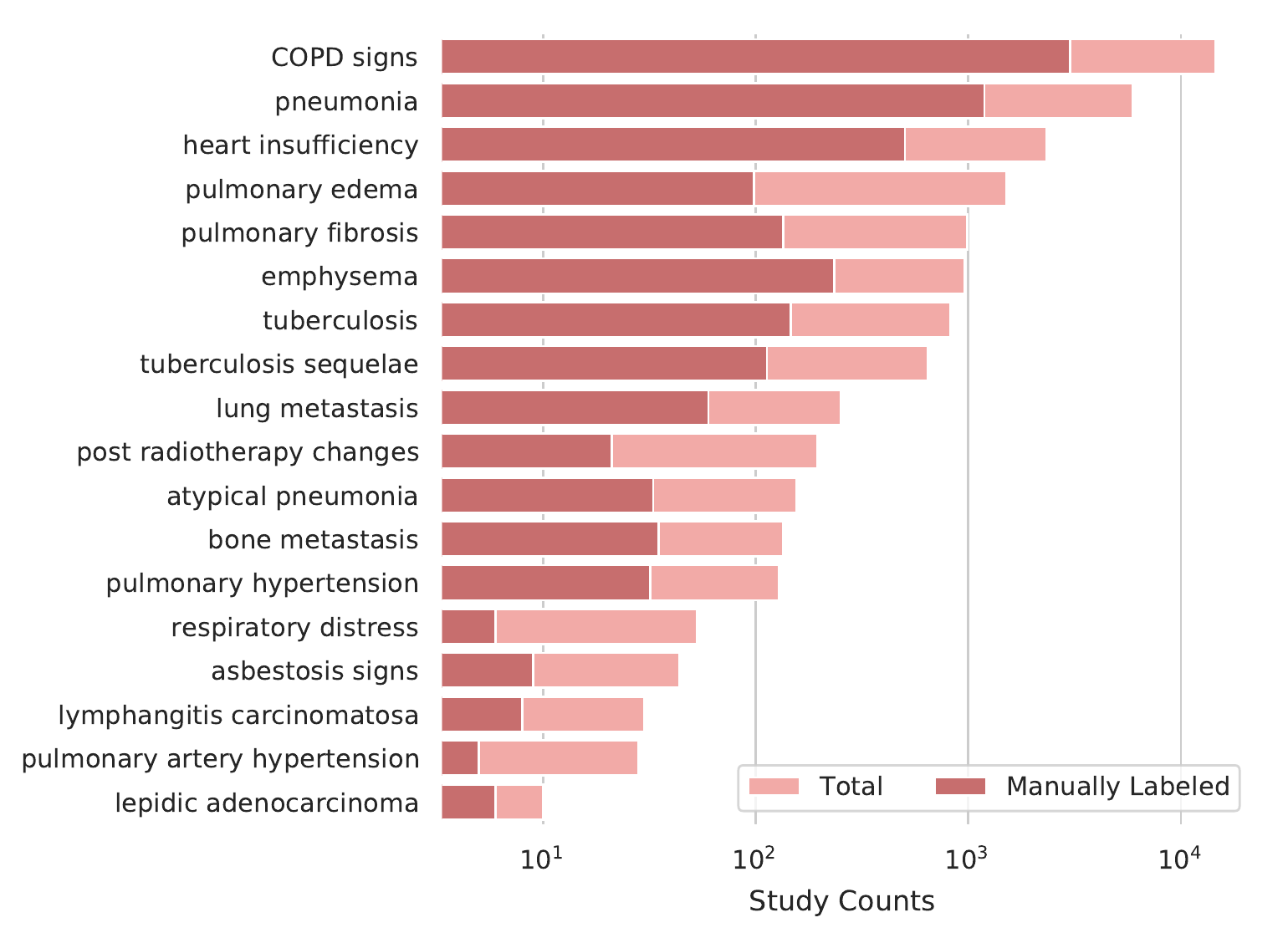}
  \caption{Most frequent differential diagnoses. Data are shown for both the manual (dark color) and automatically labeled dataset (light color). See \ref{app:DD} for counts of labels on each hierarchical tree. } 
  \label{fig:DD} 
\end{figure}

Differential diagnoses were annotated with 19 different labels. In the 27,726 differential diagnoses reported (5,472 manual), the top 4 were COPD (14,557 studies, 3,017 manual), pneumonia (5,934, 1,158 manual) followed by heart insufficiency and pulmonary emphysema, as shown in Fig. \ref{fig:DD}.

Differential diagnoses were annotated with 19 different labels. In the 27,726 differential diagnoses reported (5,472 manual), the top 4 were COPD (14,557 studies, 3,017 manual), pneumonia (5,934, 1,158 manual) followed by heart insufficiency and pulmonary emphysema as shown in Fig. \ref{fig:DD}.

\begin{figure*}[t]
  \centering
  \includegraphics[width=1\linewidth]{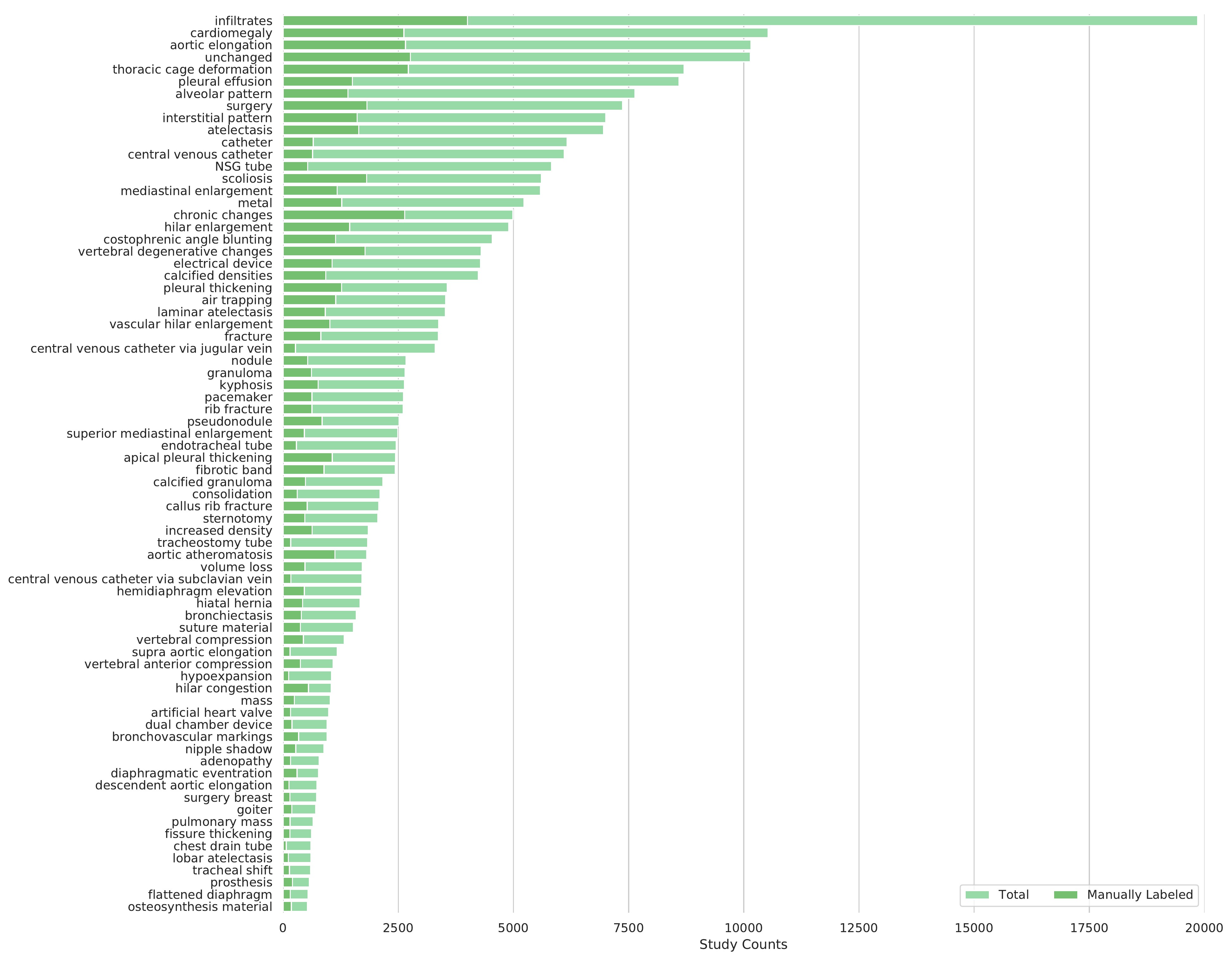}
  \caption{Most reported radiographic findings.}
  \label{fig:RxFindings}
\caption{Most frequent radiographic findings. Labels are shown for both the physician (dark color) and automatically labeled dataset (light color). See \ref{app:rxFinding} for counts of labels on each hierarchical tree.}  
\end{figure*}

Radiographic findings were annotated with 170 different labels. As shown in Fig. \ref{fig:RxFindings}, of the 177,628 (45,375 manual) findings reported, the top 10 were lung infiltrates (19,850), with an interstitial pattern in 7,001 and an alveolar pattern in 7,637 studies, followed by thoracic cage deformations (mainly owing to scoliosis), aortic elongations or dilations (10,154), pulmonary chronic changes, cardiomegaly, surgery marks, vertebral degenerative changes, atelectasis, pleural effusion (8,593) and hilar enlargements (4,895). Masses, nodules, and pseudonodules were reported in 1,026, 2,669 and 2,519 studies, respectively. Aortic atheromatosis, as detected in x-rays by means of vascular calcifications, were reported in 1,813 (1,125 manual) studies and bone fractures in 3,367 (820 manual). All artifacts, such as electric devices (4,286 with 1,063 manual), tubes (10,776 with 945 manual) and catheters (6,167 with 654 manual), were also labeled in order to include their different types. With regard to rare conditions, the Chilaiditi sign was reported in only 14 (6 manual) studies and dextrocardia was present in 14 (4 manual) studies (0.014\%).

\begin{figure*}[t]
\centering
\centering
\includegraphics[width=.8\linewidth]{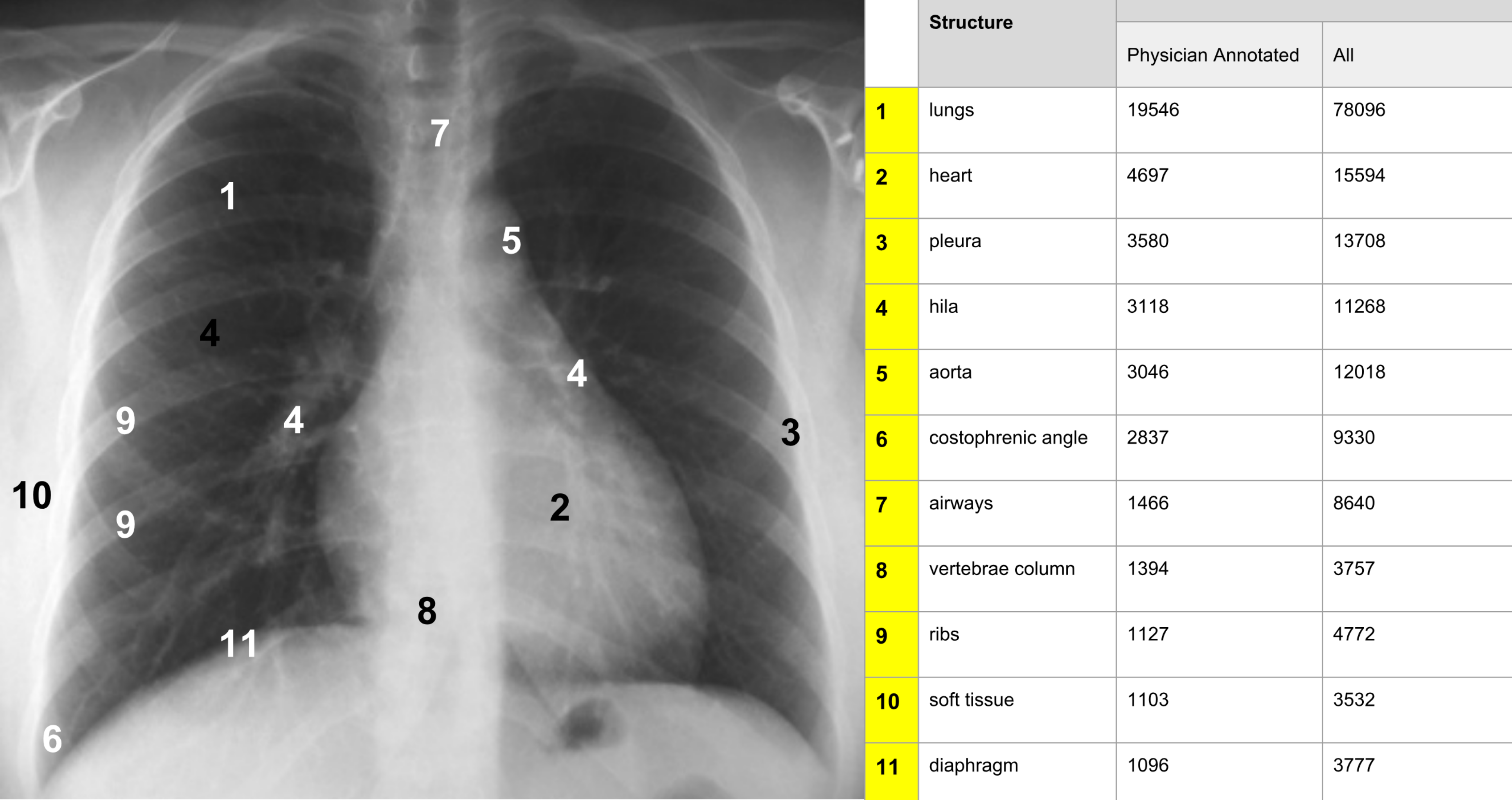}
\caption{Most common locations of radiographic findings and differential diagnoses grouped by anatomical regions. See  \ref{app:localizations} for counts on the locations tree.} 
\label{fig:LocDescription}
\end{figure*}

Anatomic locations were annotated with 104 different labels. In the 249,469 locations specified, the majority of the findings and/or diagnoses were localized in the lung fields (78,888 findings, the majority being - 34,661 - in the lower lung fields), followed by mediastinum (39,134 findings, the majority being in the lower mediastinum in the cardiac region), bone (12,862 findings) and soft tissue (3,567 findings). There were 13,935 findings reported in the pleural region, 11,349 in the hilar region, 9850 aortic findings and 3,777 diaphragmatic findings, as shown in Fig. \ref{fig:LocDescription}.

\begin{figure*}
\centering
\includegraphics[width=1\linewidth]{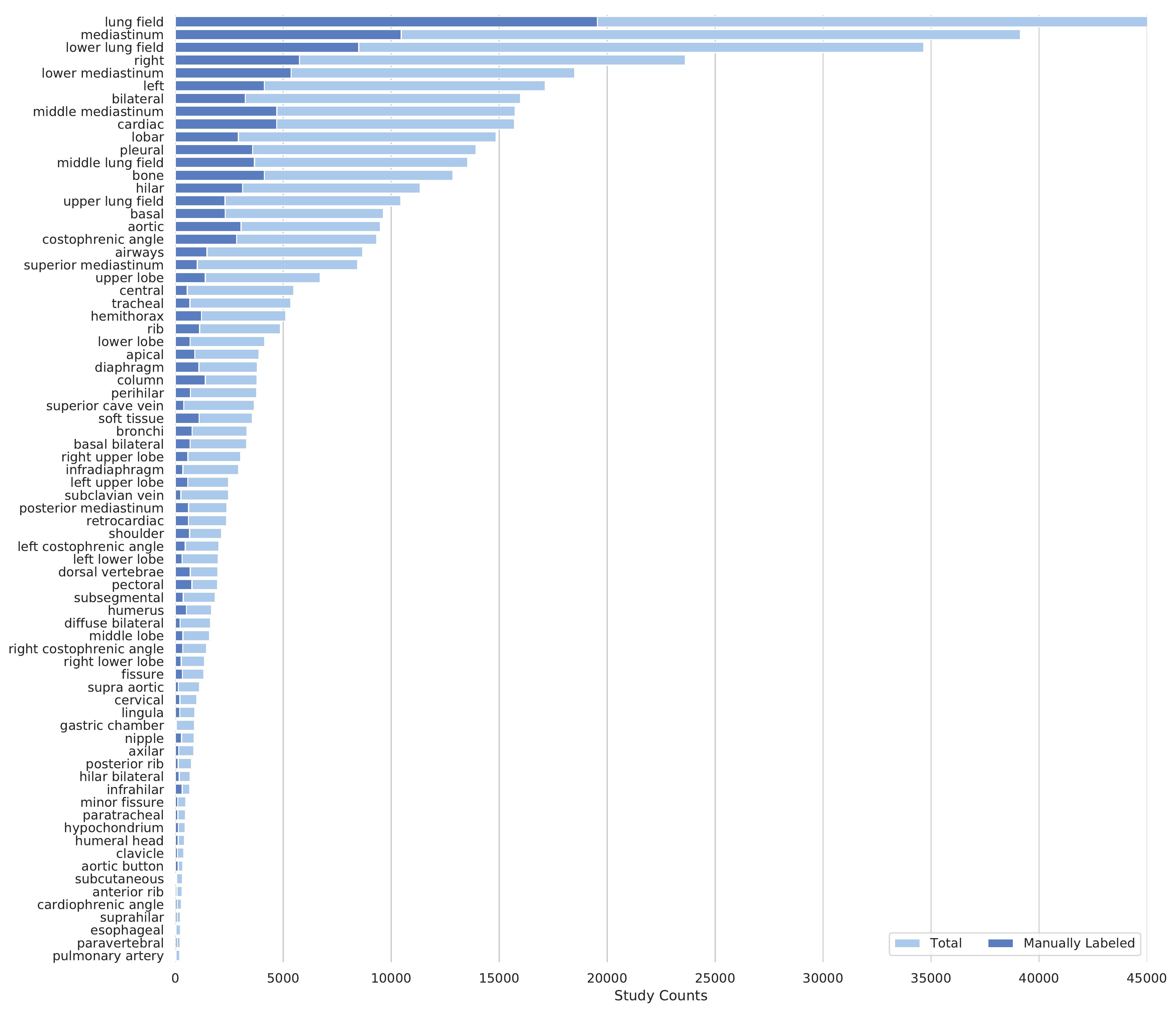}
\caption{Most common locations of radiographic findings and differential diagnoses. See \ref{app:localizations} for counts on the locations tree.} 
\label{fig:LocDescription2} 
\end{figure*}

\subsection{Image projections and additional characteristics}

The mean and standard deviation images were obtained for visualization purposes. They were calculated from a random sample of 500 training examples.  

\begin{figure*}
\centering
\begin{subfigure}{.16\textwidth}
  \centering
  \includegraphics[width=1\textwidth]{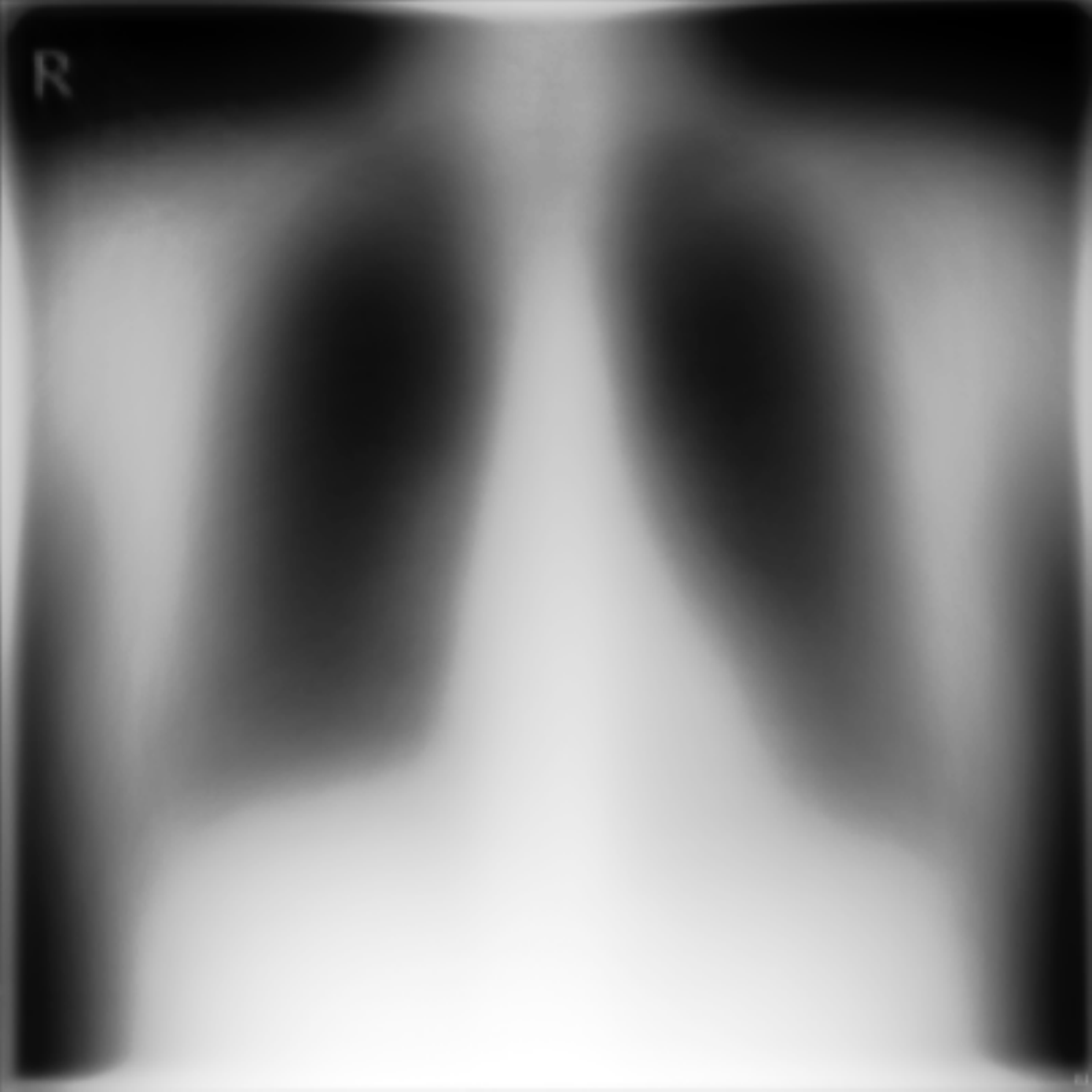}
  \caption{P-A}
  \label{fig:PAMean}
\end{subfigure}
\begin{subfigure}{.16\textwidth}
  \centering
  \includegraphics[width=1\textwidth]{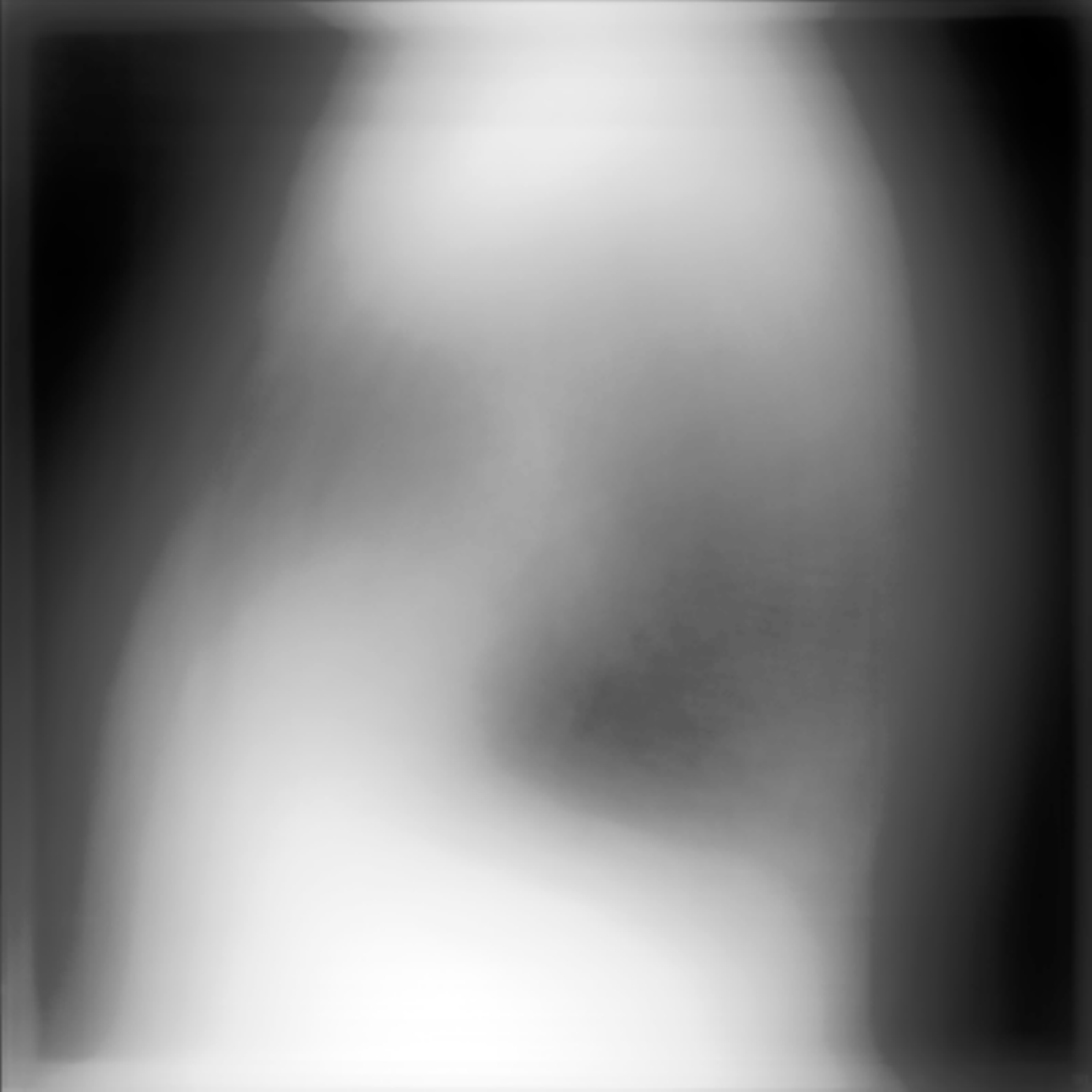}
  \caption{Lateral}
  \label{fig:LMean}
\end{subfigure}
\begin{subfigure}{.16\textwidth}
  \centering
  \includegraphics[width=1\textwidth]{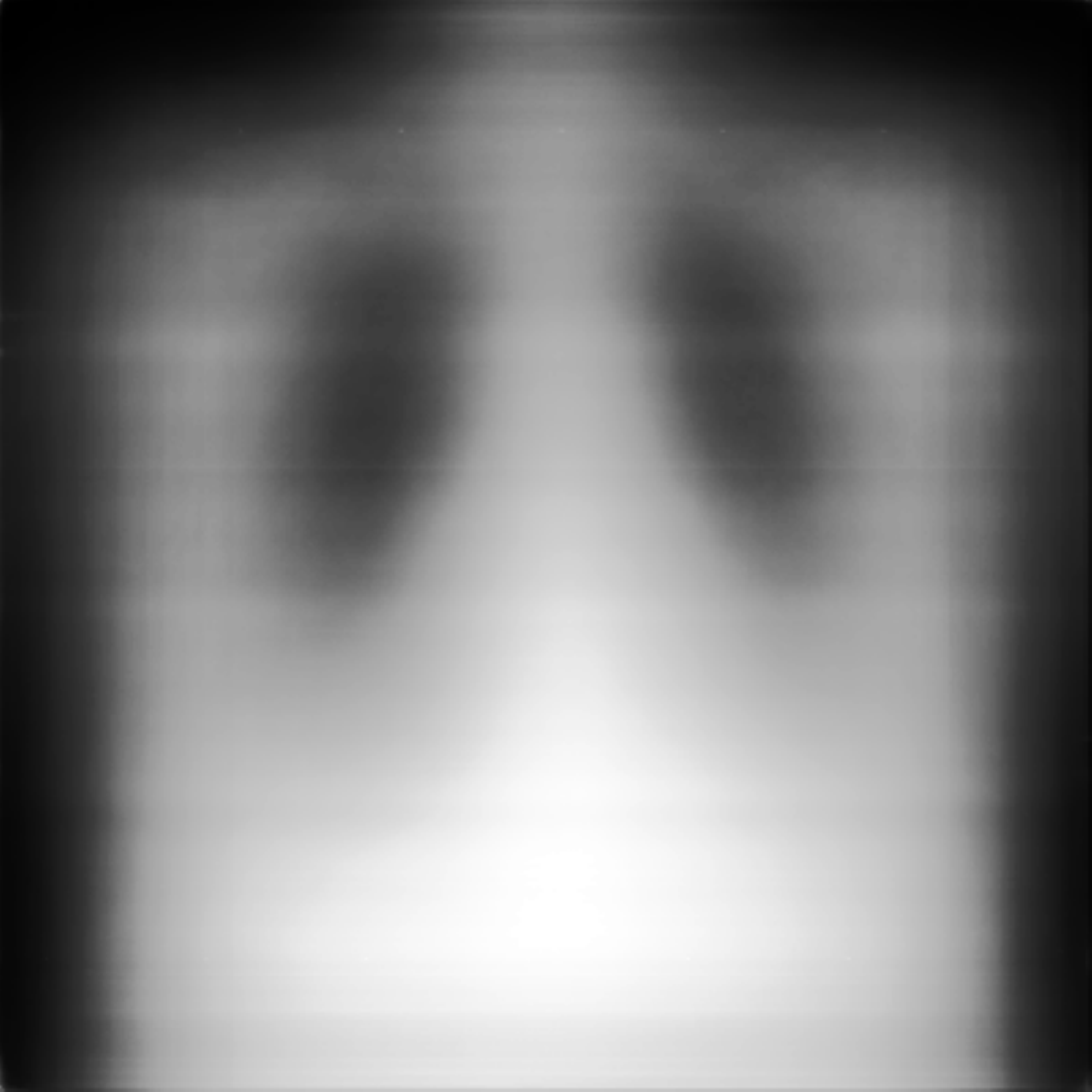}
  \caption{AP supine}
  \label{fig:APSupineMean}
\end{subfigure}
\begin{subfigure}{.16\textwidth}
  \centering
  \includegraphics[width=1\textwidth]{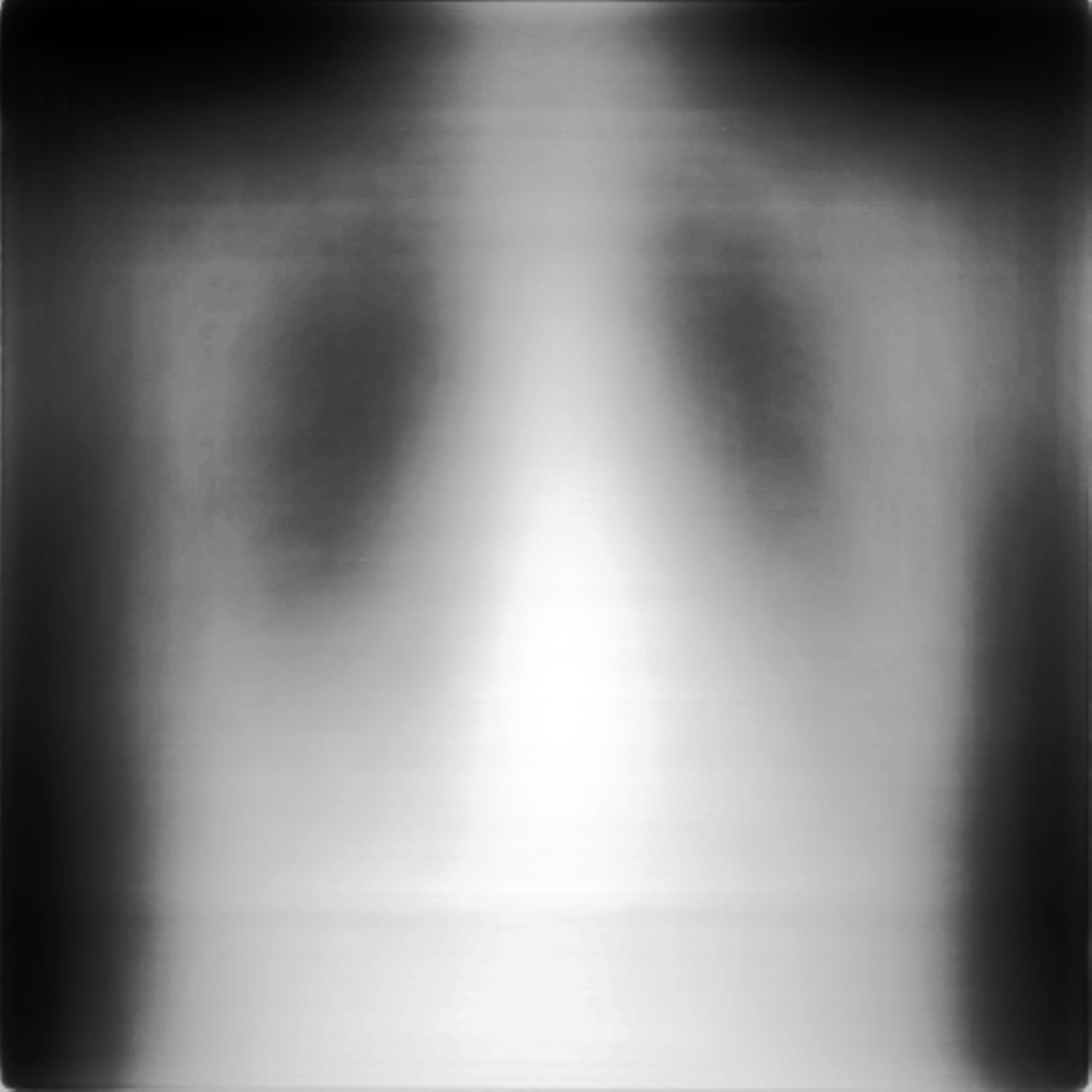}
  \caption{AP erect}
  \label{fig:APErectMean}
\end{subfigure}
\begin{subfigure}{.16\textwidth}
  \centering
  \includegraphics[width=1\textwidth]{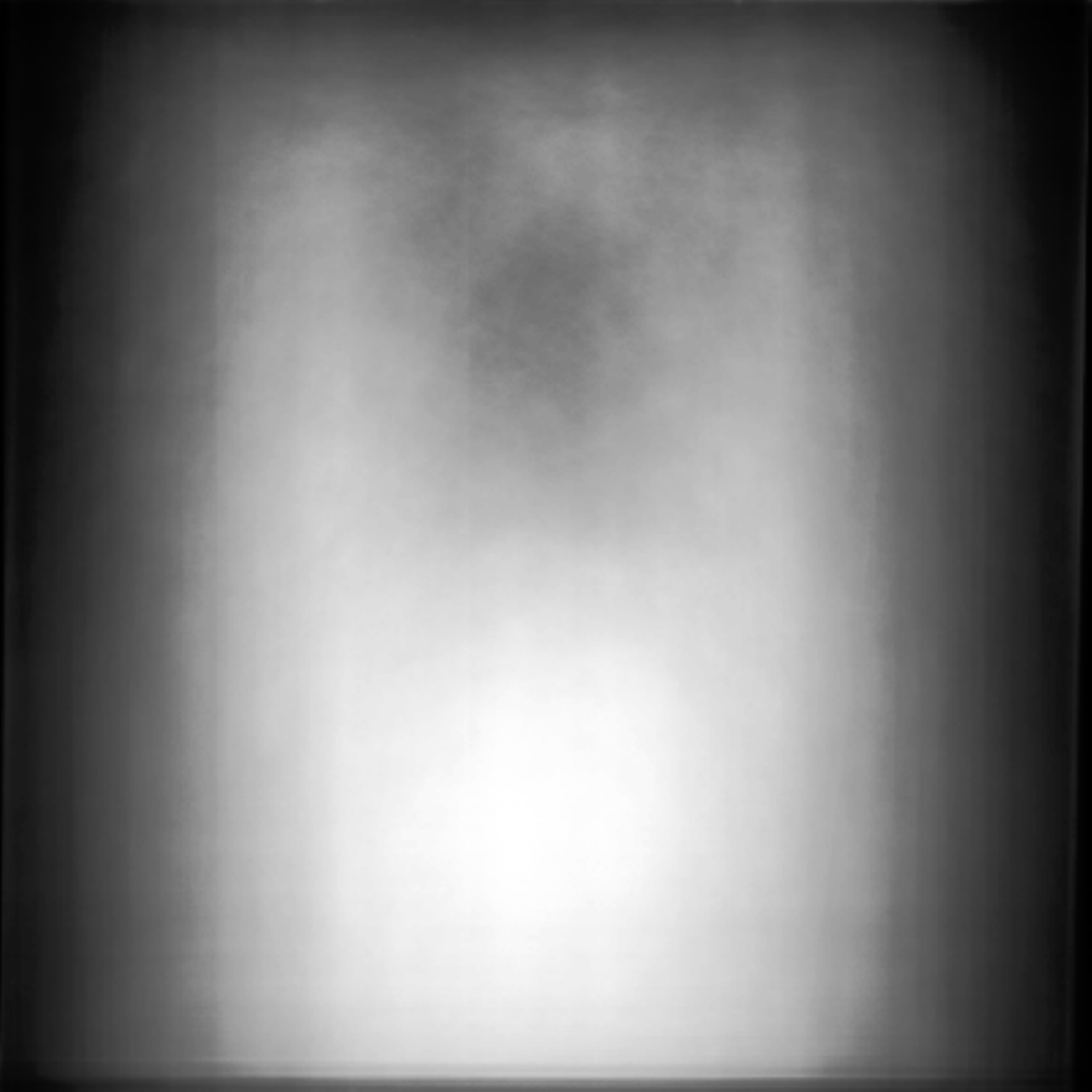}
  \caption{Ribs}
  \label{fig:CostalMean}
\end{subfigure}
\begin{subfigure}{.16\textwidth}
  \centering
  \includegraphics[width=1\linewidth]{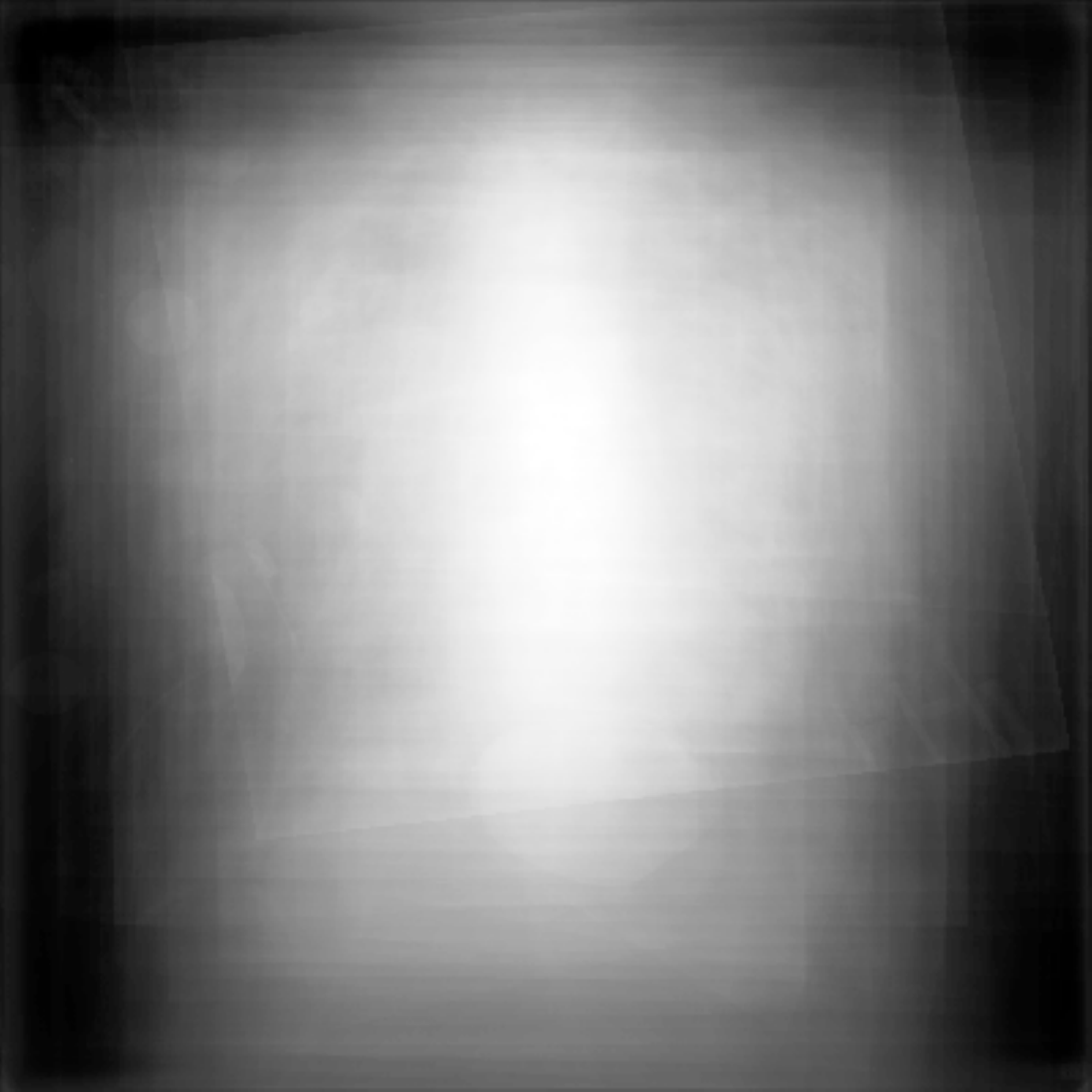}
  \caption{Pediatric}
  \label{fig:PedMean}
\end{subfigure}
\\
\centering
\begin{subfigure}{.16\textwidth}
  \centering
  \includegraphics[width=1\linewidth]{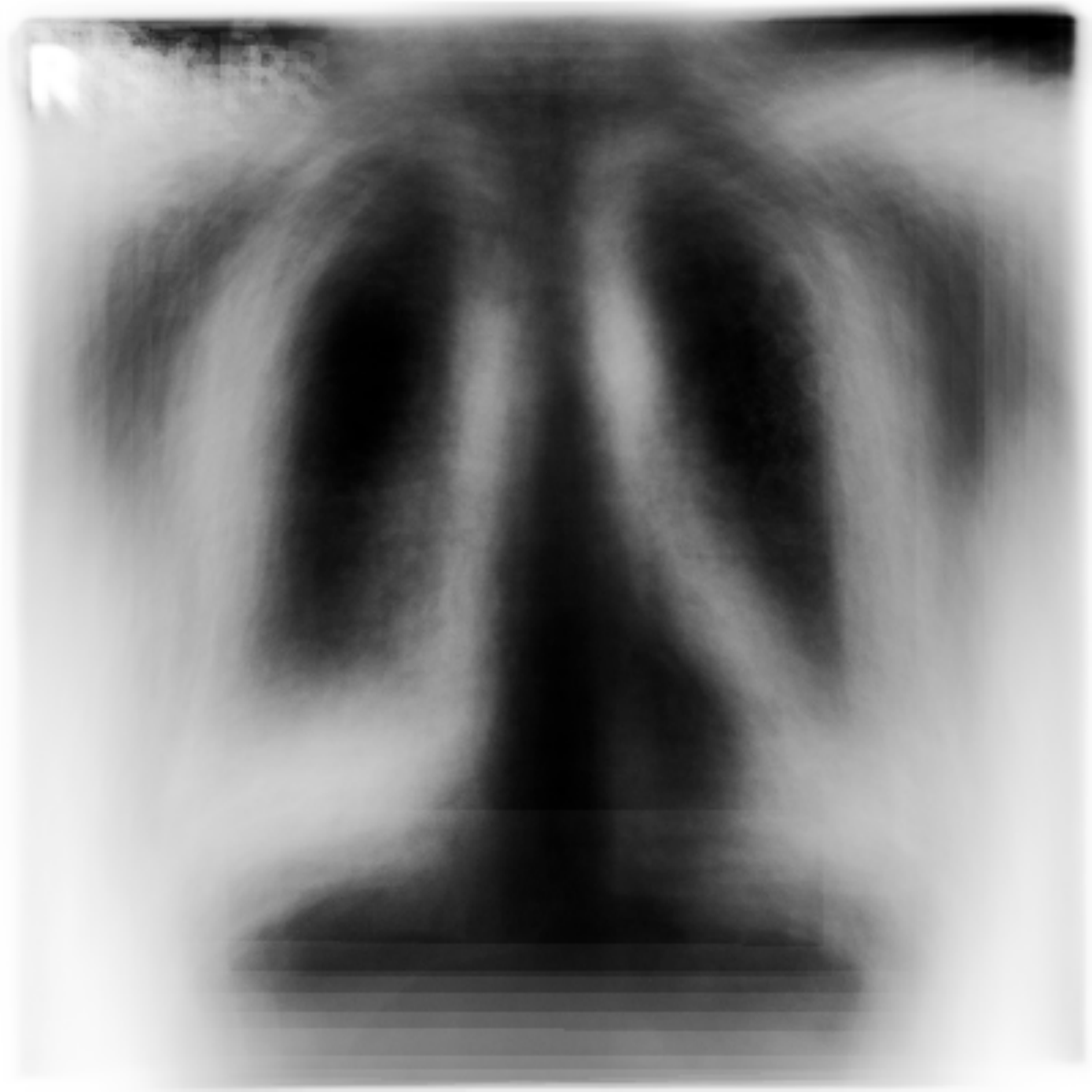}
  \caption{P-A}
  \label{fig:PAStd}
\end{subfigure}
\begin{subfigure}{.16\textwidth}
  \centering
  \includegraphics[width=1\linewidth]{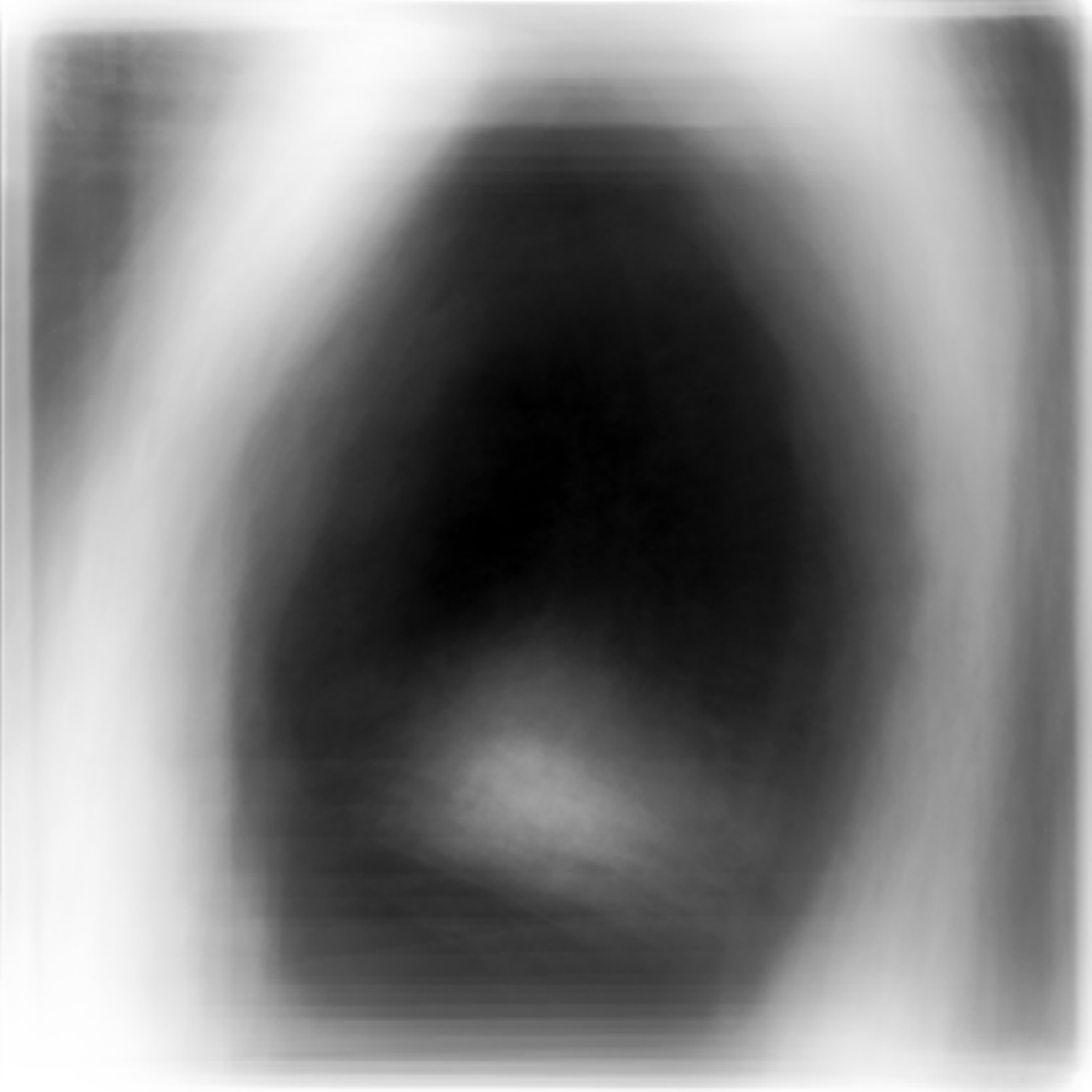}
  \caption{Lateral}
  \label{fig:LateralStd}
\end{subfigure}
\begin{subfigure}{.16\textwidth}
  \centering
  \includegraphics[width=1\linewidth]{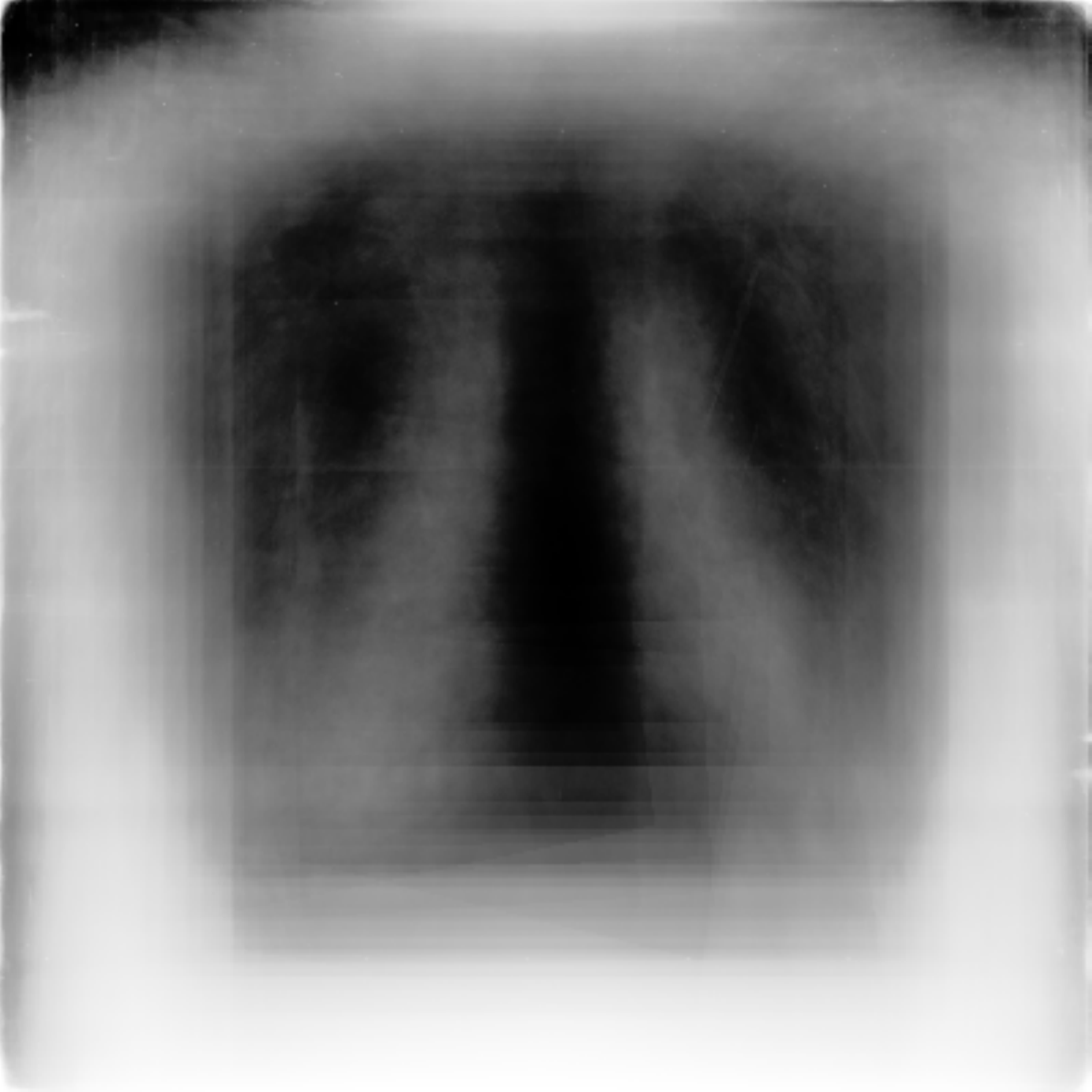}
  \caption{AP supine}
  \label{fig:APSupineStd}
\end{subfigure}
\begin{subfigure}{.16\textwidth}
  \centering
  \includegraphics[width=1\linewidth]{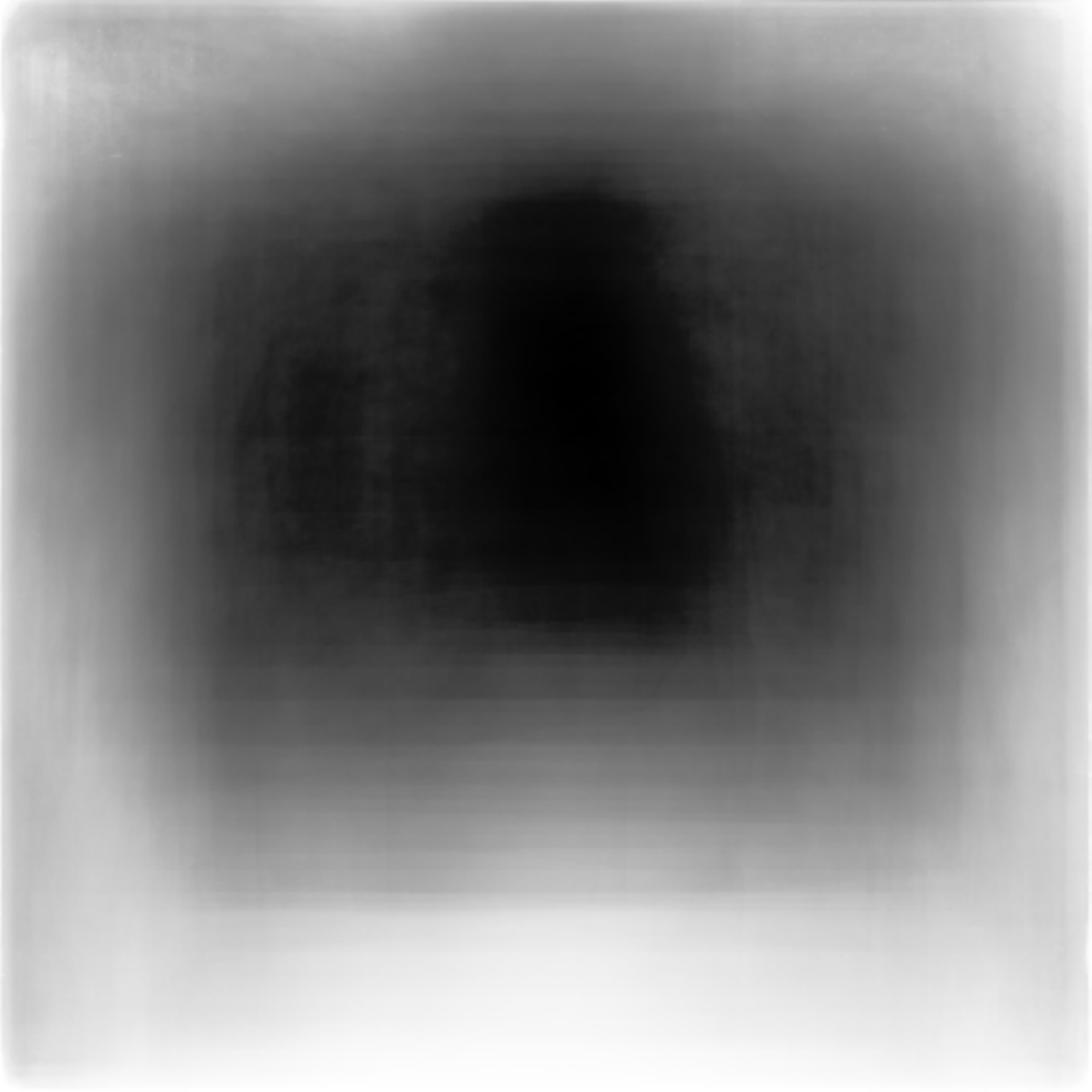}
  \caption{AP erect}
  \label{fig:APErect}
\end{subfigure}
\begin{subfigure}{.16\textwidth}
  \centering
  \includegraphics[width=1\linewidth]{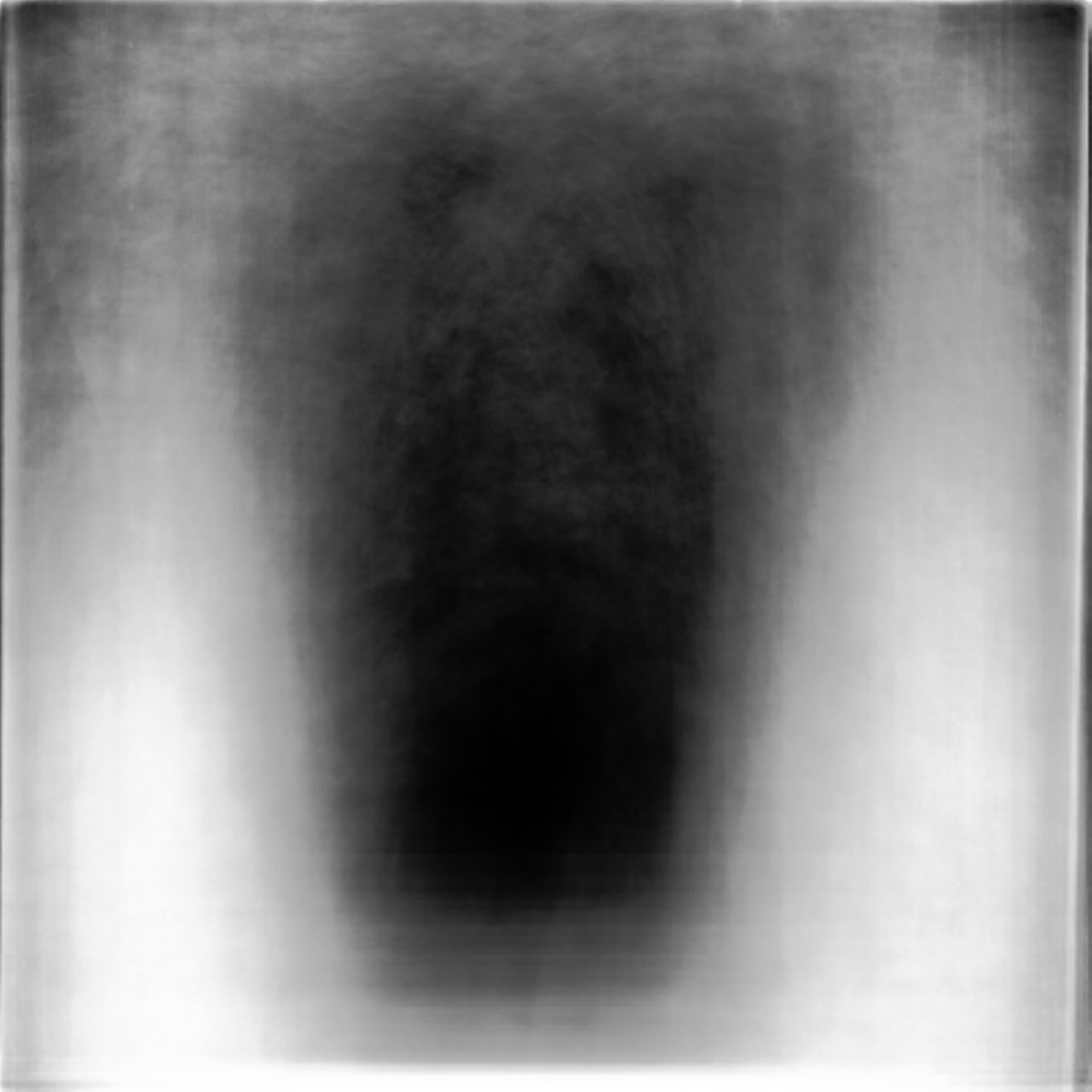}
  \caption{Ribs}
  \label{fig:RibsStd}
\end{subfigure}
\begin{subfigure}{.16\textwidth}
  \centering
  \includegraphics[width=1\linewidth]{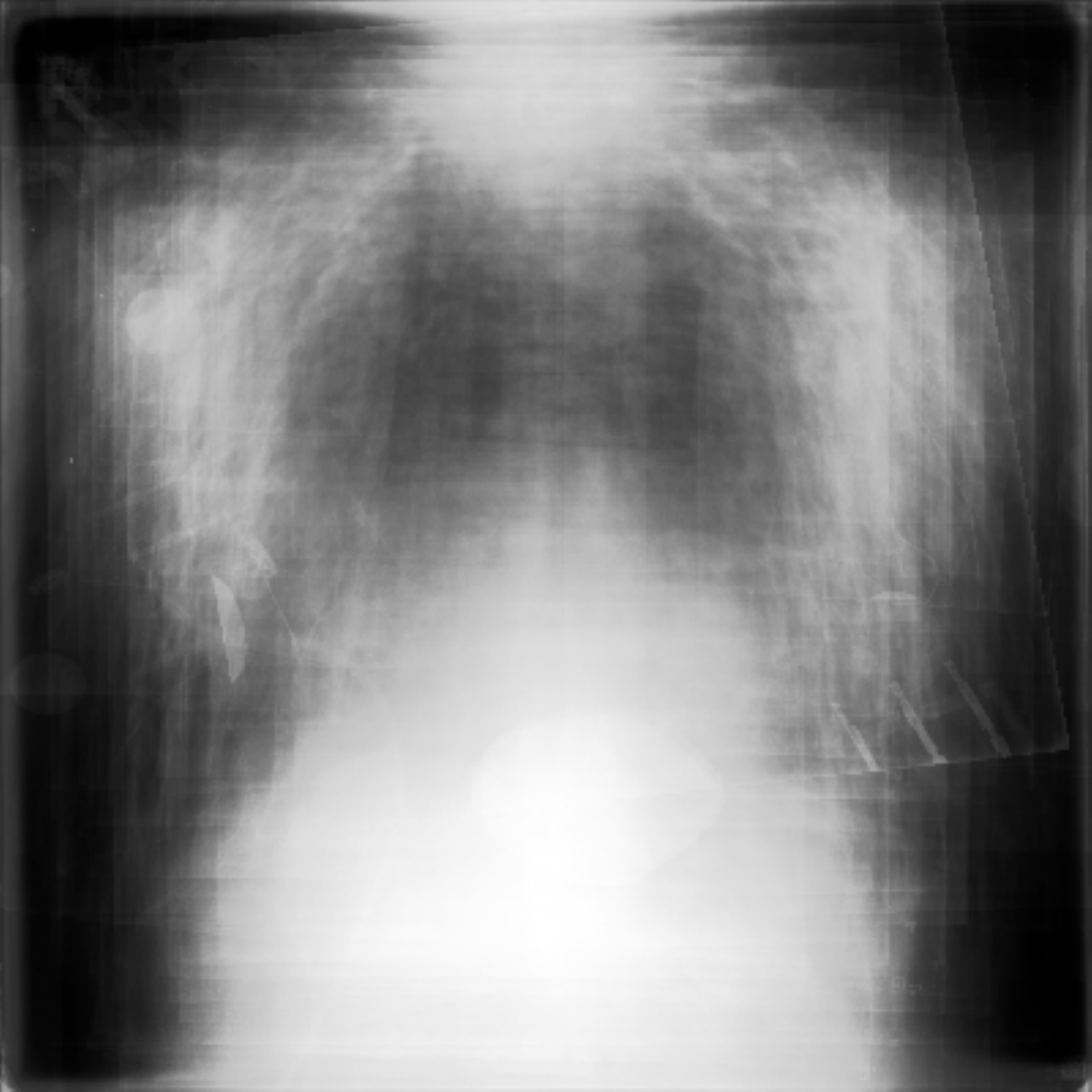}
  \caption{Pediatric}
  \label{fig:PedStd}
\end{subfigure}
\caption{From a) to f): Mean image by position view. From g) to l): Standard deviation image by position view.}
\label{fig:MeanStdByPosition}
\end{figure*}

These images showed relevant differences among the different x-ray projections (see Fig. \ref{fig:MeanStdByPosition}). In PA, the mean shows that images were fairly aligned to the center, while the standard deviation for the same random sample of 500 images, in which higher values are whiter, indicates a higher variation at the boundaries when compared to the center. In AP supine position, the mean heart silhouette was enlarged. Vertical and horizontal lines in the pediatrics standard deviation illustrate that there are many sizes of frames enclosing the chest and part of the abdomen.

\section{Conclusions}
\label{sec:conclusions}

In this work, we present one of the largest public chest x-ray dataset, which contains more than 160K images and Spanish reports labeled with 299 different medical entities mapped on to UMLS CUIs. It is expected that this dataset and the projects derived from it will ultimately assist in the development of diagnostic decision support systems, thus increasing clinical practice efficiency. The main contributions of PadChest are the following: 

\begin{itemize}
\item This is the first large-scale exploitable dataset available in Spanish. Given that it has been labeled with standardized medical codes, it is ready for further image exploitation, regardless of the language required. 
\item This is one of the largest publicly available labeled dataset that, in contrast to previously published works, includes different projection views and additional metadata, which is essential information if the models are to be trained adequately, as discussed previously. 
\item Trained physicians manually annotated 27\% of the dataset samples, and a RNN-ATT method was used to label the remaining reports. 
\item We propose hierarchical taxonomies in order to categorize radiographic findings, differential diagnoses and anatomical locations following UMLS CUI standard codes. To the best of our knowledge, available labeled datasets have a lower number of different labels.
\end{itemize}

The manually labeled dataset was used to train models for the automatic annotation of a Spanish corpus of x-ray reports. The best model was used to label 73\% of the PadChest dataset and achieved a 0.93 MicroF1 score for an independent test set, proving the robustness of the proposed approach as regards labeling the full dataset. 

One downside is that all the data were drawn from a single institution, but the fact that the training and test sets belong to different year intervals (2014-2017 and 2009-2014, respectively), thus increasing the variability of the reporting radiologists, supports that the model has the potential to be generalized. This would,however, have to be confirmed by using reports from other Spanish institutions. 

This dataset can be used to train models to predict thoracic pathologies affecting lung, cardiovascular and bone structures from x-ray images. 

We are aware that the dataset has some limitations. First limitation is the information bias inherent to employing the reports as the only source of the ground-truth. Specifically, the image could have radiographic abnormalities which are not described in the reports. For example, the presence of medical devices, such as pacemakers, catheters or prosthesis, or the presence of chronic and/or non severe conditions, such as chronic bone degenerative changes, granulomas, prior surgery marks, etc., are not systematically reported in clinical practice by radiologists and the decision to include them on the report depends on the clinical context, the requesting department and even on the radiologists’ work-load. Although these entities are not usually clinically relevant, we decided to label them so as to allow the models to assess their influence as potential bias or confounding factors when predicting clinically relevant entities. Moreover, the retrieval of images with these radiographic findings may be useful to help build more focused datasets with, for example, the goal of identifying misplaced medical devices. This would be useful when employing a chest x-ray to verify the correct position of the distal end after having inserted a central catheter. Although additional labeling would be required for this purpose,  this dataset would initially help retrieve the relevant images.

Second, the severity of medical entities when described is not currently captured in the labels. The adjectives usually employed to describe severity are: severe, intense, important, marked, light, faint, thin, slight, minimal, prominent, discreet, moderate, etc. Nonetheless, anatomic extensions such as bi-lateralization, multi-lobality, etc., which usually correlate with severity, are included in the topological labels.

Third limitation is the selection bias introduced by including only those x-ray studies with retrievable reports from PACs. Radiology explorations prescribed to answer questions of type yes/no and not for diagnosis are not necessarily accompanied by reports. This selection bias may be, along with other recognized factors such as differences as regards disease burden, accountable for the differences in the pathology frequencies in data-sets from different institutions.

One important warning is that datasets across different institutions may vary greatly, and this has been proved to greatly influence the generalization of Machine Learning Decision Support Systems (ML-DSS), particularly if confounding variables exist in internal data that do not exist in external data. A recent cross-sectional study \citep{zech2018variable} showed that CNN models trained to predict pneumonia from x-rays learned to exploit large differences in pathology prevalence between hospital systems in training data by calibrating their predictions to the baseline prevalence in each hospital system rather than exclusively discriminating on the basis of direct pathology findings. In addition, the CNNs robustly identified a hospital system and a department within a hospital, which had large differences as regards disease burden and may have confounded the predictions. The authors caution that the estimates of CNN performance based on test data from hospital systems used for model training may overstate their real-world performance. 

Before ML-DSS models can be confidently deployed in real-world clinical settings, two important requirements need to be addressed 1) controlling for confounding factors, this being particularly true in the case of DL models in which features are learned by the model rather than being engineered and predefined by researchers, and 2) prospectively evaluating the systems in multi-centric comparative studies that define target populations and adequate primary and secondary endpoints measuring health outcomes and comparing them with usual practice. 

Sound study designs would make it possible to consistently evaluate the performance of the systems, including their capacity to generalize across not only departments and hospitals, but also different populations and diversities of clinical settings.

Finally we provide the following advice for any researchers who intend to use PadChest for the task of predicting pathologies from x-rays:
\begin{enumerate}
\item To explore the use the PadChest dataset in conjunction with x-ray datasets from other institutions to help increase generalization.
\item To explore multi-input models and use all provided context information -including details of image acquisition, manufacturer, projection and clinical labels- to help increase the understanding of confounding factors exploited by ML-DSS models and to develop methods to control them, like for instance adjusting by expected prevalence of diseases. 
\item To acknowledge the different nature of radiographic findings versus differential diagnoses, which has been commonly overlooked in prior works (for instance, the task of predicting cardiomegaly has been regarded as same type of problem as predicting pneumonia) and to explore different methods and evaluation criteria for each of them.  
\item To explore models designed for dealing with labels having hierarchical relationships.
\item To acknowledge the information bias inherent to employing the reports as the only source of the ground-truth.
\end{enumerate}

\section*{Acknowledgement}

This work was supported by Medbravo, the Pattern Recognition and Artificial Intelligence Group (GRFIA) and the University Institute for Computing Research (IUII) at the University of Alicante. The Medical Image Bank of the Valencian Community as well as de-identification and anonymization services, were partially funded by the Regional Ministry of Health (FEDER program) and the Horizon 2020 Framework Programme of the European Union under grant agreement 688945 (Euro-BioImaging PrepPhase II).  
We thank Dr. Santiago F. Marco Domenech from the Radiology Department of the Castellon General Hospital (Spain), for his constructive suggestions and contributions to the hierarchies.




\ifdefined\arxiv
    \bibliographystyle{unsrt}
\else
    \bibliographystyle{model2-names}\biboptions{authoryear}
\fi

\newpage
\bibliography{bibliography.bib}

\clearpage

\appendix
\section{Hierarchies of labels}
\label{app:hierarchies}

\subsection{Counts of annotated labels in x-ray studies}

27,593 study reports were manually labeled and reviewed by trained physicians. 
Medical entities were extracted as per physician criteria and mapped to NLM Unified Medical Language System (UMLS) controlled biomedical vocabulary unique identifiers referred as CUIs. Medical entities whose exact meaning could not be found in any term of the UML metathesaurus but they were relevant for annotation purposes were also extracted using the physician's label but a CUI was not assigned. Four additional labels with special coding rules were explicitly defined: normal,  exclude, unchanged and suboptimal study. 

\subsubsection{Hierarchy of labels of radiographic findings}
\label{app:rxFinding}

A total of 177,628 radiographic findings (being 45,375 of those manually annotated in 27,593 study reports) were annotated in all the study reports. The labels ``normal", ``exclude" and ``suboptimal study" are counted separately.

\bigbreak

\begin{footnotesize}
\noindent
\textbf{label [CUI, study counts, study counts with child labels]}\\
normal [CUI:C0205307, counts:37871]\\
exclude [counts:1392]\\
suboptimal study [CUI:C2828075, counts:1813]\\
radiological finding [CUI:C0436485]\\
├── unchanged [counts:10140, 10140]\\
├── obesity [CUI:C0028754, counts:30, 30]\\
├── chronic changes [CUI:C0742362, counts:4988, 4988]\\
├── calcified densities [CUI:C2203586, counts:849, 4238]\\
│\verb!   !├── calcified granuloma [CUI:C0333404, counts:2165, 2165]\\
│\verb!   !├── calcified adenopathy [counts:434, 442]\\
│\verb!   !│\verb!   !└── calcified mediastinal adenopathy [counts:8, 8]\\
│\verb!   !├── calcified pleural thickening [counts:450, 450]\\
│\verb!   !├── calcified pleural plaques [counts:160, 160]\\
│\verb!   !├── heart valve calcified [CUI:C2073448, counts:129, 129]\\
│\verb!   !└── calcified fibroadenoma [counts:43, 43]\\
├── granuloma [CUI:C0235557, counts:481, 2646]\\
│\verb!   !└── calcified granuloma [CUI:C0333404, counts:2165, 2165]\\
├── end on vessel [counts:152, 152]\\
├── adenopathy [CUI:C0478664, counts:350, 784]\\
│\verb!   !└── calcified adenopathy [counts:434, 434]\\
├── nodule [CUI:C0034079, counts:2420, 2669]\\
│\verb!   !└── multiple nodules [CUI:C2073563, counts:249, 249]\\
├── pseudonodule [counts:1478, 2519]\\
│\verb!   !├── nipple shadow [counts:889, 889]\\
│\verb!   !└── end on vessel [counts:152, 152]\\
├── abscess [CUI:C0024110, counts:9, 9]\\
├── cyst [CUI:C2073485, counts:5, 5]\\
├── cavitation [CUI:C0578537, counts:367, 367]\\
├── fibrotic band [CUI:C0865843, counts:2435, 2435]\\
├── volume loss [CUI:C3203358, counts:1718, 1718]\\
├── hypoexpansion [counts:1053, 1053]\\
├── bullas [CUI:C0241982, counts:418, 418]\\
├── pneumothorax [CUI:C2073565, counts:411, 440]\\
│\verb!   !└── hydropneumothorax [CUI:C0020303, counts:29, 29]\\
├── pneumoperitoneo [CUI:C0032320, counts:56, 56]\\
├── pneumomediastinum [CUI:C2073636, counts:18, 18]\\
├── subcutaneous emphysema [CUI:C0038536, counts:190, 190]\\
├── hyperinflated lung [CUI:C0546312, counts:409, 409]\\
├── flattened diaphragm [CUI:C2073504, counts:539, 539]\\
├── lung vascular paucity [counts:108, 108]\\
├── air trapping [CUI:C0231819, counts:3527, 3527]\\
├── bronchiectasis [CUI:C0006267, counts:1588, 1588]\\
├── infiltrates [CUI:C0277877, counts:5212, 19850]\\
│\verb!   !├── interstitial pattern [CUI:C2073538, counts:6060, 7001]\\
│\verb!   !│\verb!   !├── ground glass pattern [CUI:C3544344, counts:213, 213]\\
│\verb!   !│\verb!   !├── reticular interstitial pattern [counts:427, 427]\\
│\verb!   !│\verb!   !├── reticulonodular interstitial pattern [CUI:C2073672, counts:175, 175]\\
│\verb!   !│\verb!   !└── miliary opacities [CUI:C2073583, counts:126, 126]\\
│\verb!   !└── alveolar pattern [CUI:C1332240, counts:5384, 7637]\\
│\verb!   !\verb!   ! ├── consolidation [CUI:C0521530, counts:1959, 2106]\\
│\verb!   !\verb!   ! │\verb!   !└── air bronchogram [CUI:C3669021, counts:147, 147]\\
│\verb!   !\verb!   ! └── air bronchogram [CUI:C3669021, counts:147, 147]\\
├── bronchovascular markings [CUI:C2073518, counts:954, 954]\\
├── air fluid level [CUI:C0740844, counts:115, 115]\\
├── increased density [CUI:C1443940, counts:1851, 1851]\\
├── atelectasis [CUI:C0004144, counts:2666, 6959]\\
│\verb!   !├── total atelectasis [CUI:C0264497, counts:24, 24]\\
│\verb!   !├── lobar atelectasis [counts:601, 601]\\
│\verb!   !├── segmental atelectasis [counts:104, 104]\\
│\verb!   !├── laminar atelectasis [counts:3525, 3525]\\
│\verb!   !├── round atelectasis [CUI:C2062952, counts:3, 3]\\
│\verb!   !└── atelectasis basal [CUI:C0746053, counts:36, 36]\\
├── mediastinal shift [CUI:C0264576, counts:39, 39]\\
├── azygos lobe [CUI:C0265794, counts:327, 327]\\
├── fissure thickening [counts:162, 620]\\
│\verb!   !├── minor fissure thickening [counts:342, 342]\\
│\verb!   !├── major fissure thickening [counts:30, 30]\\
│\verb!   !└── loculated fissural effusion [counts:86, 86]\\
├── pleural thickening [CUI:C0264545, counts:677, 3565]\\
│\verb!   !├── apical pleural thickening [counts:2438, 2438]\\
│\verb!   !└── calcified pleural thickening [counts:450, 450]\\
├── pleural plaques [CUI:C0340030, counts:15, 175]\\
│\verb!   !└── calcified pleural plaques [counts:160, 160]\\
├── pleural effusion [CUI:C2073625, counts:8289, 8593]\\
│\verb!   !├── loculated pleural effusion [CUI:C0747639, counts:185, 185]\\
│\verb!   !├── loculated fissural effusion [counts:86, 86]\\
│\verb!   !├── hydropneumothorax [CUI:C0020303, counts:29, 29]\\
│\verb!   !├── empyema [CUI:C0014009, counts:4, 4]\\
│\verb!   !└── hemothorax [CUI:C0019123, counts:0, 0]\\
├── pleural mass [CUI:C1709576, counts:2, 2]\\
├── costophrenic angle blunting [CUI:C0742855, counts:4542, 4542]\\
├── vascular redistribution [CUI:C0239041, counts:367, 499]\\
│\verb!   !└── central vascular redistribution [counts:132, 132]\\
├── hilar enlargement [CUI:C1698506, counts:1170, 4895]\\
│\verb!   !├── adenopathy [CUI:C0149711, counts:350, 350]\\
│\verb!   !└── vascular hilar enlargement [counts:3192, 3375]\\
│\verb!   !\verb!   ! └── pulmonary artery enlargement [CUI:C2072932, counts:183, 183]\\
├── hilar congestion [CUI:C0582411, counts:1047, 1047]\\
├── cardiomegaly [CUI:C0018800, counts:10527, 10527]\\
├── pericardial effusion [CUI:C0031039, counts:56, 56]\\
├── kerley lines [CUI:C0239019, counts:69, 69]\\
├── dextrocardia [CUI:C0011813, counts:14, 14]\\
├── right sided aortic arch [CUI:C0035615, counts:18, 18]\\
├── aortic atheromatosis [CUI:C1096249, counts:1813, 1813]\\
├── aortic elongation [counts:7724, 10154]\\
│\verb!   !├── descendent aortic elongation [CUI:C4476542, counts:736, 736]\\
│\verb!   !├── ascendent aortic elongation [CUI:C3889085, counts:154, 154]\\
│\verb!   !├── aortic button enlargement [CUI:C1851119, counts:368, 368]\\
│\verb!   !└── supra aortic elongation [counts:1172, 1172]\\
├── aortic aneurysm [CUI:C0003486, counts:33, 33]\\
├── mediastinal enlargement [CUI:C2021206, counts:272, 5589]\\
│\verb!   !├── superior mediastinal enlargement [CUI:C4273001, counts:610, 2488]\\
│\verb!   !│\verb!   !├── goiter [CUI:C0018021, counts:706, 706]\\
│\verb!   !│\verb!   !└── supra aortic elongation [counts:1172, 1172]\\
│\verb!   !├── descendent aortic elongation [CUI:C4476542, counts:736, 736]\\
│\verb!   !├── ascendent aortic elongation [CUI:C3889085, counts:154, 154]\\
│\verb!   !├── aortic aneurysm [CUI:C0003486, counts:33, 33]\\
│\verb!   !├── mediastinal mass [CUI:C0240318, counts:234, 234]\\
│\verb!   !└── hiatal hernia [CUI:C3489393, counts:1672, 1672]\\
├── tracheal shift [counts:600, 600]\\
├── mass [CUI:C2603353, counts:45, 1026]\\
│\verb!   !├── mediastinal mass [CUI:C0240318, counts:234, 234]\\
│\verb!   !├── breast mass [CUI:C0024103, counts:1, 1]\\
│\verb!   !├── pleural mass [CUI:C1709576, counts:2, 2]\\
│\verb!   !├── pulmonary mass [CUI:C0149726, counts:654, 654]\\
│\verb!   !└── soft tissue mass [CUI:C0457196, counts:90, 90]\\
├── esophagic dilatation [CUI:C0192389, counts:3, 3]\\
├── azygoesophageal recess shift [counts:47, 47]\\
├── pericardial effusion [CUI:C0031039, counts:56, 56]\\
├── mediastinic lipomatosis [CUI:C1333298, counts:289, 289]\\
├── thoracic cage deformation [CUI:C4538889, counts:166, 8702]\\
│\verb!   !├── scoliosis [CUI:C0036439, counts:5610, 5610]\\
│\verb!   !├── kyphosis [CUI:C2115817, counts:2631, 2631]\\
│\verb!   !├── pectum excavatum [CUI:C2051831, counts:185, 185]\\
│\verb!   !├── pectum carinatum [CUI:C2939416, counts:30, 30]\\
│\verb!   !└── cervical rib [CUI:C0158779, counts:80, 80]\\
├── vertebral degenerative changes [CUI:C4290224, counts:2974, 4298]\\
│\verb!   !└── vertebral compression [CUI:C0262431, counts:236, 1324]\\
│\verb!   !\verb!   ! └── vertebral anterior compression [counts:1088, 1088]\\
├── lytic bone lesion [CUI:C0476382, counts:56, 56]\\
├── sclerotic bone lesion [CUI:C4315325, counts:397, 450]\\
│\verb!   !└── blastic bone lesion [CUI:C2203581, counts:53, 53]\\
├── costochondral junction hypertrophy [counts:98, 98]\\
├── sternoclavicular junction hypertrophy [counts:5, 5]\\
├── axial hyperostosis [CUI:C1400000, counts:184, 184]\\
├── osteopenia [CUI:C0029453, counts:374, 374]\\
├── osteoporosis [CUI:C0029456, counts:164, 164]\\
├── non axial articular degenerative changes [counts:236, 236]\\
├── subacromial space narrowing [counts:34, 34]\\
├── fracture [CUI:C0016658, counts:14, 3367]\\
│\verb!   !├── clavicle fracture [CUI:C0159658, counts:271, 271]\\
│\verb!   !├── humeral fracture [CUI:C0020162, counts:285, 285]\\
│\verb!   !├── vertebral fracture [CUI:C0080179, counts:190, 190]\\
│\verb!   !└── rib fracture [CUI:C0035522, counts:533, 2607]\\
│\verb!   !\verb!   ! └── callus rib fracture [CUI:C0006767, counts:2074, 2074]\\
├── gynecomastia [CUI:C0018418, counts:445, 445]\\
├── hiatal hernia [CUI:C3489393, counts:1672, 1672]\\
├── Chilaiditi sign [CUI:C3178780, counts:14, 14]\\
├── hemidiaphragm elevation [CUI:C2073707, counts:1702, 1702]\\
├── diaphragmatic eventration [CUI:C0011981, counts:770, 770]\\
├── tracheostomy tube [CUI:C0184159, counts:1839, 1839]\\
├── endotracheal tube [CUI:C0336630, counts:2456, 2456]\\
├── NSG tube [counts:5827, 5827]\\
├── chest drain tube [CUI:C0008034, counts:607, 607]\\
├── ventriculoperitoneal drain tube [CUI:C0162702, counts:33, 33]\\
├── gastrostomy tube [CUI:C0150595, counts:9, 9]\\
├── nephrostomy tube [CUI:C0184149, counts:3, 3]\\
├── double J stent [CUI:C0441293, counts:11, 11]\\
├── catheter [CUI:C0085590, counts:63, 6167]\\
│\verb!   !└── central venous catheter [CUI:C1145640, counts:444, 6104]\\
│\verb!   !\verb!   ! ├── central venous catheter via subclavian vein [CUI:C0398281, counts:1712, 1712]\\
│\verb!   !\verb!   ! ├── central venous catheter via jugular vein [CUI:C0398278, counts:3299, 3299]\\
│\verb!   !\verb!   ! ├── reservoir central venous catheter [CUI:C2026143, counts:466, 466]\\
│\verb!   !\verb!   ! └── central venous catheter via umbilical vein [CUI:C0398284, counts:183, 183]\\
├── electrical device [counts:26, 4286]\\
│\verb!   !├── dual chamber device [CUI:C2732817, counts:955, 955]\\
│\verb!   !├── single chamber device [CUI:C2733312, counts:529, 529]\\
│\verb!   !├── pacemaker [CUI:C0030163, counts:2614, 2614]\\
│\verb!   !└── dai [CUI:C0972395, counts:162, 162]\\
├── artificial heart valve [CUI:C1399223, counts:488, 992]\\
│\verb!   !├── artificial mitral heart valve [CUI:C0869752, counts:299, 299]\\
│\verb!   !└── artificial aortic heart valve [CUI:C0869748, counts:205, 205]\\
├── surgery [counts:161, 7371]\\
│\verb!   !├── metal [CUI:C0025552, counts:1113, 5225]\\
│\verb!   !│\verb!   !├── osteosynthesis material [CUI:C0016642, counts:530, 530]\\
│\verb!   !│\verb!   !├── sternotomy [CUI:C0185792, counts:2056, 2056]\\
│\verb!   !│\verb!   !└── suture material [CUI:C4305366, counts:1526, 1526]\\
│\verb!   !├── bone cement [CUI:C0005934, counts:8, 8]\\
│\verb!   !├── prosthesis [CUI:C0175649, counts:17, 566]\\
│\verb!   !│\verb!   !├── humeral prosthesis [CUI:C0745058, counts:117, 117]\\
│\verb!   !│\verb!   !├── mammary prosthesis [CUI:C0917968, counts:382, 382]\\
│\verb!   !│\verb!   !└── endoprosthesis [CUI:C0005846, counts:19, 50]\\
│\verb!   !│\verb!   !\verb!   ! └── aortic endoprosthesis [counts:31, 31]\\
│\verb!   !├── surgery breast [CUI:C3714726, counts:278, 729]\\
│\verb!   !│\verb!   !└── mastectomy [CUI:C0024881, counts:451, 451]\\
│\verb!   !├── surgery neck [CUI:C0185773, counts:442, 442]\\
│\verb!   !├── surgery lung [CUI:C0038903, counts:213, 213]\\
│\verb!   !├── surgery heart [CUI:C0018821, counts:14, 14]\\
│\verb!   !└── surgery humeral [CUI:C0186326, counts:13, 13]\\
├── abnormal foreign body [CUI:C0016542, counts:49, 49]\\
└── external foreign body [counts:34, 34]\\
\end{footnotesize}

\subsubsection {Differential Diagnoses Hierarchy}
\label{app:DD}

SJ-PadChest contains a total of 27,726 differential diagnoses (5,472 were manually annotated from 27,593 study reports). 

\bigbreak

\begin{footnotesize}
\noindent
\textbf{label [CUI, study counts, study counts with child labels]}\\
differential diagnosis\\
├── pneumonia [CUI:C0032285, counts:5778, 5934]\\
│\verb!   !└── atypical pneumonia [CUI:C1412002, counts:156, 156]\\
├── tuberculosis [CUI:C0041296, counts:176, 824]\\
│\verb!   !└── tuberculosis sequelae [CUI:C0494132, counts:648, 648]\\
├── lung metastasis [CUI:C0153676, counts:251, 251]\\
├── lymphangitis carcinomatosa [CUI:C0238258, counts:30, 30]\\
├── lepidic adenocarcinoma [CUI:C4049711, counts:10, 10]\\
├── pulmonary fibrosis [CUI:C0034069, counts:751, 991]\\
│\verb!   !├── post radiotherapy changes [CUI:C1320687, counts:196, 196]\\
│\verb!   !└── asbestosis signs [CUI:C0003949, counts:44, 44]\\
├── emphysema [CUI:C0034067, counts:958, 958]\\
├── COPD signs [CUI:C0024117, counts:14557, 14557]\\
├── heart insufficiency [CUI:C0018801, counts:2339, 2339]\\
├── respiratory distress [CUI:C0476273, counts:53, 53]\\
├── pulmonary hypertension [CUI:C0020542, counts:95, 129]\\
│\verb!   !├── pulmonary artery hypertension [CUI:C2973725, counts:28, 28]\\
│\verb!   !└── pulmonary venous hypertension [CUI:C4477098, counts:6, 6]\\
├── pulmonary edema [CUI:C0034063, counts:1515, 1515]\\
└── bone metastasis [CUI:C0153690, counts:135, 135]\\
\end{footnotesize}

\subsubsection {Anatomical Locations}
\label{app:localizations}

A total of 249,469 differential diagnoses were annotated (55,472 manually tagged in 27,593 study reports).

\bigbreak


\begin{footnotesize}
\noindent
\textbf{label [CUI, study counts, study counts with child labels]}\\
localization\\
├── extracorporal [CUI:C0424529, counts:0, 0]\\
├── cervical [CUI:C0920882, counts:992, 992]\\
├── soft tissue [CUI:C0225317, counts:413, 3567]\\
│\verb!   !├── subcutaneous [CUI:C0443315, counts:330, 330]\\
│\verb!   !├── axilar [CUI:C0004454, counts:863, 863]\\
│\verb!   !└── pectoral [CUI:C0230111, counts:1081, 1961]\\
│\verb!   !\verb!   ! └── nipple [CUI:C0028109, counts:880, 880]\\
├── bone [CUI:C0262950, counts:1472, 12862]\\
│\verb!   !├── shoulder [CUI:C0037004, counts:287, 2149]\\
│\verb!   !│\verb!   !├── acromioclavicular joint [CUI:C0001208, counts:69, 69]\\
│\verb!   !│\verb!   !├── rotator cuff [CUI:C0085515, counts:67, 67]\\
│\verb!   !│\verb!   !├── supraspisnous [CUI:C0225001, counts:43, 43]\\
│\verb!   !│\verb!   !└── humerus [CUI:C0020164, counts:1051, 1683]\\
│\verb!   !│\verb!   !\verb!   ! ├── humeral head [CUI:C0223683, counts:430, 430]\\
│\verb!   !│\verb!   !\verb!   ! ├── humeral neck [CUI:C0448034, counts:88, 88]\\
│\verb!   !│\verb!   !\verb!   ! └── glenohumeral joint [CUI:C0225063, counts:114, 114]\\
│\verb!   !├── clavicle [CUI:C0008913, counts:388, 388]\\
│\verb!   !├── scapula [CUI:C0036277, counts:171, 171]\\
│\verb!   !├── costoesternal [CUI:C0450216, counts:36, 36]\\
│\verb!   !├── column [CUI:C0037949, counts:1431, 3783]\\
│\verb!   !│\verb!   !├── intersomatic space [CUI:C0223088, counts:52, 52]\\
│\verb!   !│\verb!   !├── dorsal vertebrae [CUI:C0039987, counts:1979, 1979]\\
│\verb!   !│\verb!   !├── cervical vertebrae [CUI:C3665420, counts:100, 100]\\
│\verb!   !│\verb!   !└── paravertebral [CUI:C0442150, counts:221, 221]\\
│\verb!   !└── rib [CUI:C0035561, counts:3756, 4863]\\
│\verb!   !\verb!   ! ├── anterior rib [CUI:C4323264, counts:320, 320]\\
│\verb!   !\verb!   ! ├── posterior rib [CUI:C4323265, counts:750, 750]\\
│\verb!   !\verb!   ! └── rib cartilage [CUI:C0222787, counts:37, 37]\\
├── hemithorax [CUI:C0934569, counts:5119, 5119]\\
├── extrapleural [CUI:C0442091, counts:73, 73]\\
├── extrapulmonary [counts:46, 46]\\
├── pleural [CUI:C0032225, counts:13935, 13935]\\
├── subpleural [CUI:C0225775, counts:115, 115]\\
├── fissure [CUI:C0458078, counts:734, 1318]\\
│\verb!   !├── minor fissure [CUI:C0734040, counts:486, 486]\\
│\verb!   !└── major fissure [CUI:C4253583, counts:98, 98]\\
├── lobar [CUI:C0225752, counts:2413, 14864]\\
│\verb!   !├── upper lobe [CUI:C0225756, counts:1222, 6720]\\
│\verb!   !│\verb!   !├── left upper lobe [CUI:C1261076, counts:1574, 2477]\\
│\verb!   !│\verb!   !│\verb!   !└── lingula [CUI:C0225740, counts:903, 903]\\
│\verb!   !│\verb!   !└── right upper lobe [CUI:C1261074, counts:3021, 3021]\\
│\verb!   !├── lower lobe [CUI:C0225758, counts:792, 4144]\\
│\verb!   !│\verb!   !├── left lower lobe [CUI:C1261077, counts:1994, 1994]\\
│\verb!   !│\verb!   !└── right lower lobe [CUI:C1261075, counts:1358, 1358]\\
│\verb!   !└── middle lobe [CUI:C4281590, counts:1587, 1587]\\
├── subsegmental [CUI:C0929165, counts:1850, 1850]\\
├── bronchi [CUI:C0006255, counts:3315, 3315]\\
├── peribronchi [CUI:C0225607, counts:114, 114]\\
├── lung field [CUI:C0225759, counts:819, 78888]\\
│\verb!   !├── pleural [CUI:C0032225, counts:13935, 13935]\\
│\verb!   !├── subpleural [CUI:C0225775, counts:115, 115]\\
│\verb!   !├── major fissure [CUI:C4253583, counts:98, 98]\\
│\verb!   !├── subsegmental [CUI:C0929165, counts:1850, 1850]\\
│\verb!   !├── bronchi [CUI:C0006255, counts:3315, 3315]\\
│\verb!   !├── peribronchi [CUI:C0225607, counts:114, 114]\\
│\verb!   !├── upper lung field [CUI:C0929227, counts:494, 10442]\\
│\verb!   !│\verb!   !└── upper lobe [CUI:C0225756, counts:1222, 9948]\\
│\verb!   !│\verb!   !\verb!   ! ├── left upper lobe [CUI:C1261076, counts:1574, 1574]\\
│\verb!   !│\verb!   !\verb!   ! ├── right upper lobe [CUI:C1261074, counts:3021, 3021]\\
│\verb!   !│\verb!   !\verb!   ! ├── apical [CUI:C0734296, counts:3884, 3884]\\
│\verb!   !│\verb!   !\verb!   ! └── suprahilar [CUI:C0934260, counts:247, 247]\\
│\verb!   !├── middle lung field [CUI:C0929434, counts:1591, 13539]\\
│\verb!   !│\verb!   !├── aortopulmonary window [CUI:C1282038, counts:113, 113]\\
│\verb!   !│\verb!   !├── hilar [CUI:C0205150, counts:6678, 11349]\\
│\verb!   !│\verb!   !│\verb!   !├── pulmonary artery [CUI:C0034052, counts:215, 215]\\
│\verb!   !│\verb!   !│\verb!   !├── hilar bilateral [counts:690, 690]\\
│\verb!   !│\verb!   !│\verb!   !└── perihilar [CUI:C0225702, counts:3766, 3766]\\
│\verb!   !│\verb!   !└── minor fissure [CUI:C0734040, counts:486, 486]\\
│\verb!   !└── lower lung field [counts:1352, 34661]\\
│\verb!   !\verb!   ! ├── basal [CUI:C1282378, counts:9632, 9632]\\
│\verb!   !\verb!   ! ├── lower lobe [CUI:C0225758, counts:792, 4144]\\
│\verb!   !\verb!   ! │\verb!   !├── left lower lobe [CUI:C1261077, counts:1994, 1994]\\
│\verb!   !\verb!   ! │\verb!   !└── right lower lobe [CUI:C1261075, counts:1358, 1358]\\
│\verb!   !\verb!   ! ├── middle lobe [CUI:C4281590, counts:1587, 1587]\\
│\verb!   !\verb!   ! ├── infrahilar [counts:675, 675]\\
│\verb!   !\verb!   ! ├── lingula [CUI:C0225740, counts:903, 903]\\
│\verb!   !\verb!   ! ├── supradiaphragm [counts:19, 19]\\
│\verb!   !\verb!   ! ├── diaphragm [CUI:C0011980, counts:3798, 3798]\\
│\verb!   !\verb!   ! ├── infradiaphragm [counts:2934, 2934]\\
│\verb!   !\verb!   ! ├── cardiophrenic angle [counts:287, 287]\\
│\verb!   !\verb!   ! └── costophrenic angle [CUI:C0230151, counts:5805, 9330]\\
│\verb!   !\verb!   !\verb!   !  ├── right costophrenic angle [CUI:C0504099, counts:1445, 1445]\\
│\verb!   !\verb!   !\verb!   !  ├── left costophrenic angle [CUI:C0504100, counts:2019, 2019]\\
│\verb!   !\verb!   !\verb!   !  └── bilateral costophrenic angle [counts:61, 61]\\
├── central [CUI:C0205099, counts:5488, 5488]\\
├── mediastinum [CUI:C0025066, counts:2686, 39134]\\
│\verb!   !├── superior mediastinum [CUI:C0230147, counts:645, 8449]\\
│\verb!   !│\verb!   !├── carotid artery [CUI:C0007272, counts:4, 4]\\
│\verb!   !│\verb!   !├── brachiocephalic veins [CUI:C0006095, counts:203, 203]\\
│\verb!   !│\verb!   !├── supra aortic [counts:1123, 1123]\\
│\verb!   !│\verb!   !├── aortic button [CUI:C0003489, counts:348, 348]\\
│\verb!   !│\verb!   !├── superior cave vein [CUI:C3165182, counts:3663, 3663]\\
│\verb!   !│\verb!   !└── subclavian vein [CUI:C0038532, counts:2463, 2463]\\
│\verb!   !├── lower mediastinum [CUI:C1261193, counts:200, 18497]\\
│\verb!   !│\verb!   !├── anterior mediastinum [CUI:C0230148, counts:64, 162]\\
│\verb!   !│\verb!   !│\verb!   !└── thymus [CUI:C0040113, counts:98, 98]\\
│\verb!   !│\verb!   !├── middle mediastinum [CUI:C0230149, counts:32, 15740]\\
│\verb!   !│\verb!   !│\verb!   !└── cardiac [CUI:C1522601, counts:15663, 15708]\\
│\verb!   !│\verb!   !│\verb!   !\verb!   ! └── coronary [CUI:C1522318, counts:45, 45]\\
│\verb!   !│\verb!   !└── posterior mediastinum [CUI:C0230150, counts:16, 2395]\\
│\verb!   !│\verb!   !\verb!   ! └── retrocardiac [counts:2379, 2379]\\
│\verb!   !└── aortic [CUI:C0003483, counts:9502, 9502]\\
├── paratracheal [CUI:C0442143, counts:480, 480]\\
├── airways [CUI:C0458827, counts:18, 8685]\\
│\verb!   !├── tracheal [CUI:C0040578, counts:5352, 5352]\\
│\verb!   !└── bronchi [CUI:C0006255, counts:3315, 3315]\\
├── esophageal [CUI:C1522619, counts:239, 239]\\
├── paramediastinum [counts:70, 70]\\
├── paracardiac [counts:188, 188]\\
├── epigastric [counts:42, 42]\\
├── gastric chamber [CUI:C3714551, counts:887, 887]\\
├── hypochondrium [CUI:C0230186, counts:193, 455]\\
│\verb!   !├── right hypochondrium [CUI:C0738590, counts:122, 196]\\
│\verb!   !│\verb!   !└── gallbladder [CUI:C0016976, counts:74, 74]\\
│\verb!   !└── left hypochondrium [CUI:C0738591, counts:66, 66]\\
├── right [CUI:C0444532, counts:23628, 23628]\\
├── left [CUI:C0443246, counts:17133, 17133]\\
└── bilateral [CUI:C0238767, counts:11034, 15982]\\
\verb!   ! ├── diffuse bilateral [counts:1636, 1636]\\
\verb!   ! └── basal bilateral [counts:3312, 3312]\\
\end{footnotesize}

\section{Regular expressions used to extract spatial concepts from Spanish reports}
\label{app:regexLoc}

 \tablefirsthead{\toprule \textbf{Regular expression} &\multicolumn{1}{l}{\textbf{Spatial Concept}} \\ \midrule
 }
 \tablehead{
 \multicolumn{2}{c}
 {{\bfseries  Continued from previous column}} \\
 \toprule
 \textbf{Regular expression} &\multicolumn{1}{l}{\textbf{Spatial Concept}} \\ \midrule
 }
 \tabletail{
 \midrule \multicolumn{2}{r}{{\textbf{Continued on next column}}} \\ \midrule
 }
 \tablelasttail{
   \\\midrule
   \multicolumn{2}{r}{{Concluded}} \\ 
   \bottomrule
 }
 \begin{footnotesize}
 \begin{supertabular}{l|l}
 \textbackslash bsen\textbackslash scost.*\textbackslash sderech & right costophrenic angle\\
 \textbackslash bsen\textbackslash scost.*\textbackslash sderech& right costophrenic angle\\
 \textbackslash bsen\textbackslash scost.*\textbackslash sizq& left costophrenic angle\\
 \textbackslash bsen\textbackslash scost.*\textbackslash sbil& bilateral costophrenic angle\\
 \textbackslash bsen\textbackslash scost& costophrenic angle\\
 \textbackslash bsupradiafrag& supradiaphragm\\
 \textbackslash bhemidiafrag|\textbackslash bdiafrag& diaphragm\\
 \textbackslash binfradiafrag& infradiaphragm\\
 \textbackslash blobul\textbackslash b& lobar\\
 \textbackslash bsubsegment& subsegmental\\
 \textbackslash bperibronqu& peribronchi\\
 \textbackslash bbronqui|\textbackslash bendobronqui|\textbackslash bbronc& bronchi\\
 \textbackslash blingul& lingula\\
 (\textbackslash bls\textbackslash s|lobul\textbackslash ssup)& upper lobe\\
 (\textbackslash bli\textbackslash s|lobul\textbackslash sinf)& lower lobe\\
 (\textbackslash blsi|\textbackslash bls\textbackslash sizq|lobul\textbackslash ssup.*sizq)& left upper lobe\\
 (\textbackslash blsd|\textbackslash bls\textbackslash sderech|lobul\textbackslash ssup.*sderec)& right upper lobe\\
 (\textbackslash blmd|\textbackslash blm\textbackslash b)& middle lobe\\
 \textbackslash bretrocardi& retrocardiac\\
 \textbackslash blii& left lower lobe\\
 \textbackslash blid& right lower lobe\\
 \textbackslash bcamp.*\textbackslash sinfer& lower lung field\\
 \textbackslash bcamp.*\textbackslash ssup& upper lung field\\
 \textbackslash bcamp.*\textbackslash smed& middle lung field\\
 \textbackslash bcamp.*\textbackslash spulm& lung field\\
 \textbackslash baxil& axilar\\
 (\textbackslash bcervical|\textbackslash bcuell\textbackslash b)& cervical\\
 (\textbackslash bvertic|\textbackslash bapic|\textbackslash bapex)& apical\\
 \textbackslash bbas&basal\\
 \textbackslash bcentral&central\\
 \textbackslash bcostal&rib\\
 \textbackslash bcostal anterior&anterior rib\\
 \textbackslash bcostal posterior&posterior rib\\
 \textbackslash bextrapleur&extrapleural\\
 \textbackslash bpleur& pleural\\
 \textbackslash bsubpleur& subpleural\\
 (\textbackslash bfundus gastric|\textbackslash bcam gastric)& gastric chamber\\
 \textbackslash bhili.*\textbackslash sbila&hilar bilateral\\
 \textbackslash bhili&hilar\\
 (\textbackslash bperihili|\textbackslash bparahili)&perihilar\\
 \textbackslash bsuprahili&suprahilar\\
 \textbackslash binfrahili&infrahilar\\
 \textbackslash bmediastin sup&superior mediastinum\\
 \textbackslash bmediastin anterior&anterior mediastinum\\
 \textbackslash bmediastin medi&middle mediastinum\\
 \textbackslash bmediastin posterior&posterior mediastinum\\
 \textbackslash bmediastin infer&lower mediastinum\\
 \textbackslash bmediastin& mediastinum\\
 \textbackslash bparamediastin& paramediastinum\\
 \textbackslash btim& thymus\\
 \textbackslash bventan aort& aortopulmonary window\\
 (\textbackslash bcolumn.*\textbackslash sdorsal|\textbackslash bvertebr.*\textbackslash sdorsal)& dorsal vertebrae\\
 (\textbackslash bcolumn.*\textbackslash scerv|\textbackslash bvertebr.*\textbackslash scerv)& cervical vertebrae\\
 (\textbackslash bcolumn.*\textbackslash sax|\textbackslash bcolumn.*\textbackslash svert|\textbackslash bvertebr)& column\\
 \textbackslash bparavertebr& paravertebral\\
 \textbackslash bhipocondri& hypochondrium\\
 \textbackslash bhipocondri derech& right hypochondrium\\
 \textbackslash bhipocondri izq& left hypochondrium\\
 \textbackslash bhues|\textbackslash bose& bone\\
 \textbackslash bhumer& humerus\\
 \textbackslash bcabez.*\textbackslash shumer& humeral head\\
 \textbackslash bcuell.*\textbackslash shumer& humeral neck\\
 \textbackslash bglen&glenohumeral joint\\
 \textbackslash bacromi.*clavi&acromioclavicular joint\\
 \textbackslash bclavicul&clavicle\\
 \textbackslash bescap& scapula\\
 \textbackslash bsupraespin& supraspisnous\\
 \textbackslash bhombr& shoulder\\
 \textbackslash bcostoesternal&costoesternal\\
 \textbackslash bcartilag cost& rib cartilage\\
 \textbackslash bcardiofren & cardiophrenic angle\\
 \textbackslash bsiluet cardi|\textbackslash bcardi|\textbackslash bcorazo& cardiac\\
 \textbackslash bparacardi& paracardiac\\
 \textbackslash bcoronar& coronary\\
 \textbackslash bsupraaort& supra aortic\\
 \textbackslash bboton aort& aortic button\\
 \textbackslash baort& aortic\\
 \textbackslash barteri pulmon|\^pulmonar\textbackslash b& pulmonary artery\\
 \textbackslash bcav\textbackslash ssup& superior cave vein\\
 \textbackslash bsubclavi& subclavian vein\\
 \textbackslash bbraquio& brachiocephalic veins\\
 \textbackslash bcarotid& carotid artery\\
 \textbackslash bsubcutan& subcutaneous\\
 \textbackslash bextracorp& extracorporal\\
 \textbackslash bmanguit.*\textbackslash srota& rotator cuff\\
 \textbackslash bcisur.*\textbackslash smenor\textbackslash b& minor fissure\\
 \textbackslash bcisur.*\textbackslash smayor\textbackslash b& major fissure\\
 \textbackslash bcisur\textbackslash b& fissure\\
 \textbackslash bintersomat& intersomatic space\\
 \textbackslash bextrapulmo& extrapulmonary\\
 \textbackslash bc\textbackslash d\textbackslash b& cervical vertebrae\\
 \textbackslash bd\textbackslash d?\textbackslash d\textbackslash b& dorsal vertebrae\\
 \textbackslash bl\textbackslash d\textbackslash b& lumbar vertebrae\\
 \textbackslash bepigastri& epigastric\\
 \textbackslash bparatraque& paratracheal\\
 \textbackslash bvia.*\textbackslash saere& airways\\
 \textbackslash btraque|\textbackslash bendotraque|\textbackslash bintratraque& tracheal\\
 \textbackslash besofag& esophageal\\
 \textbackslash bbland& soft tissue\\
 \textbackslash bpectoral|\textbackslash bmama|\textbackslash bginecomas& pectoral\\
 \textbackslash bpezon|\textbackslash bmamil& nipple\\
 (\textbackslash bcolecist|colelit|vesicul.*\textbackslash bbil)& gallbladder\\
 \textbackslash bhemit& hemithorax\\
 \textbackslash bderech\textbackslash b& right\\
 \textbackslash bizq& left\\
 \textbackslash bdifus.*\textbackslash sbilat& diffuse bilateral\\
 \textbackslash bbilat.*\textbackslash sdifus& diffuse bilateral\\
 (\textbackslash bbilat|\textbackslash bamb.*hemit|\textbackslash bamb.*camp)& bilateral\\
 \textbackslash bbibas& basal bilateral\\
 \label{tab:regexLoc}
 \end{supertabular}
 \end{footnotesize}

\section{Dataset example}
\label{app:DatasetExample}
\begin{figure}[b]
\begin{subfigure}{.5\columnwidth}
  \centering
  \includegraphics[width=.9\columnwidth]{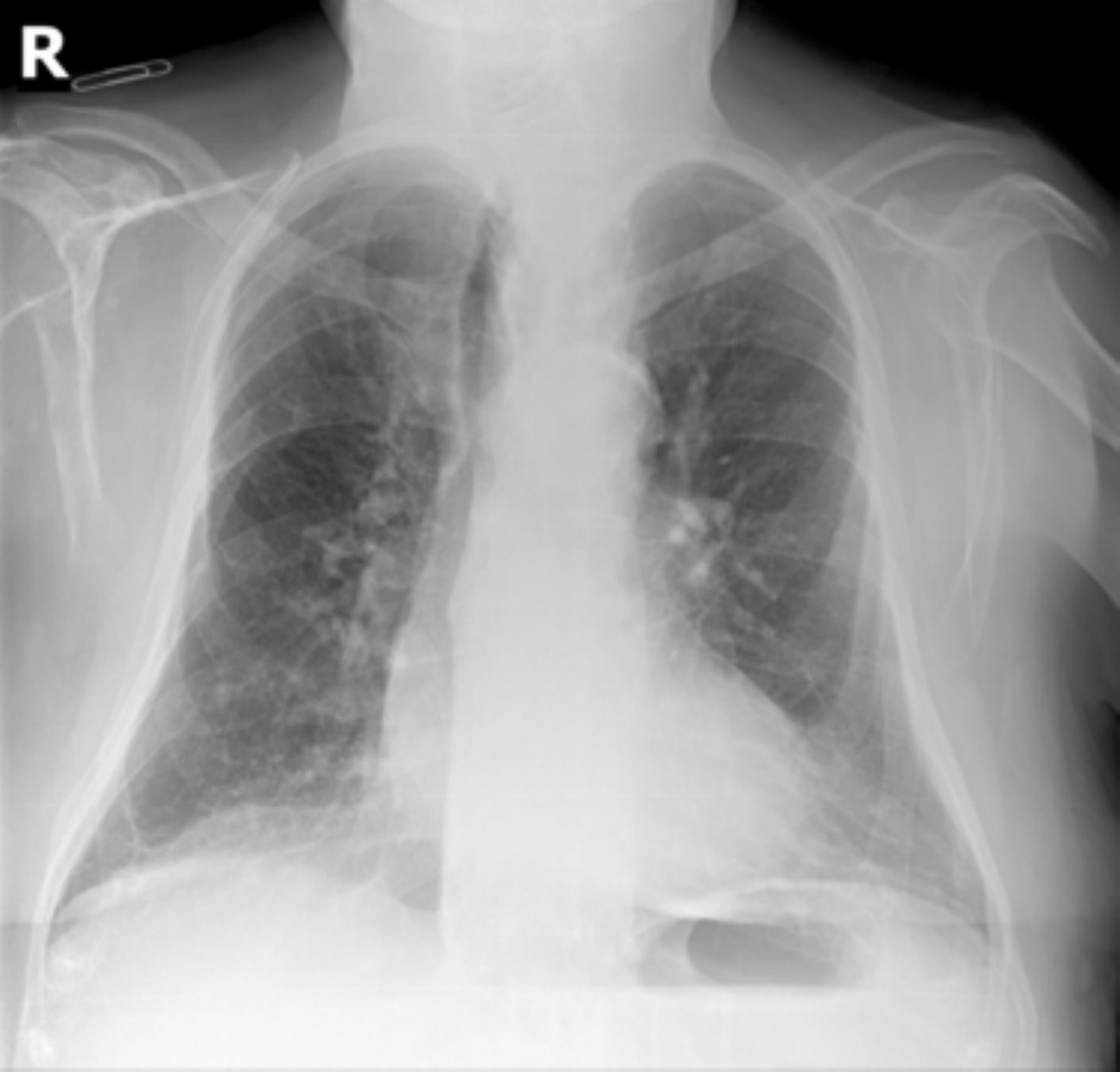}
  \caption{P-A}
  \label{fig:example_PA}
\end{subfigure}%
\begin{subfigure}{.5\columnwidth}
  \centering
  \includegraphics[width=.9\columnwidth]{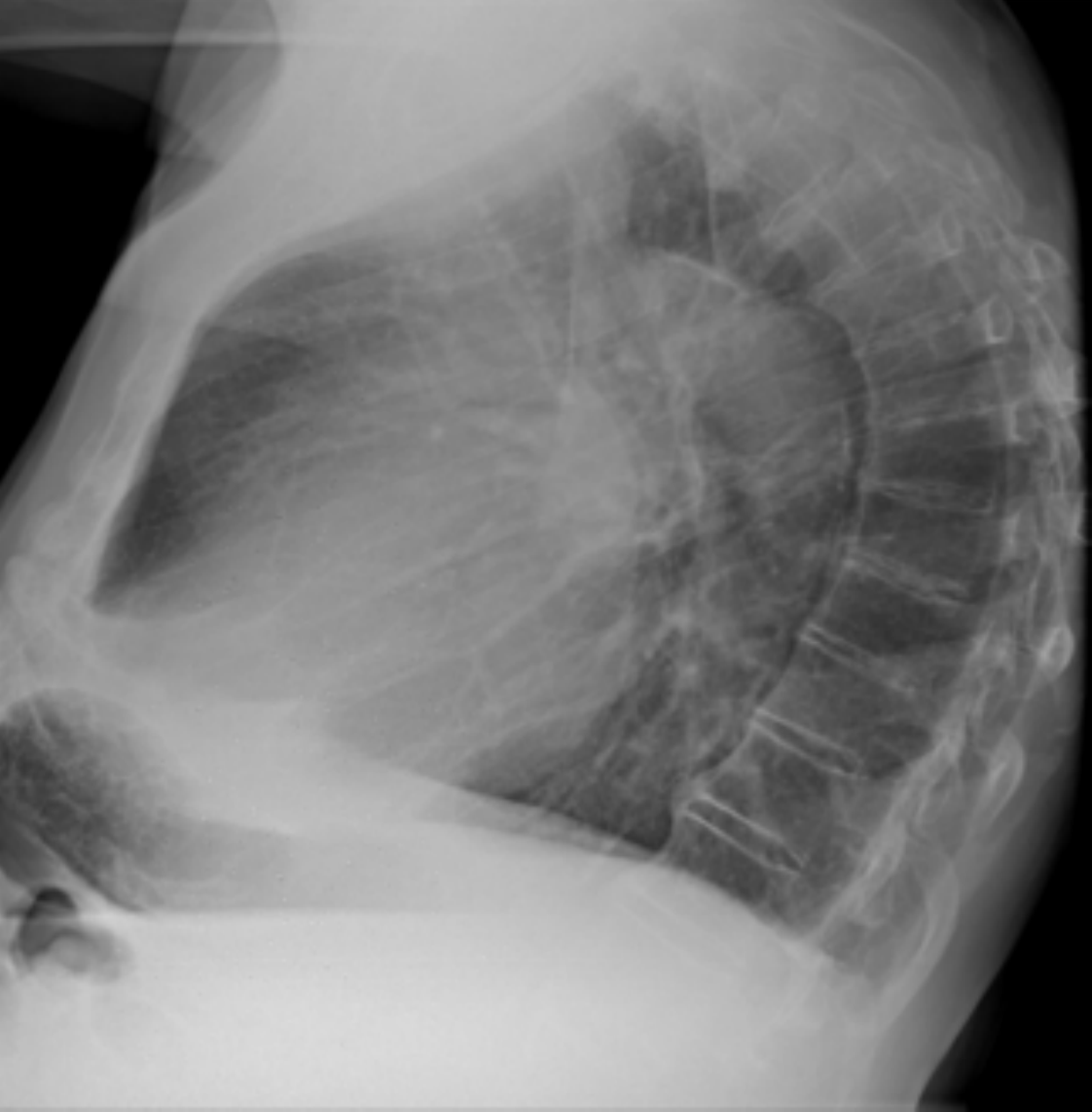}
  \caption{L}
  \label{fig:example_L}
\end{subfigure}%
\caption{Example x-ray images of the study with two projections which fields are given in Tab. \ref{tab:exampleStudy}.}
\label{fig:exampleStudy}
\end{figure}

Fig. \ref{fig:exampleStudy} shows an example of an study with two x-ray images. The associated fields in Tab. \ref{tab:exampleStudy} are shared for both projections included in the same study report. 

\begin{table*}
    \centering
    \begin{footnotesize}
    \begin{tabular}{l|p{8cm}}
         ImageID & 135803415504923515076821959678074435083\_fzis7b.png  \\
         ImageDir & 2 \\
         StudyDate\_DICOM & 2015-09-14 \\
         StudyID & 20150914, 135803415504923515076821959678074435083 \\
         PatientID & 313572750430997347502932654319389875966 \\
         PatientBirth & 1929 \\
         ReportID & 4991845 \\ 
         Report & cambi pulmonar cronic sever . sign fibrosis bibasal . sutil infiltr pseudonodul milimetr vidri deslustr localiz bas . cifosis sever . \\
         MethodLabel & Physician \\
         Labels & ['pulmonary fibrosis',  'chronic changes',  'kyphosis',  'pseudonodule',  'ground glass pattern'] \\
         LabelsLocalizationsBySentence &  [['pulmonary fibrosis',  'loc basal bilateral'],  ['chronic changes'],  ['kyphosis'],  ['pseudonodule',  'ground glass pattern',  'loc basal']] \\
         LabelCUIS & ['C0034069' 'C0742362' 'C2115817' 'C3544344'] \\
         LocalizationsCUIS & ['C1282378'] \\
         PatientSex\_DICOM & M \\
         ViewPosition\_DICOM & POSTEROANTERIOR \\
         Projection & PA \\
         MethodProjection & manual \\
         Modality\_DICOM & \\ 
         Manufacturer\_DICOM & ImagingDynamicsCompanyLtd \\
         PhotometricInterpretation\_DICOM &  MONOCHROME2  \\
         PixelRepresentation\_DICOM & \\
         PixelAspectRatio\_DICOM & \\ 
         SpatialResolution\_DICOM & \\  
         BitsStored\_DICOM & 12 \\
         WindowCenter\_DICOM & 2155 \\
         WindowWidth\_DICOM & 2880 \\
         Rows\_DICOM & 3572 \\
         Columns\_DICOM & 3732 \\
         XRayTubeCurrent\_DICOM & 320 \\
         Exposure\_DICOM & 3 \\ 
         ExposureInuAs\_DICOM & 3200 \\
         ExposureTime\_DICOM & 10 \\
         RelativeXRayExposure\_DICOM & \\ 

    \end{tabular}
    \end{footnotesize}
    \caption{Fields corresponding to the study which images are shown in Fig. \ref{fig:exampleStudy}.}
    \label{tab:exampleStudy}
\end{table*}










\end{document}